\documentclass[fleqn,usenatbib]{mnras}

\usepackage[export]{adjustbox}

\usepackage{pifont}
\newcommand{\xmark}{\ding{55}}

\usepackage{array, longtable}
\usepackage{graphicx}	% Including figure files
\usepackage{amsmath}	% Advanced maths commands
\usepackage{amssymb}	% Extra maths symbols
\usepackage{morefloats}
\title[Gas in debris disks]{Debris disks with multiple absorption features in metallic lines: circumstellar or interstellar origin?}

\author[D. Iglesias et al.]{
D. Iglesias,$^{1,2,3}$\thanks{E-mail: daniela.iglesias@postgrado.uv.cl}
A. Bayo,$^{1,3}$
J. Olofsson,$^{1,3}$
Z. Wahhaj,$^{2}$
C. Eiroa,$^{4,5}$
B. Montesinos,$^{6,5}$
\newauthor
I. Rebollido,$^{4,5}$
J. Smoker,$^{2}$
L. Sbordone,$^{2}$
M. R. Schreiber$^{1,3}$
and Th. Henning$^{7}$
\\
\\
$^{1}$Instituto  de  F\'isica  y  Astronom\'ia,  Facultad  de  Ciencias,  Universidad de Valpara\'iso, Av. Gran Breta\~na 1111, 5030 Casilla, Valpara\'iso, Chile\\
$^{2}$European Southern Observatory, Alonso de C\'ordova 3107, Vitacura, Santiago, Chile\\
$^{3}$N\'ucleo Milenio de Formaci\'on Planetaria - NPF, Universidad de Valpara\'iso, Av. Gran Breta\~na 1111, Valpara\'iso, Chile\\
$^{4}$Dpto. F\'isica Te\'orica, Universidad Aut\'onoma de Madrid, Spain\\
$^{5}$Unidad Asociada AstroUAM-CSIC\\
$^{6}$CAB (CSIC-INTA), P.O. Box 78, 28691 Villanueva de la Canada, Madrid, Spain \\
$^{7}$Max-Planck-Institut f\"ur Astronomie (MPIA), K\"onigstuhl 17, D-69117 Heidelberg, Germany \\
}

\date{Accepted XXX. Received YYY; in original form ZZZ}

\pubyear{2018}

\begin{document}
\label{firstpage}
\pagerange{\pageref{firstpage}--\pageref{lastpage}}
\maketitle

% Abstract of the paper
\begin{abstract}
Debris disks are second generation dusty disks thought to be devoid of gas. However, this idea has been challenged in the last years by gas detections in some systems. We compiled a database of 301 debris disks and collected high--resolution optical spectra for $\sim77\%$ of them. From the analysis of these data we identified a group of 23 debris disks presenting several absorption features superimposed to the photospheric  Ca\,{\sc ii} and Na\,{\sc i} doublets. These absorptions could be due to circumstellar material or interstellar clouds. In order to discriminate between the two scenarios, we characterized each feature in terms of its radial velocity, equivalent width and column density. Additionally, we searched in the literature for local clouds in the line of sight of the stars, and looked for the presence of similar absorption features in nearby stars. Our study concludes that while all the objects present interstellar absorptions in their spectra, three objects show features more compatible with circumstellar origin: HD\,110058 presents a stable circumstellar absorption, while HR\,4796 and c Aql present variable absorption features likely due to exocometary activity. The minute-scale variability we detect towards c Aql is the shortest of this kind detected so far. The detection of circumstellar features in these objects is consistent with their near edge-on inclinations. We also provide evidence challenging previous claims of circumstellar gas detections for HR\, 6507. Given the properties of the sample, we speculate that transient gaseous events must be a common phenomenon among debris disks.

\end{abstract}

% Select between one and six entries from the list of approved keywords.
% Don't make up new ones.

%%%%%%%%%%%%%%%%%%%%%%%%%%%%%%%%%%%%%%%%%%%%%%%%%%

%%%%%%%%%%%%%%%%% BODY OF PAPER %%%%%%%%%%%%%%%%%%

\section{Introduction}
\label{sec:intro}

Planets are believed to form in protoplanetary disks. The mass of these disks is initially composed of 99\% gas and 1\% dust (ISM-like ratios, \citealt{Bohlin78}; but see, among others, \citealt{Williams14}). After a few Myrs (e.g. \citealt{Hernandez07} and \citealt{Fedele10}) these protoplanetary disks evolve from optically thick gas-rich systems into transition disks, and later, at a stellar age of about 10 Myrs, they are transformed into a collection of rocks, dust, planetesimals (and maybe gaseous giant planets) known as debris disks. At this stage, the disk is supposed to be fully depleted of gas due to gas removal processes such as photoevaporation, accretion and radiation pressure (e.g. \citealt{Pontoppidan14} and \citealt{Alexander2006} for reviews on volatiles and photoevaporation in protoplanetary disks, respectively). Therefore, the current paradigm is that the majority of debris disks do not harbour gas and contain very little second generation dust produced by collisions among planetesimals; gas giants would have to form during the earlier gaseous stage of the disk; and rocky planets can form or continue to grow later during the gas-poor phase of the disk \citep{Wyatt2008}. 

The idea that debris disks should be gas free has been challenged in the past few years by the discovery of a number of debris disks containing some gas detected either in the far-infrared (FIR), infrared (IR), optical or UV wavelengths. These gaseous debris disks have been found mainly around young A type stars, like the well-studied $\beta$ Pictoris \citep{Brandeker2004}, 49 Ceti \citep{Roberge2014}, HD\,32297 \citep{Redfield2007}, HD\,172555 \citep{Riviere-Marichalar2012}, HD\,21997 \citep{Moor2011}, a few B type stars like $\sigma$ Her \citep{Chen&Jura2003} and 51 Oph \citep{Thi2013}, and very few F type stars, HD\,181327 and $\eta$ Corvi (\citealt{Marino2016}, \citealt{Marino2017}). Several tracers have been used to this end, for instance CO, C and O emissions at mm wavelengths (\citealt{Greaves2016}, \citealt{Kral2017}), Ca\,{\sc ii} and Na\,{\sc i} absorptions in the optical \citep{Kiefer2014twoAtype} and different C species (C\,{\sc i}, C\,{\sc ii}) and Fe I absorptions in the far UV, among others (\citealt{Brandeker2004}, \citealt{Roberge2006}, \citealt{Roberge2014}). 

The possible origin of the gas detected in debris disks has been widely discussed (\citealt{Moor2011}, \citealt{Wyatt2015}, \citealt{Kospal2016}, \citealt{Kral2017review}). In short, it could be residual gas that remained from the earlier gaseous stage of the disk, which would imply that the efficiency of gas removal processes may be lower than we thought. Or, it could be second generation gas, produced by icy comets that, either orbiting or as they approach the star, begin to ``evaporate" (or, more correctly, sublimate) and release small amounts of gas. Gas of secondary origin could also be produced by collisions among volatile-rich dust grains or comet-like bodies (\citealt{Higuchi2017} and references therein) or even by photon-stimulated desorption of solids \citep{Matthews2014}. These latter two processes could replenish the disk with a stable gaseous component likely to be located in the outer regions of the disk \citep{Brandeker2004}. On the other hand, the idea of the ``falling evaporating bodies" (FEBs) has been gaining more acceptance in the last few years, since FEB-like events have been detected (mostly) around A-type stars with debris disks and some shell stars (e.g., \citealt{Beust1998}, \citealt{Kiefer2014HD172555}, \citealt{Eiroa16}). These events manifest as stochastic absorption features usually at redshifted velocities with respect to the radial velocity of the star. Variable absorption features have been detected over short time windows of hours or night to night as well as over months or years (\citealt{Barnes2000}, \citealt{Thebault2001}, \citealt{Welsh2013}). One interesting by-product of studying FEBs is posing the question on the cause of such instabilities in the debris disk. A possible cause for such instabilities could be the presence of a larger body like a planetesimal or a planet interfering with the dust transport and evolution of the disk \citep{Beust1998}. %. Therefore, detecting FEB-like events could be seen as an indirect hint for the presence of at least one giant planet embedded in the debris disk system \citep{Beust1998}. 

Independently of its origin, the implications of the presence of gas in debris disks are many (see, for a recent review, \citealt{Hughes2018}). It can change our understanding of gas removal processes by setting new constraints on their efficiency \citep{WilliamsCieza2011}. Particularly for photoevaporation, thought to be the main cause of gas removal in protoplanetary disks (\citealt{Alexander2006}, \citealt{Canovas2017}). In addition, gas can influence the morphology of the dust in the disk providing us with a possible answer to the formation of the observed gaps in some debris disks \citep{LyraKuchner2013}. Gas can also imprint changes in the dynamics of the system since even small amounts of gas can drag dust and pebbles \citep{Wyatt2008}; and dust grains can couple to the gas component which acts as a fluid affecting rocky planet formation processes (\citealt{Fernandez2006}, \citealt{Cleeves2016}, \citealt{Kenyon2016}). Since the presence of gas can have a strong impact on the formation and evolution of planetary systems it is essential to understand its frequency and how gas detections relate to properties of debris disk systems, such as age, multiplicity, stellar type, metallicity, dust content and disk-planet interaction.

We are currently analysing a robust sample of 301 debris disk (Olofsson et al. in prep.) to learn what percentage of debris disks contain gas, how the gas is physically related to the dust and what properties characterize the stars that possess circumstellar gas in their surrounding debris disks. Instruments such as ALMA and APEX can provide us with a plethora of information about disks besides being able to detect gas emission. However, observing such a large sample with either facility would be extremely expensive in terms of telescope time. In comparison, the analysis of UV-optical ground-based spectra provides a very efficient way to find debris disks with gas to be followed-up with other instruments (\citealt{Montgomery2012}, \citealt{Welsh2013}, \citealt{Kiefer2014twoAtype}). Just as a simple illustration, integration times of CO surveys with ALMA or APEX are $\sim$1 hour per target (\citealt{Kospal2013}, \citealt{Hales2014}), while optical high-resolution high-signal-to-noise-ratio spectroscopic observations of similar targets conforming our sample take only a few minutes or even seconds (in class 2 to 8 meter telescopes). Thus the number of spectra taken during one night ranges from $\sim$50 to $\sim$100, depending on the telescope/instrument and the targets.

In order to optimize the search for circumstellar gas in our sample, we have used the method described in \cite{Kiefer2014HD172555}, where the presence of gas in the line of sight of debris disks with near edge-on inclinations, is inferred by the identification of (narrow) extra absorption(s) in metallic lines in the UV/optical regime (Ca\,{\sc ii} H \& K and Na\,{\sc i} D1 and D2 being particularly good tracers). %A very narrow absorption feature is observed superimposed on the photospheric line in the spectrum, indicating that some gas is passing in the line of sight of the star. 
So far we have been able to collect data for about 77\% of our sample and analysing those data, we have found a few particularly interesting objects with multiple gas features which are presented in this paper.

%One key aspect of this study is to distinguish circumstellar gas absorption features from interstellar ones. They both appear as narrow absorption features superimposed on photospheric lines such as Ca\,{\sc ii} H \& K and Na\,{\sc i} D1 \& D2 and they both can be detected at very similar radial velocities. One good discriminator between circumstellar and interstellar gas is that the interstellar medium (ISM) is not expected to present short term variability. Only minor changes have been detected in ISM over time spans of years but not within days or months \citep{McEvoy2015}. Interstellar clouds are more common at larger distances; stars are more likely to have several clouds crossing the line of sight at distances over 500 parsecs, however, some local clouds are found within distances of 15 pc of the Sun \citep{Redfield08}. Typical values for column density ratios N(Ca{\sc ii})/N(Na{\sc i}) of the ISM range between 1--10 for distances within 80 parsecs and 0.05--2 for more distant lines of sight \citep{Welsh2010}. These properties must be analysed in order to determine the origin of gas absorption features. 
%This paper is organized as follows. We present our debris disks sample in Sec. \ref{sec:sample}, describe the data we have used for this study in Sec. \ref{sec:dataobs}, detail all the different analyses performed on the objects in Sec. \ref{sec:methods}, outline the results for each particular analysis in Sec. \ref{sec:results}, discuss these results in Sec. \ref{sec:discussion} and report our conclusions in Sec. \ref{sec:conclusions}.

\section{Debris disk sample}
\label{sec:sample}

In this paper, we present a sub-sample of 23 gas-rich debris disk candidates characterized by showing gas detections at different radial velocities within our database of observations prior to December 2016. The full sample of debris disks we are observing consists of 301 systems selected from an original list compiled by Olofsson et al. (in prep). The original list, was assembled via a thorough literature search for debris disks that had been observed with the IRS instrument \citep{Houck2004} on board of the Spitzer space telescope \citep{Werner2004}. This search resulted in $\sim$ 500 objects that were then filtered down to 301 imposing different criteria on the significance of the excess in the mid-infrared and excluding debris disks that display strong emission features in their IRS spectra (e.g., \citealp{Olofsson2012}), as these objects are not really representative of ``classical'' debris disks. It follows then that the selection criteria applied to achieve our database of 301 debris disks are unbiased with respect to disk inclinations as the disks are optically thin and, since most inclinations are unknown, they can be assumed to be following a uniform distribution.
%All objects from the original list ($\sim$500) had been observed with the IRS instrument \citep{Houck2004} on board of the Spitzer space telescope \citep{Werner2004}. What we refer as the full sample, was selected based on a thorough literature search for debris disks. One should note that we excluded debris disks that display strong emission features in their Spitzer IRS spectra (e.g., \citealp{Olofsson2012}), as these objects are not really representative of ``classical'' debris disks. 

The ages of the systems in the full sample are mostly within the range 10--100 Myrs and most of them are located within distances of less than 200 pc. We are in the process of collecting high-resolution, high-signal-to-noise spectroscopic archival data and obtaining new observations of these objects to analyse the sample in the wavelength ranges covering the Ca\,{\sc ii} H and K lines at 3968.5 and 3933.7\AA, and the Na\,{\sc i} D1 and D2 lines at 5895.9 and 5889.9\AA. We have been able to collect adequate data for 234 objects in our sample, with $\sim$55\% of the selected objects coming from our own observations and the rest from the ESO archive. As mentioned before, the sub-sample presented here was extracted from our database of observations updated up to December 2016.

The main properties of the sub-sample are detailed in Tables \ref{tab:sample} and \ref{tab:info}. Unless otherwise indicated, most of the stellar parameters reported in Table\,\ref{tab:info} were retrieved from the Simbad database \citep{Wenger2000}. In particular, distances came from the Tycho-Gaia Astrometric Solution (TGAS) from the Gaia data-release 1 \citep{Gaia2016} or from the Hipparcos new reduction  of \protect\cite{VanLeeuwen2007} when the former were not available. Luminosities are estimated via SED fitting (assuming the previously mentioned distances) with Kurucz models \citep{Castelli1997} using VOSA \citep{Bayo2008}. Isochronal ages are estimated with VOSA based on the SED fitted parameters and different sets of isochrones (\citealt{Siess2000}, \citealt{Baraffe1998}). The multiplicity column highlights objects reported in the literature to be multiple systems. Note that the large uncertainties in isochronal ages are attributable to uncertainties in the distance to the objects and/or the set of isochrones assumed for the estimates (further discussion on the ages of the sample will be given in Olofsson at al. in prep.). For those objects confirmed to belong to young moving groups (i.e. $\beta$03\,Tuc, 66\,Psc, $\nu$ Hor, HD\,24966, HD\,54341, $\eta$ Cha, HD\,110058 and HR\,6507), we have adopted the literature age commonly assigned to those moving groups (in principle more precise than isochronal dating). 
%The sample was hand-picked considering several criteria to ensure that these objects are in fact classical debris disks. The selection criteria are unbiased with respect to disk inclinations as the disks are optically thin and, since most inclinations are unknown, they can be assumed as random. The sample does not contain objects with any emission features from warm silicate grains which should have been dispersed given the age of the star.
 
 \begin{table}
	\centering
	\caption{Sub-sample of gas-rich debris disk candidates with multiple absorption features and their respective coordinates. }
	\label{tab:sample}
	\begin{tabular}{lccc} % four columns, alignment for each
		\hline
		Name & HD Id & R.A. & Dec. \\
		 &  & [J2000] & [J2000]  \\
		\hline
		$\beta$03\,Tuc & HD\,3003 &  00:32:43.9 &  -63:01:53.4  \\
		66\,Psc & HD\,5267 & 00:54:35.2 & +19:11:18.3  \\
		$\nu$ Hor & HD\,17848 & 02:49:01.5 & -62:48:23.5 \\
		HD\,24966 & HD\,24966 & 03:56:29.4 & -38:57:43.8 \\
		HD\,290540 & HD\,290540 & 05:31:31.4 & -01:49:33.3 \\
		HD\,36444 & HD\,36444 & 05:31:40.5 & -01:07:33.3 \\
		HD\,290609 & HD\,290609 & 05:33:05.6 & -01:43:15.5 \\
		HR\,1919  & HD\,37306 & 05:37:08.8 & -11:46:31.9  \\
		HD\,54341 & HD\,54341  & 07:06:20.9 & -43:36:38.7 \\
		HD\,60856 & HD\,60856 & 07:35:56.9 & -14:42:39.0 \\
		HR 3300 & HD\,71043 & 08:22:55.2 & -52:07:25.4 \\
		$\eta$ Cha & HD\,75416 & 08:41:19.5 & -78:57:48.1 \\
		HD\,92536 & HD\,92536 & 10:39:22.8 & -64:06:42.4 \\
		3 Crv & HD\,105850 & 12:11:03.8 & -23:36:08.7 \\
		HD\,106036 & HD\,106036 & 12:12:10.3 & -63:27:14.8 \\
		HR\,4796  & HD\,109573 & 12:36:01.0 & -39:52:10.2  \\
		HD\,110058 & HD\,110058 & 12:39:46.2 & -49:11:55.5 \\
		HD\,112810 & HD\,112810 & 12:59:59.9 & -50:23:22.5 \\
		HD\,126135 & HD\,126135 & 14:24:43.9 & -40:45:18.6 \\
		HD\,141378 & HD\,141378 & 15:48:56.8 & -03:49:06.6 \\
		HD\,141327 & HD\,141327 & 15:49:43.1 & -32:48:29.8 \\
%		HD\,144981 & HD\,144981 & 16:09:20.9 & -19:27:25.9 \\
%		HD\,145554 & HD\,145554 & 16:12:21.8 & -19:34:44.6 \\
%		HD\,145631 & HD\,145631 & 16:12:44.1 & -19:30:10.3 \\
%		HR\,6051 & HD\,145964 & 16:14:28.9 & -21:06:27.5 \\
		HR\,6507 & HD\,158352 & 17:28:49.7 & +00:19:50.3 \\
		cAql & HD\,183324 & 19:29:00.9 & +01:57:01.6 \\
		\hline
	\end{tabular}
\end{table}

\begin{table*}
	\centering
	\caption{Stellar parameters for the sample. 
	%Most parameters (unless otherwise indicated) were retrieved from the Simbad database \citep{Wenger2000}. In particular, distances came from the Tycho-Gaia Astrometric Solution (TGAS) from the Gaia data-release 1 \citep{Gaia2016} or from the Hipparcos new reduction  of \protect\cite{VanLeeuwen2007} when the former were not available. Luminosities are estimated via SED fitting (assuming the previously mentioned distances) with Kurucz models \citep{Castelli1997} using VOSA \citep{Bayo2008}. Isochronal ages are estimated with VOSA based on the SED fitted parameters and different sets of isochrones (\citealt{Siess2000}, \citealt{Baraffe1998}). The multiplicity column highlights objects reported in the literature to be multiple systems.
	}
	\label{tab:info}
	\begin{tabular}{lccccccccc} % four columns, alignment for each
		\hline
		Name & $v$sin$i$ & Spectral & radV & distance & Isochronal Age & Literature Age & $\log(L_{bol}(L_\odot))$ & Multi-\\
		     & [km.s$^{-1}$] & Type & [km.s$^{-1}$] & [pc] & [Myr] & [Myr] & [dex] & plicity\\
		\hline
		$\beta$03\,Tuc &  93 & A0V & 7.70$\pm$0.80 & 45.56$\pm$0.394$^{b}$ & 115.48$\pm^{137.30}_{115.48}$ & 30.0$^{d}\pm^{0.0}_{20.0}$  & 1.19$\pm_{0.01}^{0.01}$  &	\checkmark$^{j}$ \\
		66\,Psc  & 144 & A1Vn & 8.50$\pm$2.80 & 108.11$\pm$7.48$^{b}$ & 5.00$\pm^{0.98}_{0.31}$ & 200$^{e}$ & 1.46$\pm_{0.07}^{0.06}$  & \checkmark$^{k}$ \\
		$\nu$ Hor  & 143.7 & A2V & 30.90$\pm$2.00 & 52.13$\pm$1.76$^{c}$ & 529.02$\pm^{103.10}_{129.08}$ & 100$^{e}$ & 1.23$\pm_{0.05}^{0.04}$ & --\\
		HD\,24966  & -- & A0V & -- & 105.82$\pm$4.03$^{b}$ & 195.41$\pm_{195.41}^{102.01}$ & 10$^{e}$ & 1.13$\pm_{0.04}^{0.04}$ & --\\
		HD\,290540  & -- & A2 & -- & 357.32$\pm$54.90$^{c}$ & 11.57$\pm_{11.57}^{338.74}$ & 112$^{f}$ & 1.23$\pm_{0.17}^{0.12}$ & --\\
		HD\,36444  & -- & B9V & -- & 458.20$\pm$98.20$^{c}$ & 4.33$\pm_{1.33}^{603.23}$ & 101$^{f}$ & 1.57$\pm_{0.26}^{0.16}$ & --\\
		HD\,290609$^{1}$  & -- & A0 & -- & 23.86$\pm$11.43$^{b}$ & -- & 5$^{g}$ & 2.41$\pm_{0.00}^{0.44}$ & --\\
		HR\,1919   & 148.1 & A1V & 23.00$\pm$0.70 & 70.76$\pm$4.15$^{c}$ & 28.52$\pm_{28.52}^{307.72}$ & 453$^{f}$ & 1.10$\pm_{0.07}^{0.06}$ & -- \\
		HD\,54341  & -- & A0V & -- & 102.35$\pm$3.77$^{b}$ & 7.94$\pm_{0.97}^{391.30}$ & 10$^{e}$ & 1.34$\pm_{0.05}^{0.05}$ & --\\
		HD\,60856 & 44 & B5V & 31.20$\pm$1.90 & 363.83$\pm$88.54$^{c}$ & 2.19$\pm_{0.20}^{595.55}$ & 196$^{f}$ & 1.84$\pm_{0.29}^{0.17}$ & --\\
		HR 3300  & 224 & A0V & 22.50$\pm$1.10 & 70.03$\pm$1.13$^{b}$ & 710.26$\pm_{701.81}^{111.03}$ &  404$^{f}$ & 1.11$\pm_{0.02}^{0.02}$  & --\\
		$\eta$ Cha  & 296$^{a}$ & B8V & 14.00$\pm$7.40 & 94.97$\pm$1.44$^{b}$ & 3.09$\pm_{0.09}^{297.86}$ & 6.0$^{d}\pm^{1.0}_{0.0}$ & 1.80$\pm_{0.03}^{0.03}$ & --\\
		HD\,92536 & -- & B8V & 10.00$\pm$1.00 & 145.13$\pm$8.75$^{c}$ & 4.04$\pm_{0.90}^{0.49}$ & 231$^{f}$ & 1.66$\pm_{0.08}^{0.07}$ & --\\
		3 Crv  & 126.8 & A1V & 11.00$\pm$4.20 & 58.82$\pm$1.94$^{b}$ & 907.37$\pm_{899.21}^{92.96}$ &  465$^{f}$ & 1.14$\pm_{0.05}^{0.04}$ & --\\
		HD\,106036  & -- & A2V & 7.70$\pm$1.30 & 99.00$\pm$3.87$^{c}$ & 463.25$\pm_{463.25}^{238.95}$ & 17$^{h}$ & 0.98$\pm_{0.05}^{0.05}$ & --\\
		HR\,4796   &  152.0 & A0V & 7.10$\pm$1.10 & 72.78$\pm$1.75$^{b}$ & 69.59$\pm_{69.59}^{30.88}$ & 378$^{f}$ & 1.56$\pm_{0.05}^{0.05}$ & \checkmark$^{l}$ \\
		HD\,110058  & -- & A0V & 5.00$\pm$1.20 & 188.76$\pm$34.11$^{c}$ & 560.29$\pm_{560.29}^{48.08}$ & 10$^{e}$ & 1.24$\pm_{0.34}^{0.19}$ & --\\
		HD\,112810  & 82 & F3/5IV/V & 4.20$\pm$1.20 & 134.60$\pm$7.22$^{c}$ & 1997.73$\pm_{1997.73}^{1012.17}$ & 10$^{i}$ & 0.46$\pm_{0.05}^{0.05}$  & --\\
		HD\,126135  & -- & B8V & 12.00$\pm$6.00 & 165.02$\pm$16.34$^{b}$ & 5.00$\pm_{0.80}^{692.27}$ & 103$^{f}$ & 1.44$\pm_{0.08}^{0.10}$ & --\\
		HD\,141378 & 107 & A5IV-V & -16.40$\pm$2.00 & 55.54$\pm$2.32$^{c}$ & 10.05$\pm_{0.15}^{0.65}$ & 587$^{f}$  & 0.99$\pm_{0.04}^{0.04}$ & --\\
		HD\,141327  & -- & B9V & -5.10$\pm$2.40 & 213.29$\pm$22.48$^{c}$ & 4.88$\pm_{0.88}^{794.54}$ & 196$^{f}$ & 1.46$\pm_{0.12}^{0.09}$ & --\\
%		HD\,144981  & 155.20 & A0V & -1.30$\pm$1.90 & 148.05$\pm$10.17$^{c}$ & 8.69$\pm_{1.53}^{1.17}$ & -- & 1.11$\pm_{0.07}^{0.08}$ & --\\
%		HD\,145554  & 340 & B9V & -6.15$\pm$2.57 & 153.85$\pm$11.26$^{c}$ & 4.00$\pm_{4.00}^{76.72}$ & -- & 1.99$\pm_{0.13}^{0.10}$ & --\\
%		HD\,145631  & 287.00 & B9V & -5.56$\pm$0.21 & 144.39$\pm$8.88$^{c}$ & 4.02$\pm_{4.02}^{93.26}$ & -- & 1.90$\pm_{0.08}^{0.07}$ & --\\
%		HR\,6051 & 306 & B9V & -7.80$\pm$1.70 & 108.70$\pm$6.85$^{b}$ & 7.16$\pm_{1.59}^{737.63}$ & -- & 1.30$\pm_{0.06}^{0.06}$ & --\\
		HR\,6507  & 180 & A8Vp & -36.10$\pm$2.00 & 59.63$\pm$0.93$^{b}$ & 7.83$\pm_{0.72}^{0.22}$ & 600$^{e}$ & 1.17$\pm_{0.01}^{0.01}$ & --\\
		cAql  & 110 & A0IVp & 12.00$\pm$4.30 & 61.20$\pm$1.35$^{b}$ & 60.55$\pm_{60.55}^{0.00}$ & 506$^{f}$ & 1.29$\pm_{0.06}^{0.05}$ & --\\
		\hline
	\end{tabular}
	\begin{description}
	  \small
      \item $^{a}$\cite{Zorec2012}, $^{b}$\cite{VanLeeuwen2007}, $^{c}$TGAS, $^{d}$Torres et al (in prep.), $^{e}$\cite{Rhee07}, $^{f}$\cite{Gontcharov12}, $^{g}$\cite{Hernandez2006}, $^{h}$\cite{Mittal2015}, $^{i}$\cite{Ballering13}, $^{j}$\cite{Dommanget2002}, $^{k}$\cite{DocoboLing2007}, $^{l}$\cite{Jura1993}.
      \item $^{1}$Distances reported for this object range from $\sim$24 pc \citep{VanLeeuwen2007} to $\sim$775 pc \citep{Kharchenko2001}; both of those estimates report huge uncertainties (above 50\%), and propagate to unrealistic luminosities and thus isochronal ages. For this reason we do not include this object in any comparative analysis that involves age, luminosity and/or distance.
	\end{description}
\end{table*}

As mentioned before, the 23 candidates presented in this paper have been chosen because they display particularly interesting gas absorption features: they all have not only one, but multiple absorptions at different radial velocities with respect to the star within the 12 years of baseline considered for this paper. Merely by statistical arguments, this multiplicity increases the chances of a circumstellar gas detection, and for the cases that a circumstellar origin for several simultaneous features could be confirmed, it would imply a very interesting disk configuration and/or geometry. For instance, a disk containing several gas rings at different distances from the star or different populations of exocomets, which would require further study and possible follow-up with high angular resolution instruments like ALMA.% for a better characterization of the morphology of the disk and the co-location of the gas with respect to the dust. 

\begin{table*}
	\centering
	\caption{Number of spectra for each star per date, instrument, ESO observing period and dates of observations. Observations from our programmes are flagged with a *.}
	\label{tab:obs}
	\begin{tabular}{lcccc} % four columns, alignment for each
		\hline
		Name & Number of Spectra & Instrument & Period & Observation dates\\
		\hline
		$\beta$03\,Tuc & 6, 4, 2, 2, 2 & HARPS  & P73, P75, P77, P77, P77 & 2004-09-30, 2005-08-19, 2006-05-[20, 25, 26] \\
		& 2, 2, 2, 2, 2, 2 & HARPS  & P84, P84, P84, P84, P84, P84 & 2009-11-[12, 13, 14, 15], 2009-12-[05, 08] \\
		 & 2, 2, 2, 2 & HARPS  & P85, P86, P86, P87 &  2010-07-07, 2011-01-[06, 07], 2011-07-23 \\
		 & 1, 1 & FEROS & P96, P96 &  2015-10-[23, 24]*\\
		66\,Psc & 1, 2 & UVES & P96, P97 &  2015-11-14*, 2016-07-23* \\ 
		$\nu$ Hor & 6, 2, 2  & HARPS & P75, P75, P77  &   2005-08-19, 2005-09-09, 2006-09-11 \\
		& 2, 2, 2, 6, 2 & HARPS & P80, P80, P80, P94, P94  &  2007-12-[05, 06, 10], 2015-01-[18, 20] \\
		HD\,24966 & 1 & FEROS & P96 &  2016-01-04*  \\
		 & 2, 2, 2 & UVES  & P97, P97, P97 &  2016-07-23*, 2016-08-[21, 29]* \\
		HD\,290540 & 1, 1 & FEROS & P96, P96 &  2016-01-[03, 04]* \\
		HD\,36444 & 1 & FEROS & P96 &  2016-01-04*\\
		HD\,290609 & 1 & FEROS & P96 &  2016-01-04* \\
		HR\,1919  & 4, 2, 2, 2, 2, 2 & HARPS & P76, P76, P76, P76, P76, P76 &  2006-02-[08, 09, 10, 11, 13], 2006-03-12  \\
		& 2, 2, 2, 2 & HARPS & P78, P80, P80, P80 &  2006-11-18, 2007-12-[05, 06, 10]  \\
		 & 2, 1, 1 & FEROS & P96, P96, P96 &  2015-10-23*, 2016-03-[28, 29]* \\
		HD\,54341 & 3, 1 & HARPS & P94, P94 &  2015-01-[19, 20]  \\
		HD\,60856 & 1 & FEROS & P96 &  2016-01-04* \\ 
		HR 3300 & 6, 2, 2 & HARPS & P94, P4, 94 &   2015-01-[18, 20, 21] \\
		& 1, 2, 3, 3 & FEROS  & P82, P82, P82, P88 &  2008-11-[18, 21, 22], 2011-12-07 \\
		 & 1, 1 & FEROS  & P96, P96 & 2015-10-23*, 2016-03-26* \\
		$\eta$ Cha & 2, 3 & UVES  & P66, P66  &  2001-02-[16, 18] \\
		 & 3, 6, 2 & HARPS  & P60, P94, P94 &  2005-02-12, 2015-01-[19, 20] \\
		 & 2, 2, 5 & FEROS & P60, P60, P60  & 2006-10-24, 2006-12-08, 2007-01-01 \\
		 & 4, 4, 4 & FEROS & P60, P60, P60  & 2007-2-[16, 23, 24] \\
		 & 1, 1, 1 & FEROS & P84, P84, P84  &  2009-11-30, 2009-12-[05, 07] \\
		 & 1, 1, 1, 1, 1, 1 & FEROS & P84, P84, P84, P84, P85, P85  &  2010-01-[27, 28, 29, 30], 2010-06-[01, 14] \\
		HD\,92536 & 3, 3, 2, 1 & HARPS & P94, P94, P94, P94 &   2015-01-[18, 19, 20, 21]\\  
		3 Crv & 6, 2, 2, 2, 2 & HARPS & P77, P77, P77, P80, P84 &  2006-05-[20, 21, 25], 2007-12-10, 2009-12-05\\
		& 4, 2, 6, 2 & HARPS & P85, P85, P86, P86 &  2010-07-[08, 09], 2011-01-[04, 07]\\
		  & 2, 2, 2, 2 & HARPS & P86, P86, P86, P87 &  2011-02-[4, 5, 6], 2011-07-21\\
		  & 1, 1 & FEROS & P96, P96 &  2016-03-[26, 29]*\\
		HD\,106036 & 2 & UVES & P96 &   2016-02-21* \\
		HR\,4796  &  74, 152, 76, 76, 92, 16 & UVES & P68, P79, P79, P79, P79, P79  &  2002-01-19, 2007-05-[07, 08, 13, 14, 15] \\ 
		 & 2, 3, 2, 3 & HARPS & P80, P94, P94, P94 &  2007-12-06, 2015-01-[19, 20, 21] \\
		 & 1, 1, 2, 1, 1 & FEROS & P79, P79, P79, P79, P79 &  2007-03-09, 2007-05-[01, 04, 12, 27] \\
		 & 2, 3, 4, 3 & FEROS & P96, P96, P96, P96 &  2016-03-[26, 27, 28, 29]* \\
		HD\,110058 & 2, 1 & FEROS & P84, P87  & 2010-01-31, 2011-04-16 \\
		& 6, 2, 2, 2 & HARPS & P94, P94, P94, P94 &   2015-01-[18, 19, 20, 21] \\		
		HD\,112810 & 2 & UVES & P96 &   2016-03-23* \\
		HD\,126135 & 1, 1, 2 & HARPS & P94, P94, P94 &  2015-01-[18, 20, 21]  \\
		HD\,141378 & 1 & FEROS & P92 &  2014-02-18 \\
		HD\,141327 & 1, 1, 2 & FEROS & P79, P81, P82 & 2007-04-02, 2008-04-01, 2009-02-06 \\
%		HD\,144981 & 1 & FEROS & P96 &   2016-03-28* \\
%		HD\,145554 & 1 & FEROS & P96 &  2016-03-29*  \\
%		HD\,145631 & 1 & FEROS & P96 &  2016-03-29*  \\
%		HR\,6051 & 1, 1, 1, 1 & FEROS & P96, P96, P96, P96 &  2016-03-[26, 27, 28, 29]*  \\
		HR\,6507 & 4, 2, 2, 2 & HARPS & P75, P77, P77, P77 &  2005-08-20, 2006-05-[20, 25], 2006-09-12 \\
		& 2, 5, 2, 2 & HARPS & P80, P80, P84, P84 &   2008-03-[17, 21], 2010-07-[08, 09] \\
		 & 6 & UVES  & P83 &  2009-04-16 \\
		c Aql & 10, 38  & UVES & P79, P87 &  2007-06-30, 2011-05-27 \\
		& 1, 1, 1 & FEROS & P83, P83, P83 &  2009-06-[02, 03], 2009-08-24 \\ 
		& 1, 1, 1 & FEROS & P85, P85, P85 &  2010-07-22, 2010-08-[23, 31]\\ 
		\hline
	\end{tabular}
\end{table*}

\section{Spectroscopic data}
\label{sec:dataobs}

We performed observations with FEROS \citep{1999Msngr..95....8K} on the MPG/ESO 2.2m telescope at the La Silla Observatory in Chile and UVES \citep{2000SPIE.4008..534D} on the VLT UT2 telescope at Paranal Observatory, Chile. We also queried the ESO archive searching for all relevant high-resolution spectra covering the blue-optical wavelength ranges. In particular data from HARPS \citep{2003Msngr.114...20M}, UVES and FEROS instruments were searched for\footnote{The list of programme IDs can be found in the Acknowledgements.}. 

\subsection{New data}
\label{sec:obs}

\subsubsection{FEROS observations}
\label{sec:ferosobs}

We performed observations with the FEROS \'Echelle spectrograph for the objects listed in Table \ref{tab:obs}\footnote{Programme IDs: 094.A-9012(A), 096.A-9018(A) and 099.A-9004(A)}. %HR\,1919, HD\,290540, 3 Crv, $\beta$03\,Tuc, HR\,6051, HD\,24966, HD\,144981, HD\,145964, HD\,144981, HD\,145554 and HD\,145631. 
The instrument choice is motivated by its characteristics: 
%high efficiency ($\sim$20\%), 
large wavelength range (the complete optical spectral region from $\sim$3500\AA~ to $\sim$9200\AA~ in only one exposure), high resolution ($R$ = 48,000) and high spectral stability, which makes it suitable for detecting narrow absorption features in a wide variety of spectral lines. Spectra of these objects were taken on the nights listed in Table \ref{tab:obs} with exposure times computed with the online FEROS Exposure Time Calculator\footnote{https://www.eso.org/observing/etc/bin/gen/form?INS.NAME =FEROS+INS.MODE=spectro} to obtain a signal-to-noise ratio (S/N hereafter) of about $\sim$150 around the blue wavelength range. Standard calibrations were taken and the ESO pipeline with the default parameters was used to reduce the data. The reduced spectra were corrected for heliocentric velocity shifts and telluric contamination (see Sec. \ref{sec:telluric}).

\subsubsection{UVES observations}
\label{sec:uvesobs}

We obtained UVES spectra in service mode during periods P96\footnote{Programme ID: 096.C-0238(A)}  and P97\footnote{Programme IDs: 097.C-0409(A) and 097.C-0409(B)} for the objects listed in Table \ref{tab:obs}. Data from P96 were observed with the blue arm centred at 4370\AA~ and with a spectral coverage of 3731--4999\AA~ and data from P97 were observed in dichroic mode with the blue arm centred at 3900\AA~ and the red arm centred at 5800\AA, covering the spectral ranges  3282--4562\AA~ and 4726--6835\AA, respectively. The narrowest slits were used in order to achieve the maximum resolution possible; the 0.4$\arcsec$ slit in the blue arm and the 0.3$\arcsec$ slit in the red arm, yielding resolutions of 80,000 and 110,000, respectively. Exposure times were computed with the online UVES Exposure Time Calculator\footnote{https://www.eso.org/observing/etc/bin/gen/form?INS.NAME =UVES+INS.MODE=spectro} aiming to obtain a S/N $\sim$150 (achieved for most of the spectra) around the blue wavelength range under thick (very low transparency) sky conditions. Standard calibrations were also taken and the ESO pipeline with the default parameters was used to reduce the data. The spectra were a posteriori corrected for telluric contamination (see Sec. \ref{sec:telluric}).

\subsection{Archival data}
\label{sec:data}

We queried the ESO archive looking for optical spectra with resolution high enough to detect narrow absorption features with widths of $\sim$0.1\AA. This restricted the instruments to HARPS, UVES and FEROS. Our targets have been observed multiple times (e.g., searching for planets via radial velocity variations), hence we found a large number of spectra (from one spectrum up to ~3000 spectra).
% (probably looking for companions of substellar or planetary masses). 
Consequently, a considerable fraction of our dataset ($\sim$35\% of the full sample) comes from different archives. In fact, 16 of our debris disks with multiple features were identified with these observations.

\subsubsection{HARPS data}
\label{sec:harps} % used for referring to this section from elsewhere

HARPS is an echelle spectrograph fed by a pair of fibres, one of them collects the star light, while the second is used to either record simultaneously a Th-Ar reference spectrum or the background sky. HARPS spectra covers the wavelength range 3780--6910\AA, has a spectral resolution of 120,000 and has been optimised for mechanical stability, which makes it ideal for our study. We retrieved several epochs of HARPS data for the debris disks listed in Table \ref{tab:obs}. All these obsevrations were already reduced with the ESO pipelines and corrected for heliocentric radial velocity shifts, and we then corrected for telluric contamination (see Sec. \ref{sec:telluric}).

\subsubsection{UVES data}
\label{sec:uvesdata}

Additional to the HARPS data, we found a considerable number of UVES observations for our sample (see Table \ref{tab:obs}). UVES is a two-arm cross-dispersed echelle spectrograph, its blue arm covers the wavelength range 3000--5000\AA~ and the red arm covers 4200--11000\AA. Overall, the spectral coverage depends on the instrumental set-up used for the observations since UVES allows the use of dichroic beam splitters, but in general we have spectra covering the ranges 3043--3916\AA, 3236--4563\AA, 3731--4999\AA, 4549--6686\AA, 4726--6835\AA~ and 6650--10426\AA. These UVES data had already been reduced with dedicated ESO pipelines but additional corrections of heliocentric radial velocity shifts and telluric contamination were necessary.
  %As mentioned before, the UVES maximum resolution is 80,000 in the blue arm and 110,000 in the red arm. 

\subsubsection{FEROS data}
\label{sec:ferosdata}

We also retrieved FEROS archival data for some of our targets (Table \ref{tab:obs}). Similarly to the HARPS data, the FEROS observations were reduced with the available instrument pipelines and corrected for heliocentric radial velocity shifts. Posterior corrections for telluric contamination were performed.

\subsection{Telluric lines correction}
\label{sec:telluric}

\begin{figure}
	% To include a figure from a file named example.*
	% Allowable file formats are eps or ps if compiling using latex
	% or pdf, png, jpg if compiling using pdflatex
	\includegraphics[width=1.1\columnwidth]{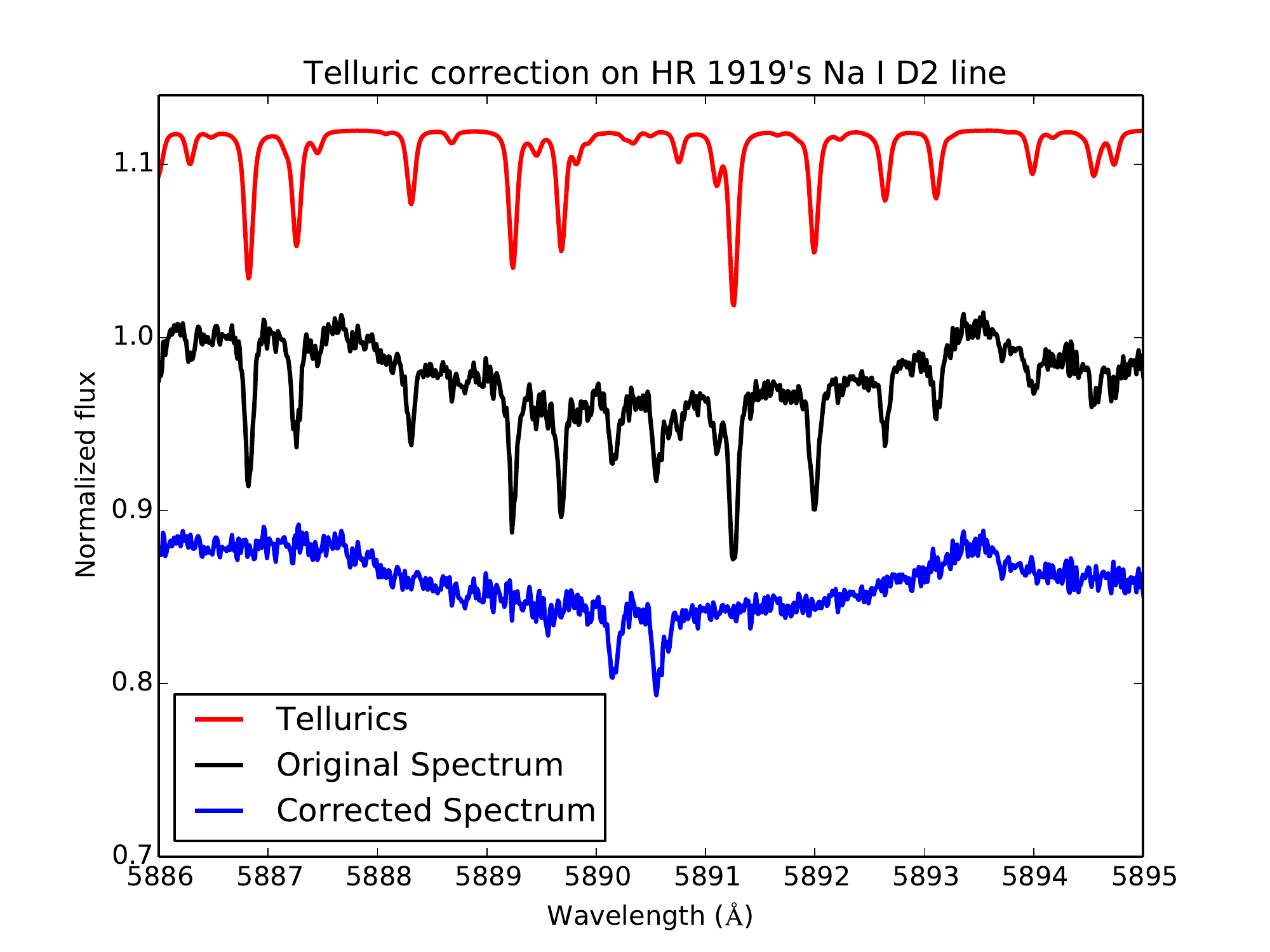}
    \caption{Example of telluric correction in HR\,1919's Na\,{\sc i} D2 line. Telluric lines modeled by molecfit in red, original spectrum in black, and corrected spectrum in blue. The different spectra are shifted with respect to each other for clarity.}
    \label{fig:telu}
\end{figure}

The red domain of the spectra contains many telluric absorption features mostly due to water vapour, $O_{2}$ and $O_{3}$. Since one of our main tracers is the sodium doublet at $5895.9$ and $5889.9$\AA, it is imperative to perform a correct subtraction of telluric contamination. To this end, we used \texttt{Molecfit}\footnote{http://www.eso.org/sci/software/pipelines/skytools/molecfit} (\citealt{Smette2015}, \citealt{Kausch2015}), a tool developed to correct observations for telluric absorption which can be used for any kind of spectra without the need to observe a standard star. We used the wavelength range 5902.5-5927.0\AA~to fit the continuum but excluded gas absorption features to ensure that they do not affect our best-fitting result. We applied these corrections to every epoch for all the objects except for HD\,106036 and HD\,112810 which spectra did not cover the red wavelength range. We successfully removed telluric absorptions and reduced them to the noise level (i.e. they were reduced by about 99\%). An example of the telluric correction is shown in Figure \ref{fig:telu}.

\section{Methods and results}
\label{sec:methods_results}

We analysed the calcium H \& K lines at 3968.47 and 3933.66\AA~and the sodium D1 \& D2 lines at 5895.92 and 5889.95\AA, respectively. The aim is to detect narrow absorption lines superimposed on the photospheric line. These ``extra'' absorption lines indicate the presence of gas in the line of sight of the star. To determine the nature of the gas we followed different approaches, that we detail in the following sub-sections. 

The first step in our analysis is to measure the radial velocity of the stars. Afterwards, we determined the photospheric contribution for each line either by performing spectral synthesis or by finding a ``spectral twin''. Then, we identified additional (stable or transient) components by removing the photospheric contribution before characterizing their properties.
%ated, we modelled the features to perform measurements of equivalent widths, apparent column density and apparent column density ratios N(Ca{\sc ii}/Na{\sc i}) of the absorption components. Then we studied the variability of the absorption features in terms of intensity and velocity and looked for additional FEB-like events in particular epochs.

Additionally, we also searched for signatures of Diffuse Interstellar Bands (DIBs), compared the radial velocity of the absorption features to the radial velocities of known local clouds and searched for similar extra absorption lines in nearby stars to better assess their nature.

%Further description of each step follows.

\begin{table*}
	\centering
	\caption{Stellar parameters determined from the synthetic model fitting and radial velocity estimates computed as described in Section\,\ref{RV}. Radial velocity of stars in multiple or binary systems are flagged with *. The typical uncertainties for the estimates of $v$sin$i$ and T$_{\rm eff}$(Ca\,{\sc ii} K), are of $\sim$5 km.s$^{-1}$ and 5-10\%, respectively. Dispersion values come from the measurements in the four different lines or from the grid step size (entries marked with $^{g}$) when the four lines yielded the same values.}
	\label{tab:param}
	\begin{tabular}{lccccccc} % four columns, alignment for each
		\hline
		Name & $v$sin$i$ & radV & T$_{\rm eff}$(Ca\,{\sc ii} K) & $\log g$ & [Fe/H] \\
		  & [km.s$^{-1}$] & [km.s$^{-1}$] & [K] & [dex] &  \\
		\hline
		$\beta$03\,Tuc & 100  & 6.05 $\pm$ 1.60* & 9550  & 4.00 $\pm$ 0.50 & +0.17 $\pm$ 0.20 \\
		66\,Psc & 150  & 4.32 $\pm$ 2.66* & 10750  & 3.90 $\pm$ 0.23 & -0.38 $\pm$ 0.54 \\
		$\nu$ Hor & 140  & 13.58 $\pm$ 1.67 & 8300   & 4.25 $\pm$ 0.25 & -0.20 $\pm$ 0.31 \\
		HD\,24966 & 210  & 15.70 $\pm$ 3.15 & 9250  & 4.38 $\pm$ 0.22 & -0.03 $\pm$ 0.58 \\
		HD\,290540 & 200  & 27.26 $\pm$ 3.12  & 10500   & 4.25 $\pm$ 0.25 & -0.88 $\pm$ 0.22 \\
		HD\,36444 & 360  & 26.72 $\pm$ 4.50 & 10250   & 4.00 $\pm$ 0.35 & -0.62 $\pm$ 0.65 \\
		HD\,290609 & 100  & 25.87 $\pm$ 1.65 & 10500  & 3.88 $\pm$ 0.41 & -0.07 $\pm$ 0.26 \\
		HR\,1919 & 140  & 23.49 $\pm$ 1.28 &  8800   & 4.15 $\pm$ 0.38 & -0.12 $\pm$ 0.22 \\
		HD\,54341 & 140  & 41.03 $\pm$ 0.95 & 10500   & 4.50 $\pm$ $0.25^{g}$ & -0.12 $\pm$ 0.22 \\
		HD\,60856 & 40  & 34.78 $\pm$ 3.58 & 14000   & 4.12 $\pm$ 0.22 & -0.38 $\pm$ 0.41 \\
		HR 3300 & 210  & 22.35 $\pm$ 0.72 & 9550  & 4.25 $\pm$ 0.25 & +0.10 $\pm$ 0.37 \\
		$\eta$ Cha & 280  & 15.21 $\pm$ 1.41 & 11750   & 3.75 $\pm$ 0.25 & -0.88 $\pm$ 0.22 \\
		HD\,92536 & 180  & 15.45 $\pm$ 0.44 & 11150   & 3.88 $\pm$ 0.41 & -0.45 $\pm$ 0.55 \\
		3 Crv & 130  & 14.41 $\pm$ 1.09 & 8500  & 4.12 $\pm$ 0.22 & +0.17 $\pm$ 0.41 \\
		HD\,106036 & 160  & 9.37 $\pm$ 2.34 & 9000   & 4.50 $\pm$ $0.25^{g}$ & +0.00 $\pm$ $0.25^{g}$\\
		HR\,4796 & 150  & 5.35 $\pm$ 2.94*  & 9800   & 4.25 $\pm$ 0.25 & -0.07 $\pm$ 0.26 \\
		HD\,110058 & 150  & 11.20 $\pm$ 0.81 & 9000   & 4.03 $\pm$ 0.36 & -0.33 $\pm$ 0.47 \\
		HD\,112810 & -- & 5.25 $\pm$ 2.24 & -- & -- & --\\
		HD\,126135 & 310  & 11.73 $\pm$ 0.67 & 11250   & 3.88 $\pm$ 0.41 & -0.62 $\pm$ 0.65 \\
		HD\,141378 & 80  & -14.68 $\pm$ 2.62 & 8750   & 4.50 $\pm$ $0.25^{g}$ & +0.42 $\pm$ 0.13 \\
		HD\,141327 & 250  & -4.65 $\pm$ 2.40 & 10550   & 4.00 $\pm$ 0.35 & -0.50 $\pm$ 0.50 \\
%		HD\,144981 & 210 $\pm$ 5 & -6.47 $\pm$ 1.59 & 9625.00 $\pm$ 279.51 & 4.38 $\pm$ 0.22 & -0.12 $\pm$ 0.41 \\
%		HD\,145554 & 250 $\pm$ 5 & -9.66 $\pm$ 1.49 & 10687.50 $\pm$ 108.25 & 3.88 $\pm$ 0.41 & -1.00 $\pm$ $0.25^{g}$ \\
%		HD\,145631 & 300 $\pm$ 5 & -9.86 $\pm$ 1.90 & 10087.50 $\pm$ 341.64 & 4.00 $\pm$ 0.35 & -1.00 $\pm$ $0.25^{g}$ \\
%		HR\,6051 & 320 $\pm$ 5 & -7.26 $\pm$ 4.11 & 10775.00 $\pm$ 309.23 & 3.77 $\pm$ 0.28 & -1.00 $\pm$ $0.25^{g}$ \\
		HR\,6507 & 140  & -36.13 $\pm$ 2.49 & 7750   & 4.25 $\pm$ 0.25 & +0.28 $\pm$ 0.13 \\
		c Aql & 90  & 17.29 $\pm$ 2.61 & 9700  & 4.35 $\pm$ 0.26 & -0.57 $\pm$ 0.49 \\
		\hline
	\end{tabular}
\end{table*}

\subsection{Radial velocities}
\label{RV}

Since five of our objects did not have any reported radial velocity measurements in the literature, we performed our own estimates for all objects aiming at a homogeneously determined set of values and to assess the accuracy of our results. 

In a first attempt to obtain the radial velocities, we computed the cross correlation function for every epoch of each object using a synthetic model as a template. Unfortunately, since most of our objects are fast rotators, their absorption lines have very wide profiles, and we did not obtain consistent results between all the epochs, with dispersions up to $30$\,km.s$^{-1}$.
Therefore, we decided to take a different approach and use a simpler but, in this case, more suitable technique. For every epoch of each object we fit Lorentzian profiles to the most prominent absorption lines in our spectra: H$\alpha$, H$\beta$, H$\gamma$ and H$\delta$. We excluded H$\epsilon$ because it is blended with the Ca\,{\sc ii} H line. We used a range of 1000 km.s$^{-1}$ for the profile fitting of each (previously normalized) line in velocity space and obtained the radial velocity of the line from the position of the profile with respect to the rest frame. 

In order to address possible changes in the estimates of radial velocities due to activity, we checked all our objects for emission features in the Balmer lines (not only for emission dominated lines, but also for shallow core emissions). Only one object, namely HD\,60856, presents emission features in these lines. This emission is dominating the full line in the case of H$\alpha$ and thus it was not possible to model the photospheric profile. Therefore, only for this object, we decided to exclude H$\alpha$ from the radial velocity measurements. In addition, we bootstrapped each of the remaining Balmer lines to estimate the impact of a core emission in our fitting procedure. The standard deviation for the radial velocity from this procedure with 1000 realizations was $\sim$1.6\,km.s$^{-1}$. This object has only one epoch of observations, therefore our estimated uncertainty for the radial velocity takes into account the individual line fitting uncertainties and the dispersion found among the different lines.

For all the other objects, since the cores of the lines appeared purely photospheric, we averaged over all the epochs and lines, and the uncertainties were derived propagating the estimated errors.

In most cases, our radial velocity measurements are in good agreement with the ones found in the literature (see Table \ref{tab:info}), having average differences of $\sim$3 km.s$^{-1}$. However, the radial velocity value we determined for $\nu$ Hor differed by $\sim$17 km.s$^{-1}$ with respect to the literature. We compared both radial velocity values when fitting Kurucz \citep{Castelli1997} models and concluded that our estimate provides a better match to our data. In general, although differences with the literature were relatively small, in the model fitting process we always obtained a better fit when shifting the model to our own estimates of radial velocity. In any case, the large difference found for $\nu$ Hor is not that surprising given the high dispersion in the measurements provided from different datasets in \cite{Wilson1953}, the reference adopted in Simbad. Finally, we must also note that, for the whole sample, no significant shifts were found between epochs, obtaining velocity dispersions per object of the order of $\sim$2 km.s$^{-1}$. Our resulting radial velocity estimates are shown in Table \ref{tab:param}.

%\begin{figure}  %-> funciona bien con pdf
%\includegraphics[width=\columnwidth]{Spectype_Teff.pdf}
%\caption{Spectral types versus T$_{\rm eff}$ estimates. Spectral types are adopted from Simbad, T$_{\rm eff}$ shown in the ordinate axes correspond to our spectral synthesis estimates, while the color bar displays the values obtained via SED fit of the same collection of Kurucz models, using VOSA \citep{Bayo2008}. Errors for the T$_{\rm eff}$ from SED fitting are $\sim$125 K. HD\,112810 is not shown in this Figure given that its parameters were not obtained via synthetic modelling. }
%\label{fig:Teff}
%\end{figure}

\subsection{Spectral synthesis}
\label{sec:synthe} 

For most objects (except HD\,112810, see Sec. \ref{sec:twin}), we used Kurucz models \citep{Castelli1997} to fit and normalize the photospheric absorption, thus isolating the additional absorption lines. The models were computed using the spectral synthesis codes \texttt{SYNTHE} and \texttt{ATLAS 9} \citep{Sbordone04}. 

For each line we computed the normalized median spectrum from all the epochs, to use it as a robust reference for the fitting process. Since the radial velocity dispersion along all the epochs is small, the median can be used as a good reference. For each of the median spectrum, the uncertainties are derived using the standard deviation of all epochs if there are more than 2 epochs. For objects with only one or two epochs, the pipelines do not always provide uncertainties. Therefore, for each wavelength point, we estimate the standard deviation in a moving box with a width of $33$ wavelength points.
% To compute this reference spectrum we have normalized each section of the spectrum containing the absorption lines and their wings. The normalization was performed fitting a line to continuum segments of 10\AA~  width at both sides of the absorption and dividing the whole section by this line so all the epochs have their continuum levels at 1 and the absorptions have a comparable flux. We thus obtained the median spectrum from all the normalized epochs. 
Then, we computed a grid of models with different stellar parameters for the wavelength ranges containing the Ca\,{\sc ii} and Na\,{\sc i} doublets. %We computed the models with a very high resolution, R=1'000'000, in order to have comprehensive line profiles. 
The free parameters that can be investigated in the modelling process are: the effective temperature T$_{\rm eff}$, the surface gravity $\log g$, the metallicity [Fe/H], turbulent velocity, additional turbulence, opacity threshold for the lines, and projected rotational velocity $v$sin$i$. We adopted standard values for the turbulent velocity, additional turbulence and opacity threshold; and explored the remaining parameters: T$_{\rm eff}$, $\log g$, [Fe/H] and $v$sin$i$. Table \ref{tab:synthe} summarizes the fixed values and range of parameters that we explored in the fitting process.
% Adopted values for the fixed parameters and ranges explored for the free ones are detailed in Table \ref{tab:synthe}. 
Those ranges were chosen according to previous estimates available in the literature. Our model fitting procedure consists of a simple two-step $\chi^{2}$ minimization. The fits were performed for each line independently since, possibly due to non-Local Thermodynamic Equilibrium (non-LTE) effects (\citealt{Mashonkina2000}, \citealt{Plez2013} and \citealt{Sitnova2017}), it is hardly possible to obtain good matches for all of them simultaneously. The only parameter determined using all the lines at once is $v$sin$i$. The first step consists of estimating approximate values for all the parameters, within a coarse grid of models. The step sizes are reported in Table \ref{tab:synthe}. Afterwards, we used a simplex downhill method with finer interpolations for T$_{\rm eff}$ and $\log g$, with steps of $50$\,K and $0.1$\,dex, respectively.
% The two-step sequence consists of a first estimate of the parameters obtained by a brute force approach with the step grids reported in Table \ref{tab:synthe}, followed by a simplex downhill method with finer interpolations for T$_{\rm eff}$ and $\log g$ of 50K and 0.1 dex, respectively.
To avoid local minima, we repeated the simplex downhill algorithm several times, initializing it from different regions of the parameter space that yielded similar (a factor 2 with respect to the minimum) $\chi^{2}$ values in the coarse grid. The convergence criterion for the downhill algorithm was set to an improvement in the goodness of fit by $10^{-4}$. Given that Ca\,{\sc ii} K is the most sensitive photospheric temperature tracer among all the lines studied \citep{GrayCorbally2009}, in Table \ref{tab:param} we report as the best fitting temperature the one obtained for that line. A rather conservative confidence interval is estimated from all the models that returned a relative change in $\chi^{2}$ smaller than $50$\% compared to the best fitting model. On the other hand, for $\log g$ and [Fe/H], we provide an average of the values obtained for the four lines, and the associated uncertainties correspond to their standard deviations.

Once the best fitting model per line is found, it is used to isolate the extra absorption lines from the photospheric profile. The best fits for each objects and lines are displayed in Figures\,\ref{fig:mod1} to \ref{fig:mod4}.

\begin{table}
	\centering
	\caption{Adopted and explored values for the parameters of the Kurucz models.}
	\label{tab:synthe}
	\begin{tabular*}{\columnwidth}{lc} % four columns, alignment for each
		\hline
		Parameter & Values \\
		\hline
		Turbulent Velocity & 2.0 km.s$^{-1}$\\
		Additional Turbulence & 0.0 km.s$^{-1}$ \\
		Opacity Threshold & 0.001  \\
		T$_{\rm eff}$ &  6000--13000 K, with $\Delta$=250 K \\
		$\log g$ & [3.5, 4.0, 4.5] \\
		$[Fe/H]$  & [-1.0, -0.5, 0.0, +0.2, +0.5]  \\
		$v$sin$i$  & 20--400 km.s$^{-1}$, with $\Delta$=10 km.s$^{-1}$ \\
		\hline
	\end{tabular*}
\end{table}

We were able to find matching photospheric models for all the objects, and the resulting parameters are consistent with their spectral types from the literature. The only exception is HD\,112810, which is an F3/5IV/V spectral type according to the literature and the only F-type star within the sample presented in this paper (we further discuss this object in Section\,\ref{sec:twin}).  %As an example, in Figure \ref{fig:Teff} we show the good correlation obtained between spectral types and their estimated T$_{\rm eff}$. 
Otherwise, for each individual object, the dispersion in T$_{\rm eff}$ for the four different lines is of the order of $\sim 300$\,K, which is expected when accounting for non-LTE effects (\citealt{Przybilla2011}, \citealt{Plez2013} and \citealt{Sitnova2017}). %In Table \ref{tab:param} we present the T$_{\rm eff}$ values obtained from the Ca\,{\sc ii} K line as this line is a good T$_{\rm eff}$ indicator. In the case of $\log g$ and [Fe/H] we present averaged values from the four lines we have analysed considering the best model fit parameters obtained for each one. We also show the projected rotational velocity estimates in Table \ref{tab:param}.

\begin{figure*}  %-> funciona bien con pdf
\includegraphics[width=\textwidth]{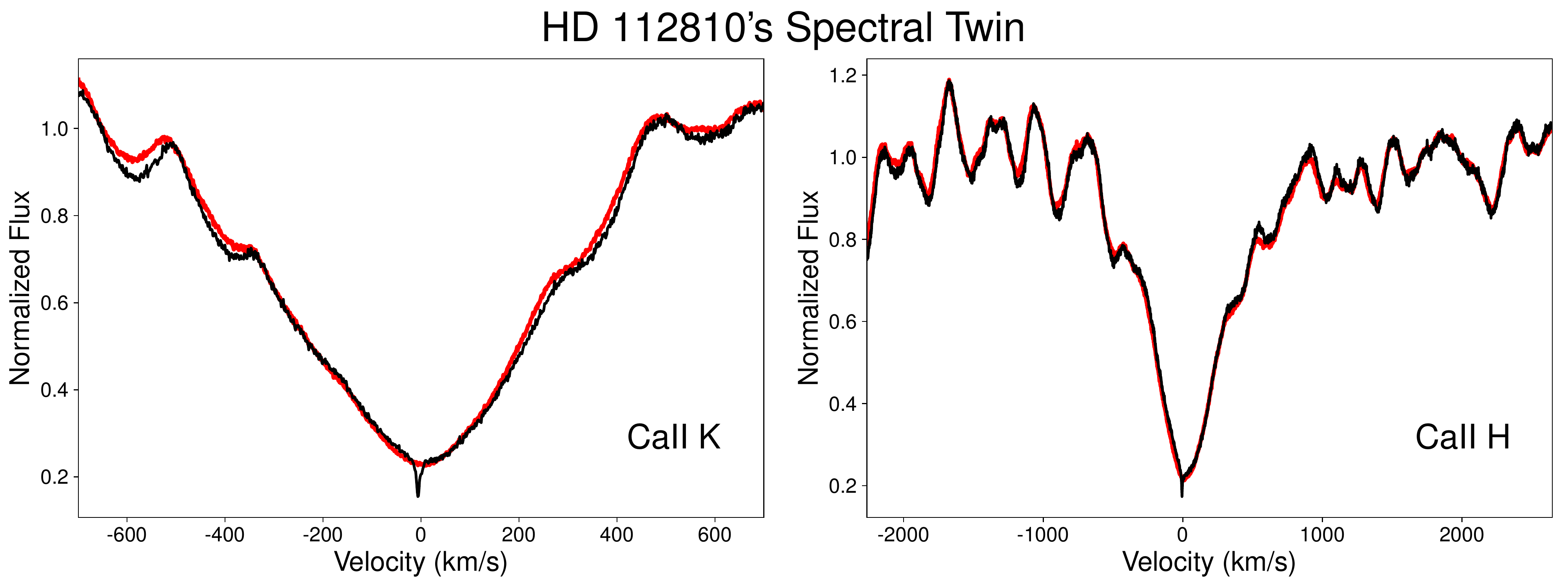}
\caption{Spectral twin found for HD\,112810 (in black): the F4IV type star HD\,15115 (in red). The radial velocities of both stars have been shifted to zero for a better comparison. No additional broadening has been added to the spectra of any of the two objects. }
\label{fig:HD112810twin}
\end{figure*}

\subsection{Spectral twins}
\label{sec:twin}

As an alternative to the synthetic spectrum, we also performed a search for the closest spectral match within all the objects in our sample for which we have spectra (i.e. 234 objects). In particular, we compared the median spectrum (the same as the reference spectrum) of each candidate against the median spectrum of each object in the sample with similar spectral types. We used a range of 11 subtypes (e.g. between A0 and F0 for an A5 candidate) for the spectral twin search. 

Similar to the synthetic model fit, we selected the best fitting ``template'' in terms of minimum $\chi^{2}$. The only difference is that in the $\chi^{2}$ calculation we neglected the wavelength regions containing the $10$\% most distant data points between the two spectra being compared, in order to avoid a bias induced by the presence of extra features in either spectra.

In the case of HD\,112810, since we were not able to find a satisfactory Kurucz model, we used the spectral twin we found for the object as a photospheric model to isolate the absorption feature. As can be seen in Fig. \ref{fig:HD112810twin}, HD\,15115 is a good match to the spectrum of HD\,112810.

\subsection{Identification and characterization of features}
\label{sec:feat}

We started by normalizing the reference (median) spectrum by the synthetic model (or spectral twin for HD\,112810) to isolate the extra absorption features. In some cases when the model is not a perfect match (often the case in the wings of the photospheric lines), the normalized spectrum shows a ``wavy'' pattern, that makes the characterization of the extra features more challenging. In those cases we performed a polynomial fit to the normalized spectrum to remove this wavy pattern.

Afterwards, we performed Gaussian fitting to each of the extra absorption features in order to derive radial velocities, equivalent widths and apparent column densities. In the case of blended absorption lines, we modelled a combined Gaussian profile with the minimum number of Gaussians that would finely fit the profile. We only considered as ``real absorption features'' those with significance above $3\sigma$ over the residual spectrum. We considered a feature to be ``the same'' as that present in another line of the same object when the absolute difference of the radial velocity of both features is $\leqslant 2\sigma$, where $\sigma=\sqrt{\sigma^{2}_{1}+\sigma^{2}_{2}}$, being $\sigma_{1}$ and $\sigma_{2}$ the uncertainty of the radial velocity of each feature. %Statistically speaking, the probability that the two signatures are compatible if $|v_{1} - v_{2}| > 2\sigma$ is less than $\sim$2.5\%.}

Radial velocities and equivalent widths were estimated from the best Gaussian(s) fit. In addition, apparent column densities were estimated following \cite{Savage1991} and using the oscillator strength values $f$ from \cite{Morton1991}. We checked our own estimates for three stars against results produced by the \texttt{Vapid} code \citep[Voigt Absorption Profile Interstellar Dabbler][which can model interstellar absorption lines]{Howarth2002} and they agreed within the uncertainties. Apparent column density (Ca{\sc ii}/Na{\sc i}) ratios were also computed as they can help to discriminate the origin of the features.

We noticed that for some of the objects with Na\,{\sc i} column densities exceeding $12.0$, the column density in Na\,{\sc i} D1 is larger by about $0.2$ dex than the one for Na\,{\sc i} D2. This is likely caused by saturation, i.e. the line with the weaker transition yields a higher column density. Although this problem affects the precision of the measurements in these particular cases, in the end it does not strongly bias the final verdict on the origin of the gas, since we have performed several independent kinds of analysis in order to reach our conclusions.

The parameters for each line and feature are presented in Table \ref{tab:feat}. The average radial velocity of each feature and their N(Ca{\sc ii}/Na{\sc i}) are shown in Table \ref{tab:origin}. As can be seen in the Tables, the range of properties is very wide, including blue and red-shifted components, weak and intense features, either Ca\,{\sc ii} or Na\,{\sc i} rich. The detailed discussion on the impact of these parameter in determining the origin of the gas responsible for the feature is left to Section \ref{sec:discussion}.

\onecolumn

\begin{center}
\begin{longtable}{l| @{\extracolsep{\fill}}ccc|ccc|ccc|ccc}
	\caption{Absorption feature parameters. Heliocentric radial velocity, apparent column density and equivalent width of the features present in each line. Uncertainties for the radial velocities are in the order of $2.3$\,km.s$^{-1}$ for the Ca{\sc ii} lines and $1.5$\,km.s$^{-1}$ for the Na{\sc i} lines. Apparent column densities and equivalent widths have uncertainties of 3--4\% for the Ca{\sc ii} lines and 1--2\% for the Na{\sc i} lines.}	\label{tab:feat} \\
	\hline 
	& & Ca{\sc ii} K   &  & & Ca{\sc ii} H   &  & & Na{\sc i} D1   &  & & Na{\sc i} D2  &\\ 
 	Name & radV & $\log_{10} N$  & $EW$ & radV & $\log_{10} N$ & $EW$ & radV & $\log_{10} N$ & $EW$ & radV & $\log_{10} N$ & $EW$  \\ 
 			   & [km.s$^{-1}$] & [cm$^{-2}$]  & [m\AA] & [km.s$^{-1}$] & [cm$^{-2}$] & [m\AA] & [km.s$^{-1}$] & [cm$^{-2}$] & [m\AA] & [km.s$^{-1}$] & [cm$^{-2}$] & [m\AA]  \\ 
 	\hline
 	\endfirsthead
 	
 	\multicolumn{5}{l}{ \tablename\ \thetable{} -- continued from previous page} \\
 		\hline 
	& & Ca{\sc ii} K   &  & & Ca{\sc ii} H   &  & & Na{\sc i} D1   &  & & Na{\sc i} D2  &\\ 
 	Name & radV & $\log_{10} N$  & $EW$ & radV & $\log_{10} N$ & $EW$ & radV & $\log_{10} N$ & $EW$ & radV & $\log_{10} N$ & $EW$  \\ 
 			   & [km.s$^{-1}$] & [cm$^{-2}$]  & [m\AA] & [km.s$^{-1}$] & [cm$^{-2}$] & [m\AA] & [km.s$^{-1}$] & [cm$^{-2}$] & [m\AA] & [km.s$^{-1}$] & [cm$^{-2}$] & [m\AA]  \\ 
 	\hline
 	\endhead
 	
 	\hline \multicolumn{13}{r}{{Continued on next page}} \\ 
	\endfoot
	
	\hline 
	\endlastfoot
 	
$\beta$03\,Tuc & -12.23 & 9.78 & 0.52 & -- & -- & -- & -- & -- & -- & -10.21 & 9.24 & 0.34 \\ 
 & -3.52 & 10.03 & 0.91 & -4.41 & 10.14 & 0.60 & -3.50 & 10.50 & 3.10 & -3.51 & 10.51 & 6.25 \\ 
 & 1.89 & 10.57 & 3.19 & 1.72 & 10.29 & 0.84 & -- & -- & -- & 3.18 & 9.64 & 0.85 \\ 
66\,Psc & -5.14 & 11.10 & 10.38 & -6.01 & 11.05 & 4.72 & -5.83 & 11.28 & 17.49 & -5.85 & 11.27 & 31.64 \\ 
 & 2.58 & 10.25 & 1.51 & -- & -- & -- & -1.44 & 10.48 & 2.93 & 0.32 & 10.18 & 2.93 \\ 
 & 10.74 & 10.56 & 3.09 & -- & -- & -- & 12.23 & 9.71 & 0.51 & -- & -- & -- \\ 
$\nu$\,Hor & 5.41 & 10.55 & 3.04 & 4.18 & 10.35 & 0.96 & 3.86 & 9.52 & 0.33 & 2.97 & 9.78 & 1.20 \\ 
 & 13.03 & 10.46 & 2.46 & 12.35 & 10.36 & 1.00 & 13.02 & 10.02 & 1.03 & 12.57 & 9.86 & 1.43 \\ 
 & -- & -- & -- & -- & -- & -- & -8.91 & 9.67 & 0.46 & -9.27 & 9.41 & 0.51 \\ 
HD\,24966 & -13.10 & 10.52 & 2.79 & -- & -- & -- & -- & -- & -- & -- & -- & -- \\ 
 & 17.00 & 10.89 & 6.55 & 15.04 & 10.71 & 2.23 & 15.34 & 9.90 & 0.78 & 15.32 & 9.87 & 1.45 \\ 
 & 30.21 & 10.37 & 1.98 & -- & -- & -- & -- & -- & -- & 34.08 & 9.55 & 0.69 \\ 
HD\,290540 & 9.51 & 11.61 & 32.77 & 9.27 & 11.50 & 13.29 & 9.09 & 11.45 & 26.23 & 9.26 & 11.44 & 49.67 \\ 
 & 23.96 & 12.00 & 64.17 & 24.15 & 12.14 & 49.91 & 23.47 & 12.13 & 98.75 & 23.56 & 11.96 & 123.98 \\ 
 & 33.78 & 11.13 & 11.33 & 37.80 & 10.67 & 2.02 & 35.16 & 11.30 & 18.97 & 35.30 & 11.29 & 35.55 \\ 
 & -- & -- & -- & -- & -- & -- & -4.17 & 10.28 & 1.86 & -3.82 & 10.44 & 5.31 \\ 
HD\,36444 & 7.27 & 11.16 & 12.03 & 6.91 & 11.28 & 7.96 & 9.02 & 11.44 & 25.95 & 8.79 & 11.30 & 36.31 \\ 
 & 21.96 & 11.70 & 38.88 & 21.72 & 11.82 & 27.16 & 22.13 & 11.95 & 72.57 & 22.15 & 11.87 & 109.98 \\ 
 & 31.40 & 11.11 & 10.57 & 31.98 & 11.13 & 5.79 & 30.81 & 11.23 & 16.26 & 31.00 & 10.98 & 18.06 \\ 
 & 38.79 & 11.21 & 13.46 & 42.23 & 11.33 & 9.06 & -- & -- & -- & -- & -- & -- \\ 
HD\,290609 & -9.23 & 10.91 & 6.86 & -- & -- & -- & -8.56 & 10.43 & 2.61 & -- & -- & -- \\ 
 & 9.65 & 11.98 & 71.21 & 9.35 & 12.04 & 44.61 & 8.46 & 11.62 & 38.92 & 8.50 & 11.54 & 61.20 \\ 
 & 24.00 & 12.01 & 69.05 & 24.08 & 12.06 & 43.66 & 22.92 & 12.26 & 125.13 & 23.02 & 12.08 & 149.96 \\ 
 & 34.38 & 11.62 & 33.45 & 37.31 & 11.85 & 29.57 & 35.39 & 11.33 & 20.65 & 32.52 & 11.12 & 24.87 \\ 
 & 49.65 & 11.63 & 35.35 & 55.24 & 11.63 & 17.92 & 53.69 & 10.78 & 5.84 & 46.45 & 10.75 & 11.01 \\ 
HR\,1919 & 10.92 & 11.20 & 12.71 & 10.56 & 11.11 & 5.48 & 10.39 & 10.44 & 2.71 & 10.63 & 10.36 & 4.44 \\ 
 & 32.17 & 11.57 & 28.28 & 30.84 & 11.37 & 9.74 & 30.92 & 10.46 & 2.80 & 30.51 & 10.34 & 4.27 \\ 
 & 38.79 & 10.64 & 3.66 & 35.43 & 11.12 & 5.61 & 36.77 & 9.69 & 0.48 & 35.58 & 9.97 & 1.82 \\ 
HD\,54341 & 10.32 & 10.07 & 1.01 & -- & -- & -- & -- & -- & -- & -- & -- & -- \\ 
 & 24.49 & 10.36 & 1.96 & 26.17 & 10.50 & 1.35 & 24.95 & 9.91 & 0.79 & 25.91 & 10.05 & 2.22 \\ 
 & -- & -- & -- & -- & -- & -- & -- & -- & -- & 3.31 & 9.85 & 1.39 \\ 
HD\,60856 & 20.39 & 11.69 & 36.77 & 20.14 & 11.82 & 26.63 & 20.38 & 12.25 & 118.41 & 20.38 & 12.03 & 134.94 \\ 
 & 30.65 & 10.71 & 4.37 & 31.10 & 10.82 & 2.85 & 28.72 & 10.70 & 4.86 & 29.47 & 10.42 & 5.13 \\ 
HR\,3300 & 5.62 & 10.77 & 4.99 & 4.95 & 10.50 & 1.38 & 7.52 & 10.18 & 1.49 & 5.72 & 10.19 & 3.01 \\ 
 & 15.83 & 10.21 & 1.37 & 16.04 & 10.28 & 0.83 & 17.66 & 10.20 & 1.57 & 17.45 & 10.22 & 3.27 \\ 
 & 20.82 & 10.30 & 1.69 & -- & -- & -- & -- & -- & -- & -- & -- & -- \\ 
$\eta$\,Cha & -3.37 & 10.13 & 1.16 & -3.46 & 10.11 & 0.55 & -1.05 & 9.57 & 0.36 & -3.46 & 9.84 & 1.36 \\ 
 & 10.09 & 10.77 & 4.97 & 9.55 & 10.76 & 2.49 & 11.92 & 10.46 & 2.85 & 12.20 & 10.44 & 5.34 \\ 
HD\,92536 & 2.81 & 10.60 & 3.36 & 1.57 & 10.22 & 0.72 & -0.23 & 9.50 & 0.31 & -0.56 & 9.74 & 1.07 \\ 
 & 9.94 & 11.40 & 19.84 & 9.59 & 11.45 & 11.85 & 9.19 & 11.55 & 31.90 & 9.18 & 11.53 & 56.67 \\ 
 & 18.21 & 11.27 & 15.18 & 18.31 & 11.22 & 6.96 & 17.59 & 10.90 & 7.77 & 17.25 & 10.90 & 15.15 \\ 
3\,Crv & -6.64 & 11.05 & 9.23 & -7.69 & 10.83 & 2.89 & -6.81 & 10.76 & 5.53 & -6.86 & 10.71 & 9.60 \\ 
 & -1.45 & 10.67 & 3.86 & -3.19 & 11.01 & 4.35 & -1.22 & 9.83 & 0.67 & -1.71 & 9.91 & 1.60 \\ 
 & 2.21 & 10.54 & 2.95 & 3.70 & 9.69 & 0.21 & -- & -- & -- & -- & -- & -- \\ 
HD\,106036 & -6.17 & 11.03 & 9.13 & -3.40 & 11.20 & 6.78 & -- & -- & -- & -- & -- & -- \\ 
 & 9.97 & 11.30 & 16.57 & 10.39 & 11.08 & 5.13 & -- & -- & -- & -- & -- & -- \\
HR\,4796 & -14.40 & 10.41 & 2.18 & -14.60 & 10.49 & 1.33 & -11.10 & 9.89 & 0.76 & -11.76 & 9.85 & 1.40 \\ 
 & -5.57 & 10.40 & 2.14 & -5.84 & 10.26 & 0.79 & -5.01 & 10.51 & 3.19 & -5.11 & 10.54 & 6.75 \\ 
 & 5.20 & 9.78 & 0.52 & -- & -- & -- & -- & -- & -- & -- & -- & -- \\ 
HD\,110058 & 1.85 & 11.47 & 24.38 & 0.97 & 11.45 & 12.04 & 0.56 & 11.23 & 16.31 & 0.42 & 11.24 & 32.67 \\ 
 & 12.50 & 11.50 & 21.89 & 12.19 & 11.77 & 20.79 & 12.36 & 11.76 & 41.72 & 12.34 & 11.49 & 43.52 \\
HD\,112810 & -12.11 & 11.07 & 9.76 & -10.86 & 10.88 & 3.23 & -- & -- & -- & -- & -- & -- \\ 
 & -3.76 & 11.60 & 29.69 & -4.49 & 11.57 & 14.99 & -- & -- & -- & -- & -- & -- \\ 
 & 3.77 & 11.01 & 8.46 & 2.28 & 11.25 & 7.58 & -- & -- & -- & -- & -- & -- \\ 
HD\,126135 & -22.78 & 10.25 & 1.51 & -- & -- & -- & -22.49 & 10.01 & 1.01 & -22.12 & 10.06 & 2.23 \\ 
 & -14.89 & 10.73 & 4.45 & -15.59 & 10.60 & 1.71 & -13.47 & 10.84 & 6.73 & -14.36 & 10.67 & 9.03 \\ 
 & -8.32 & 10.59 & 3.27 & -9.90 & 10.56 & 1.57 & -7.83 & 10.02 & 1.04 & -10.23 & 10.46 & 5.64 \\ 
 & 4.16 & 10.92 & 6.83 & 3.83 & 10.85 & 3.05 & 4.25 & 11.59 & 33.50 & 4.36 & 11.57 & 58.21 \\ 
 & -- & -- & -- & -- & -- & -- & -- & -- & -- & 26.30 & 9.63 & 0.84 \\ 
HD\,141378 & -30.03 & 10.79 & 5.16 & -- & -- & -- & -- & -- & -- & -- & -- & -- \\ 
 & -15.02 & 10.36 & 1.93 & -- & -- & -- & -- & -- & -- & -- & -- & -- \\ 
HD\,141327 & -18.37 & 11.33 & 18.11 & -21.23 & 11.22 & 7.07 & -23.48 & 10.42 & 2.58 & -23.69 & 10.41 & 4.97 \\ 
 & -1.43 & 11.95 & 66.92 & -2.40 & 11.97 & 37.57 & -4.17 & 12.51 & 203.27 & -4.20 & 12.43 & 260.97 \\ 
 & 14.78 & 11.80 & 47.59 & 14.23 & 11.82 & 26.94 & 14.66 & 10.77 & 5.73 & 14.16 & 10.77 & 11.18 \\ 
 & 22.92 & 11.04 & 9.25 & 21.83 & 10.97 & 4.06 & 23.11 & 10.60 & 3.90 & 22.75 & 10.75 & 10.90 \\ 
HR\,6507 & -38.93 & 10.32 & 1.78 & -39.26 & 11.02 & 4.49 & -38.62 & 10.39 & 2.40 & -39.01 & 9.97 & 1.82 \\ 
 & -30.41 & 11.12 & 10.98 & -29.52 & 11.29 & 8.22 & -29.80 & 10.26 & 1.80 & -29.10 & 10.42 & 5.09 \\ 
 & -24.86 & 11.25 & 14.00 & -25.05 & 11.22 & 6.84 & -25.08 & 10.97 & 8.71 & -25.09 & 10.91 & 14.68 \\ 
c\,Aql & -31.19 & 10.49 & 2.64 & -31.67 & 10.52 & 1.44 & -- & -- & -- & -- & -- & -- \\ 
 & -19.67 & 10.34 & 1.88 & -19.88 & 10.44 & 1.21 & -- & -- & -- & -- & -- & -- \\ 

\end{longtable}
\end{center}
\twocolumn

\subsection{Variability of the extra absorption features}
\label{sec:vari} 

We investigated the variability of additional absorption lines in two ways: first by analysing their stability when they are detected in all the epochs and second by looking for transient absorption features that appear in a handful of epochs. For the first method, we performed the same Gaussian fitting described above, but on each individual epoch and we searched for variations in flux and velocity of those ``stable'' components (since they are present in all the epochs they also appear in the reference spectrum). For the second method, we searched for additional variable detections above a $3\sigma$ level that might appear in some of the epochs. Such transient detections could be related to FEB-like events.

\begin{figure*}  %-> funciona bien con pdf
\includegraphics[width=0.49\textwidth]{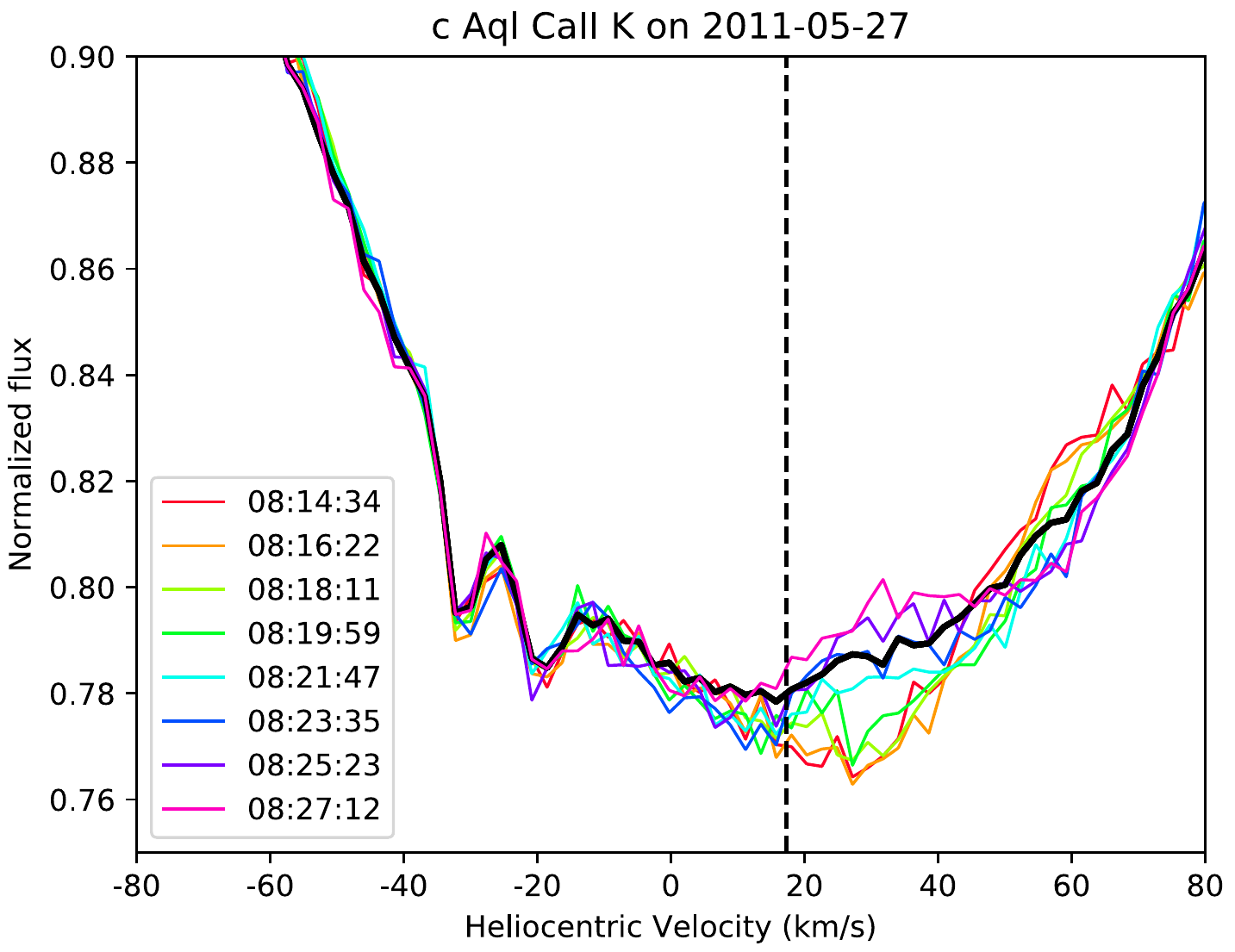}
\includegraphics[width=0.49\textwidth]{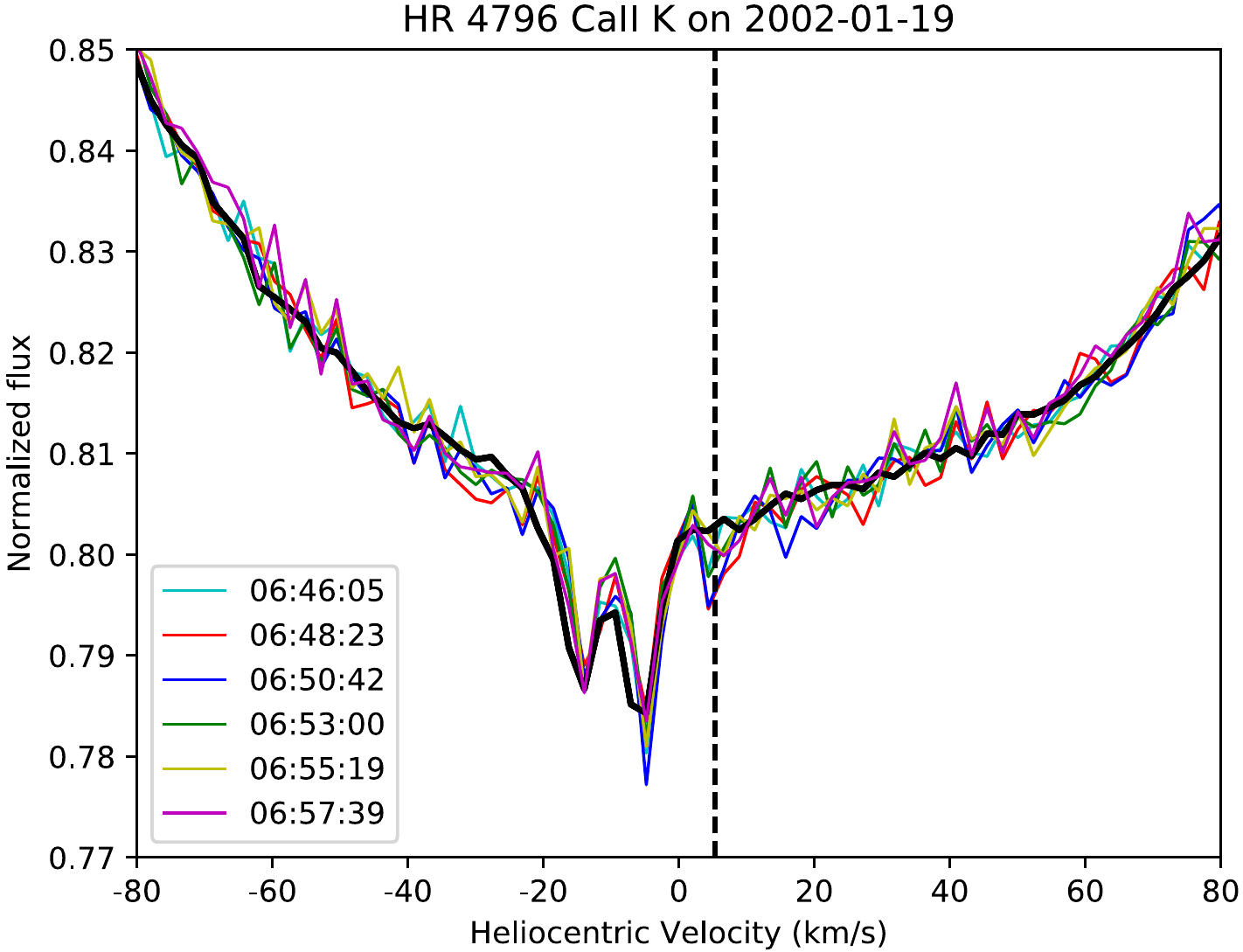}
\caption{Left: Variability in the Ca{\sc ii} K line of c\,Aql during the night of 2011-05-27, the UT of each observation is shown in the legend. Right: Example of variability detected in HR\,4796 at the stellar radial velocity along a selection of spectra taken on 2002-01-19. The thick black line in each Figure shows the median spectrum for comparison. The radial velocity of the star is marked with a dashed black line in both Figures. }
\label{fig:var}
\end{figure*}

We found variable absorption features attributable to FEB-like events in specific epochs of the objects c\,Aql and HR\,4796. In particular, in the case of c\,Aql we detected very short-term variations from within a few nights to within a few minutes. Variations detected on the night of 2011-05-27 are shown in Fig. \ref{fig:var} for the Ca\,{\sc ii} K line, and they are also present in the Ca\,{\sc ii} H and Na\,{\sc i} lines for some of the observations. These variations, likely attributable to intense exocometary activity, are detected at $\sim$35 km.s$^{-1}$, red-shifted with respect to the radial velocity of the star. %{\bf (comment on the Na corresponding presence / absence of features too?)}. 

We have also detected variability in the Ca{\sc ii} K line of HR\,4796. The observed variations appear as a small feature detected at 5.30 km.s$^{-1}$, matching the radial velocity of the star (5.35 km.s$^{-1}$). Since we have collected over 200 individual spectra of HR\,4796, in Fig.\,\ref{fig:var} we only show a selection of a few epochs as examples of the variability observed around this star. These detections are narrow and only slightly over 3$\sigma$. Since the strength of Ca\,{\sc ii} H line is roughly half that of the Ca\,{\sc ii} K line, we do not expect to have a significant detection in the latter (as was the case). However, it is reassuring (in the circumstellar gas scenario) that all the detections in the Ca\,{\sc ii} K line match the radial velocity of the star.% Further analysis of the variability detected in these two objects will be presented in Iglesias et al. (in prep).

\begin{figure}  %-> funciona bien con pdf
\includegraphics[width=0.49\textwidth]{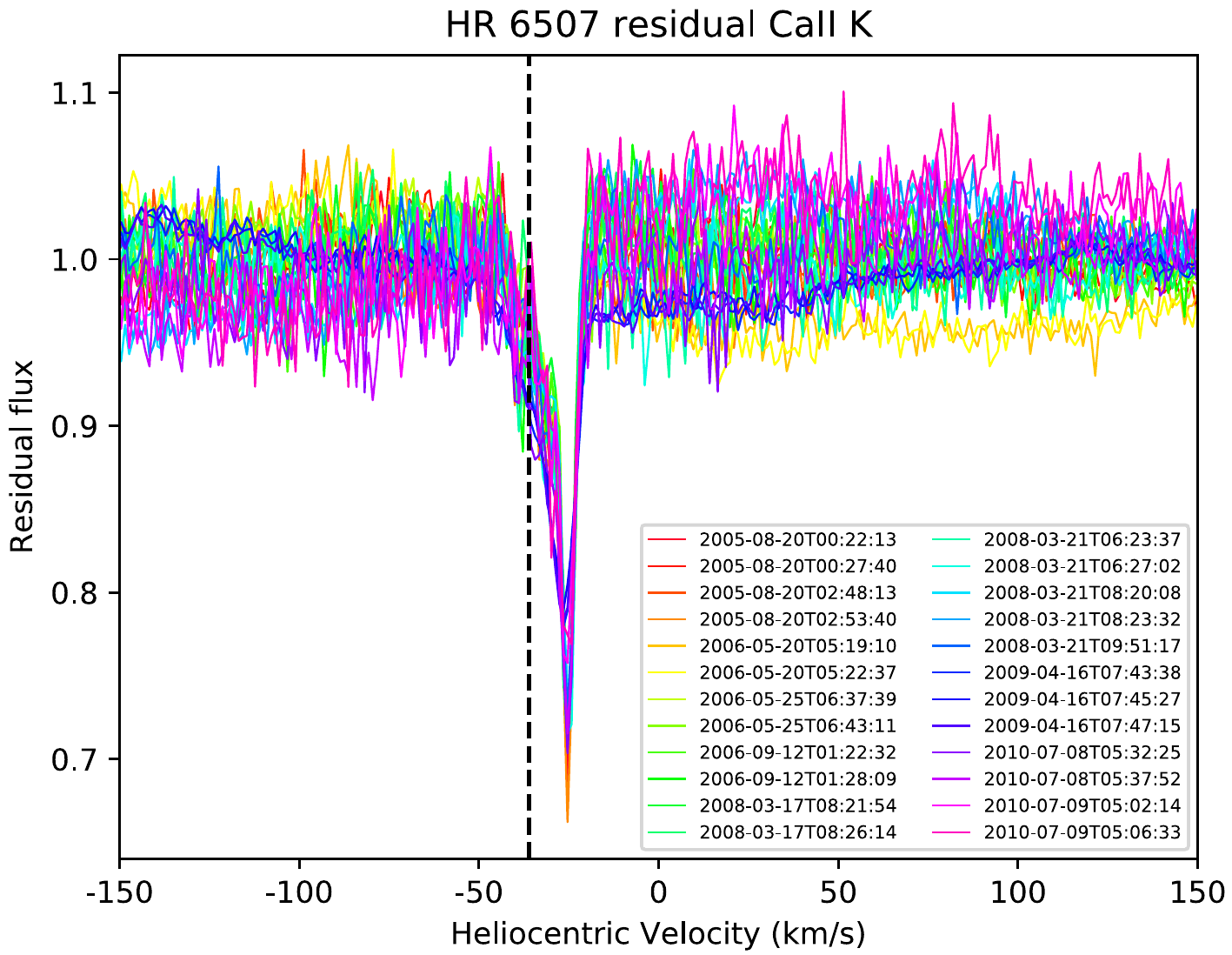}
\caption{Variability in the residuals of the Ca{\sc ii} K line of HR\,6507 after normalizing all the epochs by the reference spectrum. The date of each observation is shown. The radial velocity of the star is marked with a dashed black line. }
\label{fig:varHR6507}
\end{figure}

In addition, we have detected low-level ($\sim2\sigma$) variability in the residual spectra of HR\,6507. However, this variability presents itself as a very broad component covering the full range of velocities of the photospheric line, as can be seen in Fig.\,\ref{fig:varHR6507}.
%is not localized within certain radial velocities but it is distributed along the overall residuals, showing a wavy-like variation. 
In order to determine if this variability was produced by circumstellar gas or the star itself, we also analysed the $H_{\alpha}$ line. %, where gas is more commonly detected in emission.
Neither narrow emission nor absorption were detected in this line. However, the same broad variability was found, more consistent with photospheric variation. This star is classified as a shell star in (\citealt{HauckJaschek2000}, \citealt{Jaschek1991}), but even the shell classification is questioned in \cite{Jaschek1988} and \cite{Jaschek1998}. From the velocity field involved, this variability is more likely due to the presence of spots (as described in \citealt{Figueira2013}). %Since the radial velocity dispersion of the star along the different epochs is small ($\sim$1.6 km/s), well within the uncertainty of individual epochs, we attribute this variability to photospheric activity.

\subsection{Local Interstellar Medium features}

\subsubsection{Objects with known clouds in the line of sight}

We looked for local interstellar clouds in  the line of sight of the stars, as this clouds could explain the presence of the extra absorption lines that we observe. We used the online Local InterStellar Medium (LISM) Kinematic Calculator\footnote{http://lism.wesleyan.edu/LISMdynamics.html} \citep{Redfield08} which predicts the radial and transverse velocities of LISM clouds in any direction and calculates which clouds are traversed by any given line of sight. 

We found traversing known clouds from the \cite{Redfield08} catalogue for $18$ of our objects and, in most cases, the radial velocity of the clouds matched the velocity of some of the absorption features. In Table \ref{tab:origin} we present the clouds traversing the line of sight of each object, their heliocentric radial velocities and whether they match one of the absorption lines or not. As can be seen in the Table, a significant number of our features are attributable to gas located in the G cloud, which is an interstellar cloud located next to the Local Interstellar Cloud (LIC).

\subsubsection{Objects with Diffuse Interstellar Bands}

We have analysed the Diffuse Interstellar Bands (DIBs) at wavelengths 5780.5\AA~and 5797.1\AA. DIBs are absorption features caused by the ISM and they can be detected in the UV, optical and IR wavelengths. DIBs are much broader than the atomic interstellar lines, having full width at half maximum ranging from $\sim$0.8-30\,\AA, presumably due to unresolved rotational structure of large carbon-bearing molecules, which are common in the interstellar medium \citep{Herbig1995}. The DIBs we have chosen to analyse 
%(the previously mentioned, at 5780.5\AA~ and 5797.1\AA) 
are some of the strongest DIBs detectable in optical spectra. The presence of absorption features at any of these particular wavelengths might indicate the presence of ISM in the line of sight of the star, as DIBs are hardly attributable to circumstellar gas around pre-main sequence or main sequence stars (as opposed to objects that have departed the main sequence, see for e.g. \citealt{DiazLuis2015}).

We detected the presence of absorption lines likely to be due to DIBs in the $12$ objects listed in Table \ref{tab:DIBs}. These absorption lines are broad and diffuse, making it difficult to obtain precise measurements of their radial velocities. Therefore we use this criteria mostly to confirm the presence of ISM within a certain velocity range. We note that although in most cases we have identified diffuse bands at both wavelength locations, for HR\,3300, HD\,92536, and HD\,126135 we have detected DIBs at only one of the wavelengths. This can be explained by the fact that the intensity of the bands detected for those three sources is much lower than in the other cases; therefore we interpret the difference as a sensitivity issue rather than a physical one.

\begin{table}
	\centering
	\caption{Summary of the detection of absorptions consistent with DIBs at either 5780.5\AA or 5797.1\AA.}
	\label{tab:DIBs}
	\begin{tabular*}{\columnwidth}{lcc} % four columns, alignment for each
		\hline
		Name & DIB at 5780.5 \AA & DIB at 5797.1 \AA\\
		\hline
		$\beta$03\,Tuc & \xmark & \xmark \\
		66\,Psc & \xmark & \xmark \\
		$\nu$ Hor & \xmark & \xmark \\
		HD\,24966 & \xmark & \xmark \\
		HD\,290540 & \checkmark & \checkmark \\
		HD\,36444 & \checkmark & \checkmark \\
		HD\,290609 & \checkmark & \checkmark \\
		HR\,1919  & \xmark & \xmark \\
		HD\,54341 & \xmark & \xmark \\
		HD\,60856 & \checkmark & \checkmark \\
		HR 3300 & \xmark & \checkmark \\
		$\eta$ Cha & \xmark & \xmark \\
		HD\,92536 & \checkmark &\xmark \\
		3 Crv & \xmark & \xmark \\
		HD\,106036 & \xmark & \xmark \\
		HR\,4796  & \xmark & \xmark \\
		HD\,110058 & \xmark & \xmark \\
		HD\,112810 & \xmark & \xmark \\
		HD\,126135 & \checkmark & \xmark\\
		HD\,141378 & \xmark & \xmark \\
		HD\,141327 & \checkmark & \checkmark  \\
%		HD\,144981 & \checkmark & \checkmark \\
%		HD\,145554 & \checkmark & \checkmark  \\
%		HD\,145631 & \checkmark & \checkmark \\
%		HR\,6051 & \checkmark & \checkmark \\
		HR\,6507 & \xmark & \xmark \\
		c Aql & \xmark & \xmark \\
		\hline
	\end{tabular*}
\end{table}

\subsubsection{Nearby stars analysis}

Similarly to the analysis performed on the objects in our sample, we analysed the Ca\,{\sc ii} K lines of nearby stars searching for the presence of absorption features at similar velocities to the ones observed in our objects. Finding these similar absorption lines in the line of sight towards nearby stars would strongly suggest an ISM origin for the gas feature(s). We chose to analyse the Ca\,{\sc ii} K line since its absorption is more intense and easier to detect than the H line and it is the main tracer of circumstellar gas in the optical. 

For each star of our sample, we searched for high resolution spectra of nearby stars within a search-box of up to 6 degrees ($\sim$ 3 degrees radius). Considering the distances of the objects, the equivalent projected separations between the targets and their neighbours range between $\sim$0.1\,pc and $\sim$30\,pc. From the gathered data, we only considered objects having early spectral types, ideally between B0 and F5, since absorption features are harder to detect in later spectral types. Non-photospheric gas absorptions are easily spotted when superimposed on fast rotators having wider (and fewer) spectral lines. 

We found suitable nearby stars for all the candidates except for 66\,Psc and HD\,24966. The observations used for this analysis are described in Table \ref{tab:nearby}. We found absorption features present in all the nearby stars and they match most of the absorption features found in our objects, confirming the interstellar origin for the majority of the features. In the case of HD\,110058, we found that three nearby stars presented one absorption feature matching HD\,110058 absorption line at $\sim$1 km.s$^{-1}$, one star shows a weak absorption matching the radial velocity of the G cloud, but none of the nearby stars shows any signs of absorption lines matching the one at $\sim$12 km.s$^{-1}$, which also happens to be near the estimated radial velocity of this star. Therefore we propose a circumstellar origin for this feature. 

A comparison of the nearby stars absorption lines against our objects and their respective angular separations are shown in Figs.\,\ref{fig:near1} to \ref{fig:near4}. 

\subsection{General results}
\label{GralRes}

We present a summary of our results regarding stable features in Table\,\ref{tab:origin}, in which we report the radial velocity of the star, the traversing clouds, their radial velocities and whether they match one of the observed features, the average velocities of each absorption feature, its Ca{\sc ii}/Na{\sc i} density ratio, whether it has a matching absorption in a nearby star and our verdict on its origin; ISM (InterStellar Medium) or CS (CircumStellar).

Most of the stable features are likely produced by clouds in the line of sight and not by the circumstellar medium, except in the cases of HR\,4796 and HD\,110058. We find that two objects present variability: HR\,4796 shows flux variations in its feature located at the same velocity as the star and c\,Aql exhibits transient red-shifted absorption lines with characteristics of FEB-like events. Another interesting case is that of HR\,6507, for which a clear diagnostic cannot be attained with the available data as discussed in Sec.~\ref{sec:varFeat}. 

\onecolumn

\begin{center}
\begin{longtable}{lc|ccc|ccccc}
%\textbf{
	\caption{Absorption components and their mean radial velocity, Ca{\sc ii}/Na{\sc i} density ratio,  absorption feature detection in a nearby star and proposed origin.} \label{tab:origin} \\
 % four columns, alignment for each
		\hline
		Name &  Stellar RV & Cloud & Cloud RV & Matching & Feature RV  & $\frac{N(\mathrm{Ca}\textsc{ii})}{N(\mathrm{Na}\textsc{i})}$ & Nearby & Origin\\
		& [km.s$^{-1}$]  & Name & [km.s$^{-1}$] & Feature? & [km.s$^{-1}$] &  & star &\\
		\hline
		\endfirsthead
			
 	\multicolumn{5}{c}{{\bfseries \tablename\ \thetable{} -- continued from previous page}} \\
 		\hline 
		Name &  Stellar RV & Cloud & Cloud RV & Matching & Feature RV  & $\frac{N(\mathrm{Ca}\textsc{ii})}{N(\mathrm{Na}\textsc{i})}$ & Nearby & Origin\\
		& [km.s$^{-1}$]  & Name & [km.s$^{-1}$] & Feature? & [km.s$^{-1}$] &  & star &\\
 	\hline
 	\endhead
 	
 	\hline \multicolumn{9}{r}{{Continued on next page}} \\ 
	\endfoot
	
	\hline 
	\endlastfoot
$\beta$03\,Tuc & 6.05 $\pm$ 1.60 & -- & -- & --  & -11.22 & 3.47 & \checkmark & ISM \\
 & & Dor & 13.85 $\pm$ 0.65 & \xmark & -3.74 & 0.38 & \checkmark & ISM \\
 & & Vel & 2.54 $\pm$ 0.78 & \checkmark & 2.26 & 13.08 & \checkmark & ISM \\
66\,Psc & 4.32 $\pm$ 2.66 & -- & -- & -- & -5.71 & 0.63 & -- & ISM \\
&  & -- & -- &  --& 0.49 & 0.40 & -- & ISM \\
&  & LIC  & 11.44 $\pm$ 1.29 & \checkmark & 11.49 & 7.03 & -- & ISM \\
$\nu$ Hor & 13.58 $\pm$ 1.67 & G &  5.33 $\pm$ 1.52 & \checkmark & 4.11 & 6.16 & \checkmark & ISM \\
&  & Vel & 12.65 $\pm$ 0.93 & \checkmark & 12.74 & 2.94 & \checkmark & ISM \\
&  & Cet & 9.85 $\pm$ 0.63 & \xmark & -9.09 & $< 1$  & \checkmark & ISM \\
HD\,24966 & 15.70 $\pm$ 3.15 & Blue  &  10.59 $\pm$ 1.30 & \xmark & -13.10 & $> 5$ & -- & ISM \\
&  & G  & 17.51 $\pm$ 1.38 & \checkmark & 15.68 & 8.40 & -- & ISM \\
&  & Dor  & 32.11 $\pm$ 0.85 & \checkmark & 32.15 & 6.58 & -- & ISM \\
HD\,290540 & 27.26 $\pm$ 3.12 & -- & -- & -- & 9.28 & 1.31 & \checkmark & ISM \\
&  & -- & --  & -- & 23.78 & 1.04 & \checkmark & ISM \\
&  & -- & -- & -- & 35.51 & 0.46 & \checkmark & ISM \\
&  & -- & -- & -- & -3.99 & $< 1 $ & \checkmark & ISM \\
HD\,36444 & 26.72 $\pm$ 4.50 & -- & -- & -- & 8.00 & 0.70 & \checkmark & ISM \\
&  & LIC & 22.75 $\pm$ 0.96 & \checkmark & 21.99 & 0.72 & \checkmark & ISM \\
&  & -- & -- & -- & 31.30 & 1.00 & \checkmark & ISM \\
&  & -- & -- & -- & 40.51 & $> 5$  & \checkmark & ISM \\
 HD\,290609 & 25.87 $\pm$ 1.65 & -- & -- & -- & -8.89 & 3.05 & \checkmark & ISM \\
&  & -- & -- & -- & 8.99 & 2.70 & \checkmark & ISM \\
&  & -- & -- & -- & 23.50 & 0.73 & \checkmark & ISM \\
&  & -- & -- & -- & 34.90 & 3.24 & \checkmark & ISM \\
&  & -- & -- & -- & 51.26 & 7.33 & \checkmark & ISM \\
HR\,1919 & 23.49 $\pm$ 1.28 & -- & -- & -- & 10.63 & 5.59 & \checkmark & ISM \\
&  & -- & -- & -- & 31.11 & 11.97 & \checkmark & ISM \\
&  & -- & -- & -- & 36.64 & 12.35 & \checkmark & ISM \\
HD\,54341 & 41.03 $\pm$ 0.95 & -- & -- & -- & 3.31 & $< 1$ & \checkmark & ISM \\
&  & Blue & 9.59 $\pm$ 0.93 & \checkmark & 10.32 & $> 5$ & \checkmark & ISM \\
&  & -- & -- & -- & 25.38 & 2.80 & \checkmark & ISM \\
HD\,60856 & 31.89 $\pm$ 1.59 & LIC  & 16.37 $\pm$ 1.18 & \xmark & 20.32 & 0.40 & \checkmark & ISM \\
&  & -- & -- & -- & 29.98 & 1.55 & \checkmark & ISM \\
HR 3300 & 22.35 $\pm$ 0.72 & G & 4.62 $\pm$ 0.94 & \checkmark & 5.95 & 2.97 & \checkmark & ISM \\
&  & Vel & 15.00 $\pm$ 0.97 & \checkmark & 16.75 & 1.08 & \checkmark & ISM \\
&  & Cet & 20.50 $\pm$ 0.87 & \checkmark & 20.82 & $> 5$  & \checkmark & ISM \\
$\eta$ Cha & 15.21 $\pm$ 1.41 & G  & -4.07 $\pm$ 1.17 &\checkmark  & -2.84 & 2.47 & \checkmark & ISM \\
&  & Vel  & 0.02 $\pm$ 0.75 & \xmark & 10.94 & 2.07 & \checkmark & ISM \\
HD\,92536 & 15.45 $\pm$ 0.44 & G  & -6.09 $\pm$ 0.97 & \xmark & 0.90 & 6.52 & \checkmark & ISM \\
&  & -- & -- & -- & 9.47 & 0.77 & \checkmark & ISM \\
&  & -- & -- & -- & 17.84 & 2.21 & \checkmark & ISM \\
3 Crv & 14.41 $\pm$ 1.09 & Leo  & -5.36 $\pm$ 1.05 & \checkmark & -7.00 & 1.65 & \checkmark & ISM \\
&  & -- & -- & -- & -1.89 & 9.91 & \checkmark & ISM \\
&  & Gem  & 2.54 $\pm$ 0.95 & \checkmark & 2.96 & $> 5$  & \checkmark & ISM \\
 HD\,106036 & 9.37 $\pm$ 2.34 & G &  -11.13 $\pm$ 0.98 & \xmark & -4.78 & -- & \checkmark & ISM \\
&  & -- & -- & -- & 10.18 & -- & \checkmark & ISM \\
HR\,4796 & 5.35 $\pm$ 2.94 & -- & -- & -- & -12.97 & 3.79 & \checkmark & ISM \\
&  & -- & -- & -- & -5.38 & 0.64 & \checkmark & ISM \\
&  & -- & -- & -- & 5.20 & $> 5$  & \xmark & CS \\
HD\,110058 & 11.20 $\pm$ 0.81 & G & -14.46 $\pm$ 0.97 & \xmark & 0.95 & 1.68 & \checkmark & ISM \\
&  & -- & -- & -- & 12.35 & 1.02 & \xmark & CS \\
 HD\,112810 & 5.25 $\pm$ 2.24 & G &  -15.70 $\pm$ 0.97 & \xmark & -11.48 & -- & \checkmark & ISM \\
&  & -- & -- & -- & -4.12 & -- & \checkmark & ISM \\
&  & -- & -- & -- & 3.02 & -- & \checkmark & ISM \\
HD\,126135 & 11.73 $\pm$ 0.67 & NGP  &  -24.28 $\pm$ 1.22 & \checkmark & -22.46 & 0.82 & \checkmark & ISM \\
&  & Gem  & -14.33 $\pm$ 1.01 & \checkmark & -14.58 & 0.80 & \checkmark & ISM \\
&  & -- & -- & -- & -9.07 & 1.89 & \checkmark & ISM \\
&  & -- & -- & -- & 4.15 & 0.20 & \checkmark & ISM \\
&  & -- & -- & -- & 26.30 & $< 1$  & \checkmark & ISM \\
HD\,141378 & -14.68 $\pm$ 2.62 & G & -28.37 $\pm$ 1.18 & \checkmark & -30.03 & $> 5$ & \checkmark & ISM \\
&  & -- & -- & -- & -15.02 & $> 5$ & \checkmark & ISM \\
HD\,141327 & -4.65 $\pm$ 2.40 & G & -27.31 $\pm$ 1.08 & \xmark & -21.69 & 7.32 & \checkmark & ISM \\
&  & -- & -- & -- & -3.05 & 0.31 & \checkmark & ISM \\
&  & -- & -- & -- & 14.46 & 11.08 & \checkmark & ISM \\
&  & -- & -- & -- & 22.65 & 2.12 & \checkmark & ISM \\
%HD\,144981 & -6.47 $\pm$ 1.59 & G & -29.16 $\pm$ 1.10 & \checkmark & -26.16 & 0.83 & \checkmark & ISM \\
%&  & -- & -- & -- & -10.32 & 0.20 & \checkmark & ISM \\
%HD\,145554 & -9.66 $\pm$ 1.49 & G  &  -29.20 $\pm$ 1.10 & \checkmark & -28.23 & 0.33 & \checkmark & ISM \\
%&  & -- & -- & -- & -12.23 & 0.36 & \checkmark & ISM \\
%HD\,145631 & -9.86 $\pm$ 1.90 & G & -29.21 $\pm$ 1.10 & \checkmark & -28.51 & 0.50 & \checkmark & ISM \\
%&  & -- & -- & -- & -11.21 & 0.32 & \checkmark & ISM \\
%HR\,6051 & -7.26 $\pm$ 4.11 & G & -29.14 $\pm$ 1.10 & \checkmark & -26.55 & 0.29 & \checkmark & ISM \\
%&  & -- & -- & -- & -11.49 & 0.13 & \checkmark & ISM \\
HR\,6507 & -36.13 $\pm$ 2.49 & -- & -- & -- & -38.95 & 3.72 & \checkmark & ISM \\
&  & -- & -- & -- & -29.71 & 7.33 & \checkmark & ISM \\
&  & -- & -- & -- & -25.02 & 1.99 & \checkmark & ISM \\
c Aql & 17.29 $\pm$ 2.61 & Mic, Aql & -26.86, -25.26 & \xmark & -31.43 & $> 5$ &\xmark& ISM \\
&   & Eri & -20.11 $\pm$ 1.14 & \checkmark & -19.77 & $> 5$ & \checkmark & ISM \\
%		\hline
%}
\end{longtable}
\end{center} 

\twocolumn

\section{Discussion}
\label{sec:discussion}

Gas absorption features superimposed on photospheric lines look fairly similar whether they are caused by clouds in the line of sight of the star or by the presence of stable gas in the circumstellar environment. Therefore, a detailed analysis, involving multiple criteria, has to be performed in order to discriminate between the two scenarios. %There are several criteria that can be helpful to determine the origin of these features which we have covered in our analysis.
Below we discuss our results regarding the origin of the features in ``interstellar'' and ``circumstellar'' categories.

\subsection{Interstellar-like features}

Most of the absorption features found in this study are classified as ``interstellar'' as they do not present significant time variability beyond the noise level or attributable to different instrument or resolution. In addition, all these features posses other characteristics such as having a composition consistent with typical ISM values, matching clouds in their line of sight, or detection of a similar feature in a nearby star with velocities matching within $3\sigma$.

%\subsubsection{HD\,144981, HD\,145554 and HD\,145631}
%Similar to the previous case, HD\,144981, HD\,145554 and HD\,145631 are located within 0.72$^{\circ}$ of each other and present similar absorption features. As shown in Fig. \ref{fig:near3}, the three objects along with the nearby star HR\,6026, present remarkably similar absorption lines in terms of depth and velocity. In this case, the three objects present two features with all the characteristics pointing towards an interstellar origin. The radial velocities of the absorption lines are blueshifted with respect to the radial velocities of the stars, all the features present column density Ca{\sc ii}/Na{\sc i} ratios <1, the three objects present absorptions attributable to DIBs and, more importantly, the three of them have matching features in other nearby stars. In addition, the component at $\sim$-28 km/s matches well with the radial velocity of the G cloud, which traverses the line of sight at $\sim$-29 km/s \citep{Redfield08}. We did not find any nearby cloud at a radial velocity similar to the deep absorption feature at $\sim$-11 km/s, but the similarity of the feature present in the three objects and also in the fourth nearby star is clear, therefore we conclude interstellar origin with high certainty.
%\subsubsection{Features matching a known cloud}
Overall, we found $21$ absorption features matching the radial velocity of known clouds traversing the lines of sight of the stars. A summary of the traversing clouds, their radial velocities and the matching absorptions is provided in Table \ref{tab:origin}. 
The evidence for these features to be caused by those clouds in the line of sight, is strengthened by the fact that they are also detected in nearby stars around our science targets. A particular case of this phenomenon is observed in the objects HD\,290540, HD\,36444 and HD\,290609.

These three stars are located within an angular separation of 0.7$^{\circ}$ of each other. As can be seen in Fig.\,\ref{fig:near1}, the three objects present similar absorption features at similar velocities, which are also detected for three other nearby stars within an angular radius of 1$^{\circ}$. We obtained comparable radial velocities for these three objects, around $26$\,km.s$^{-1}$. As noticeable in Fig.\,\ref{fig:near1}, they all have a deep absorption line close to $23$\,km.s$^{-1}$, which corresponds to the Local Interstellar Cloud \citep[LIC,][]{Redfield08}. Although this absorption line is found to be close to the radial velocity of the stars, its interstellar origin is clear as it is confirmed by being present in other three stars with a similar line of sight and having a N(Ca{\sc ii}/Na{\sc i}) ratio consistent with ISM ($\lesssim$1). The other absorption lines seen in the three objects at $\sim 9$ and $\sim 34$\,km.s$^{-1}$ also seem to have a common interstellar origin. The one at $\sim 9$\,km.s$^{-1}$ is possibly attributable to the Hyades cloud at $\sim 11$\,km.s$^{-1}$, which according to \cite{Redfield08} crosses near ($<20^{\circ}$) the line of sight of these three stars. We did not find any known cloud traversing a similar line of sight at a radial velocity close to $\sim 34$\,km.s$^{-1}$, however, since this absorption line is present in several stars at a similar velocity we also conclude that it is of interstellar origin. 

There is a fourth feature at $\sim 41$\,km.s$^{-1}$ for HD\,36444 detected only in the Ca\,{\sc ii} lines which, due to its mostly Calcium composition (N(Ca{\sc ii}/Na{\sc i})$>5$) could be consistent with having a circumstellar origin. However, a similar feature is observed in the nearby star HR\,1863, and therefore it is likely to be another feature of interstellar origin, possibly warm ISM, which has been reported to have a composition richer in Calcium than cold ISM \citep{Bertin1993}. Unfortunately we have obtained only one epoch for HD\,36444 therefore we were not able to investigate the variability of this feature. Further data is thus necessary in order to fully rule out circumstellar origin.%Nevertheless, in order to fully didespite its apparent interstellar origin it would be worth obtaining more epochs for this object to analyse variability of this feature and discard or confirm a possible circumstellar origin.

HD\,290609 also presents a fourth feature, but it is detected around $\sim 51$\,km.s$^{-1}$. However, we point out that this group of stars is located within the Orion OB1 association \citep{Hernandez2006} making it likely that environmental nebular gas is observed at different velocities \citep{Brown1994}. In any case, although we conclude an ISM origin for HD\,290609's fourth feature because of the high frequency of interstellar clouds observed in the surroundings; as in the previous case, there is only one spectrum available for HD\,290609 and it would be interesting to perform further analysis gathering more epochs in order to better assess the origin of this feature.
%These stable absorptions matching the radial velocity of a known cloud are confirmed by being also detected in nearby stars around the studied debris disks system. The particularity of having a known cloud traversing the same sight line at a similar velocity and its confirmation by detecting a similar absorption in another star within a small angular separation leads us to determine an interstellar origin for all the features listed in Table \ref{tab:cloud} which coincides with a cloud described in \cite{Redfield08} and thus are marked with "\checkmark" in this Table.

%\subsubsection{Features with matching absorptions in nearby stars}

%We found suitable nearby stars data in the ESO archive for all the targets with the exception of 66\,Psc and HD\,24966. We found absorption features present in all the nearby stars and they match most of the absorption features found in our objects, confirming the interstellar origin for the majority of the features.

\subsection{Circumstellar-like features}

\subsubsection{Stable features with no matching absorptions in nearby stars}

In the case of HD\,110058, we found that three nearby stars present one absorption feature matching HD\,110058's absorption at $\sim 1$\,km.s$^{-1}$, thus we propose an interstellar origin for this feature. This interstellar feature was also reported by \cite{Hales2017}, who analysed MIKE spectra of HD\,110058 and three nearby stars (at angular separations between $1.2^{\circ}$ and $2.8^{\circ}$) and found matching features for the absorption at $\sim 1$\,km.s$^{-1}$. We obtained further spectra for four different nearby stars in the ESO archive (at angular separations between $0.74^{\circ}$ and $2.16^{\circ}$) and confirmed the matching absorption lines in three stars, in agreement with the findings by \cite{Hales2017}. Given the distance to this star ($188.7 \pm 34.1$\,pc), both studies cover a region of $2.4-9.2$\,pc in radius. Considering a typical radius of $1.5$\,pc for the warm local ISM material located within $15$\,pc from the Sun \citep{Redfield08}, the projected coverage at the distance of HD\,110058 would be about $19$\,pc, thus the local ISM material would likely cover the region in which the nearby stars are located. On the other hand, our measurements of the equivalent widths of this feature is in agreement with a more recent work by \cite{Rebollido2018}, where the authors report that the strength measured for this blue-shifted component varies with respect to \cite{Hales2017} measurement, proposing a possible circumstellar origin for the blue-shifted feature at $\sim 1$\,km.s$^{-1}$. Considering the scenario of variability in the blue-shifted component and a possible overlap of circumstellar feature over the interstellar it would be worth performing follow-up observations of this object to better assess the origin of this feature. There is a fourth nearby star analyzed which does not present a feature at said velocity, but shows a weak absorption line at $\sim -15$\,km.s$^{-1}$ matching the radial velocity of the G cloud. 
\\
HD\,110058 presents an additional absorption feature at $\sim 12$\,km.s$^{-1}$ which is very near our estimate of the radial velocity of this star ($11.20$\,km.s$^{-1}$). None of the nearby stars analyzed show any sign of absorption matching HD\,110058's absorption line at $\sim 12$\,km.s$^{-1}$. Therefore we propose a circumstellar origin for this feature. This circumstellar feature was also proposed by \cite{Hales2017} and confirmed by \cite{Rebollido2018}, thus our analysis is in agreement with their conclusions.

\subsubsection{Variable features}
\label{sec:varFeat}

We detected variable absorption features attributable to FEB-like events in the objects c\,Aql and HR\,4796. The detection of FEB-like events in c\,Aql was previously reported by \cite{Montgomery2017}, where they detected high variations from night to night and attributed them to exocometary activity. Furthermore \cite{Welsh2013} reported some nightly changes but no FEBs-like events. With the data collected from the ESO archive we found, in addition to night to night variations, strong variability within very short time scales of only a couple of minutes. To our knowledge, this is the shortest-term variability detected to date in such systems. This object is known to be a pulsating star with a period of 30.39 minutes \citep{Kuschnig1994} but a phase analysis of the Ca{\sc ii} K and the H$\alpha$ lines does not indicate any such periodicity. In addition, the residual absorption events are not associated with a counterpart in emission at a mirrored velocity with respect to the radial velocity of the star (even taking into account the uncertainty in the latter), as one would expect from pulsations. In the left panel of Fig.\,\ref{fig:var}, we show the variability detected in c\,Aql through eight individual spectra taken within a time span of $\sim 20$\,minutes. Similar short term variability has only been observed so far in $\beta$ Pictoris \citep{Kiefer2014BetaPicNature} and the shell star $\phi$ Leo \citep{Eiroa16}, with reported variability within hours. 

Variability in the Ca{\sc ii} K line of HR\,4796 at the same radial velocity of the star is reported here for the first time. Previously, only a sporadic absorption at $\sim 60$\,km.s$^{-1}$ during the night of 2007-05-04 was reported by \cite{WelshMontg2015}. A more detailed analysis of the variability detected in these two objects will be presented in Iglesias et al. (in prep).

Regarding other objects in our sample with claims of variability in the literature, there are also HR\,6507 \citep{WelshMontg2015} and HD\,24966 \citep{Welsh2018}. % For the case of HR\,6051, \cite{Welsh2013} reported the detection of weak absorption features on specific nights, which the authors attributed to the evaporation of exocomets. Their spectra were taken on May 2011 with the Sandiford Echelle Spectrograph at the McDonald Observatory, Texas. In our data of HR\,6051 taken on March 2016 with FEROS, we do not detect any significant additional absorption line besides the stable interstellar ones. Since both observations are similar in terms of S/N and resolution, we suspect our non detections could be due to the lack of events during those particular nights. It would be worth analysing more observations to confirm the activity observed in this object by previous works.
%\\
In the case of HR\,6507, \cite{WelshMontg2015} modelled the absorption lines using two components at radial velocities $\sim -37$\,km.s$^{-1}$ and $\sim -28$\,km.s$^{-1}$. The observations were taken with the Sandiford Echelle Spectrograph at the McDonald Observatory, Texas, with $\sim 60000$ resolution. We combined HARPS and UVES spectra (with higher resolution, $\sim 100000$ and $\sim 80000$, respectively) and were able to distinguish and fit three absorption features at radial velocities $\sim -39$\,km.s$^{-1}$, $\sim -29$\,km.s$^{-1}$ and $\sim -25$\,km.s$^{-1}$. \cite{WelshMontg2015} attributed a circumstellar origin for the feature observed at $\sim -37$\,km.s$^{-1}$ because of its proximity to the radial velocity of the star ($\sim -36$\,km.s$^{-1}$). However, for the corresponding feature, which we measure at $\sim -39$\,km.s$^{-1}$, we found absorption lines of similar velocity and intensity in nearby stars, suggesting ISM origin. 

HR\,6507 has also been reported to posses shell star signatures (\citealt{HauckJaschek2000}, \citealt{Jaschek1991}), although in other works no clear indication of a shell has been found, attributing this to a weakening or disappearing of the shell (\citealt{Jaschek1988}, \citealt{Jaschek1998}). Nevertheless, considering the possibility of HR\,6507 being a shell star, it is likely that it possesses circumstellar gas and therefore shows gas signatures at its radial velocity. As mentioned in \ref{sec:vari}, we found small variability in the overall residuals of all the lines observed for HR\,6507. In \cite{HauckJaschek2000}, they report the star as variable and possibly micro-variable, which might explain the observed variations. Taking all this into account, we do not attribute the variability to FEBs-like events, but a more detailed study of this source is needed to achieve stronger conclusions.% we need to further study this object since there is the possibility of having both, circumstellar and interstellar gas near the stellar radial velocity.

In the case of HD\,24966, \cite{Welsh2018} recently proposed the detection of exocomets at different velocity ranges in two out of three observations. The significance and interpretation of such variable transient FEB absorption features will be further investigated in Iglesias et al. (in prep).

\subsubsection{Relationship between circumstellar-like features and system properties}
\label{sec:DiscRelation}

Gas absorptions features of presumed circumstellar origin were found in the systems HD\,110058, HR\,4796 and c Aql. 

The debris disk around HD\,110058 has been resolved with SPHERE by \cite{Kasper2015}, where they determined an inclination of $\sim90^{\circ}$. This edge-on orientation reinforces the circumstellar verdict on the gas origin, as this is the most favourable orientation for potentially detecting gas lines in absorption. Since the absorption line that we detect is deep, narrow, stable, and close to the radial velocity of the star, it is consistent with a stable gas component, possibly located in the outer regions of the disk (\citealt{Beust1998}, \citealt{Brandeker2004}).

HR\,4796 has an inclination of 76.5$^{\circ}$ (\citealt{Milli2017}, \citealt{Kennedy2018}) which is fairly close to edge-on. Since somewhat small misalignments in cometary orbits with respect to the parental disk are common \citep{Nesvorny2017}, it is not necessarily unlikely to detect FEB events.

Regarding the disk around c\,Aql, it has only been marginally resolved with Herschel by \cite{Morales2016}. They estimated an inclination of 21$^{\circ}$ $\pm$ 42$^{\circ}$. Although this inclination does not seem favourable for circumstellar gas detections using optical spectroscopy, the estimated uncertainty on the inclination is very large as the disk was only marginally resolved. The possibility of a much higher inclination cannot be ruled out. Even if the shallower inclination is confirmed, we could still be spotting the activity of bodies with highly inclined orbits. Overall, it is reassuring that two of our candidates with gas detections are close to edge-on, with very robust inclination determinations.

Regarding the rest of the sample, only two other objects have been marginally resolved (with Herschel): $\nu$ Hor, modelled with 73.4$^{\circ}$ $\pm$ 6.5$^{\circ}$ inclination \citep{Moor2015}, and HD\,141378, with an estimated inclination of 60$^{\circ}$ $\pm$ 37$^{\circ}$ \citep{Morales2016}. We did not detect signs of circumstellar gas in these objects. Nevertheless, it would be worth performing follow-up studies to be able to analyse more epochs and therefore increase the chances of detecting stochastic activity or provide more robust evidence of the lack of such activity.

\section{Conclusions}
\label{sec:conclusions}

In this work we have analysed the multiple absorption features present in the Ca\,{\sc ii} H \& K and Na\,{\sc i} D1 and D2 lines of $23$ debris disks systems using optical high-resolution spectroscopy in order to determine if their origin is of circumstellar or interstellar nature. 

We found gas absorptions of circumstellar nature in three objects: HD\,110058, HR\,4796 and c\,Aql. HD\,110058 presents a strong stable absorption consistent with a gaseous disk, possibly residual gas leftover from the earlier gas-rich stage of the disk or from very active planetesimal collision episodes. 

Variable absorption features were found in the spectra of HR\,4796 and c\,Aql. A weak circumstellar absorption was found in HR\,4796 at the same radial velocity as the star with flux variations over $3\sigma$, possibly due to photo-dissociation processes or collisions of icy bodies producing changes in the gas content of the disk. 
Highly variable red-shifted absorptions were detected in c\,Aql, with substantial variations observed on time scales shorter than two minutes, which is the shortest variability detected so far in this type of lines. These fast changing signatures are likely due to exocometary activity within the disk surrounding c\,Aql. For these two objects, HR\,4796 and c\,Aql, we will present a more detailed analysis of the variable features in a future work. %Iglesias et al. (in prep).
The circumstellar gas detections are in agreement with the near edge-on inclinations of the two objects with robust inclination measurements: HD\,110058 with an inclination of $\sim90^{\circ}$ and HR\,4796 with an inclination of 76.5$^{\circ}$.

Given $h, r$ and $i$ the scale-height, radial distance and inclination of a circumstellar disk, respectively, and assuming a typical scale-height/distance ratio of $h/r \sim$0.1 for debris disks \citep{Thebault2009}, the typical angle subtended by the disk should be $\sim 5.7^{\circ}$. For a uniform distribution of $\sin(i)$ between 0 and 1, the probability of i $\geq (90^{\circ}- 5.7^{\circ})$ is $\sim$10\%, and therefore, the probability for a randomly inclined system to be found close to edge-on or with an inclination suitable to detect circumstellar gas absorptions is $\sim$10\%. 

However, the sample analyzed in this paper cannot, a priori, be considered ``random'' because of the selection criterion of ``having multiple absorption features'' (that may bias the sample towards objects with circumstellar gas on close to edge-on orientation). On the other hand, this sample is actually unbiased with respect to stochastic detections such as FEBs (we remind the reader that the selection function was performed on the reference spectra, i.e. only stable components are considered).

Bearing in mind that the inclination constraints could be more relaxed regarding the detection of FEBs (as discussed in Sec. \ref{sec:DiscRelation}); that FEBs, given their stochastic nature, may not be detected by mere chance at our epochs of observations; and that, in any case, we do detect circumstellar gas in three cases out of 23 objects (one stable and two variables); our results definitely point towards gas in debris disks not being a rare phenomenon. We will, however, have more quantitative and robust results on this matter once the full sample of 301 debris disks is analyzed.

%The latter statement, together with the assumption of random orientations for our sample, suggests that the presence of gas (stable or FEB-like) in debris disks may be a very common phenomenon.
%Given the fact that, for our bonefide gas detections, two out of the three objects (in our sample of $23$) have well constrained near edge-on inclinations, we suggest that the presence of gas in debris disks is not such an uncommon phenomenon.

%The last numbered section should briefly summarise what has been done, and describe
%the final conclusions which the authors draw from their work.

\section*{Acknowledgements}

DI would like to thank Iv\'an Lacerna, support astronomer at the MPG/ESO 2.2m telescope, La Silla Observatory, for his significant help and support during the observing runs with FEROS and Alain Smette, the author of Molecfit, for his great help on improving telluric corrections. AB and DI acknowledge financial support from the Proyecto Fondecyt Iniciaci\'on 11140572. DI, AB and JO acknowledge support from the Millennium Science Initiative (Chilean Ministry of Economy), through grant N\'ucleo Milenio de Formaci\'on Planetaria. This work has made use of data from the ESA space mission Gaia (http://www.cosmos.esa.int/gaia), processed by the Gaia Data Processing and Analysis Consortium (DPAC, http://www.cosmos.esa.int/web/gaia/dpac/consortium). Funding for the DPAC has been provided by national institutions, in particular the institutions participating in the Gaia Multilateral Agreement. This publication makes use of VOSA, developed under the Spanish Virtual Observatory project supported from the Spanish MICINN through grant AyA2011-24052.
%We also thank our granted telescope time of programs
Data from the following programmes have been used in this work: CNTAC 096.A-9018(A) and 094.A-9012(A); and ESO 096.C-0238(A), 097.C-0409(A), 097.C-0409(B), 073.C-0733(E), 075.C-0689(A), 077.C-0295(A), 077.C-0295(B), 184.C-0815(A), 184.C-0815(E), 184.C-0815(F), 096.C-0238(A), 075.C-0689(B), 077.C-0295(D), 080.C-0712(A), 094.C-0946(A), 076.C-0279(A), 076.C-0279(B), 076.C-0279(C), 078.C-0209(A), 082.A-9004(A), 088.A-9029(A), 66.D-0284(A), 60.A-9036(A), 60.A-9120(B), 60.A-9122(B), 084.A-9003(A), 085.A-9027(G), 184.C-0815(C), 68.C-0548(A), 078.A-9059(A), 079.A-9007(A), 079.A-9009(A), 079.C-0789(A), 085.A-9027(B), 084.D-0067(A), 087.A-9013(A), 092.A-9006(A), 179.C-0197(A), 179.C-0197(C), 082.D-0061(A), 077.C-0295(C), 083.C-0676(A), 079.D-0567(A), 083.A-9014(A), 083.A-9014(B), 087.B-0308(A), 073.C-0733(E), 075.C-0689(A), 077.C-0295(A), 077.C-0295(B), 094.A-9012(A), 179.C-0197(B), 088.C-0498(A), 074.B-0455(A), 266.D-5655(A), 194.C-0833(C), 096.A-9030(A), 096.A-9024(A), 082.C-0831(A), 084.C-1008(A), 084.A-9004(B), 091.D-0414(B), 088.A-9003(A), 072.D-0410(A), 098.C-0463(A), 093.D-0852(A), 094.C-0946(A), 094.A-9012(A), 084.A-9003(A), 086.A-9006(A), 078.D-0549(A), 084.A-9003(A), 073.D-0291(A), 074.D-0300(A), 076.C-0503(A), 077.C-0547(A), 078.D-0080(A), 194.C-0833(A), 086.D-0449(A), 086.D-0449(A), 179.C-0197(D), 090.D-0358(A), 60.A-9700(A), 60.A-9036(A), 60.A-9700(G), 078.D-0080(A), 087.D-0010(A), 083.C-0139(A), 078.D-0080(A), 087.C-0227(C), 088.C-0353(A), 089.C-0006(A), 082.B-0484(A), 084.B-0029(A), 266.D-5655(A), 185.D-0056(A), 185.D-0056(C), 266.D-5655(A), 073.D-0504(A), 075.C-0234(A), 079.C-0170(A), 081.C-0475(A), 097.D-0035(A), 072.D-0021(A), 073.D-0049(A), 082.D-0061(A), 075.C-0637(A), 079.C-0789(A), 083.A-9003(A), 085.A-9027(B), 076.B-0055(A), 077.C-0295(A), 077.C-0295(C), 083.A-9014(A), 083.A-9011(B), 083.A-9014(B), 084.A-9011(B), 085.A-9027(G), 089.D-0097(B), 090.D-0061(B), 091.D-0145(A), 179.C-0197(C), 091.C-0713(A), 075.D-0342(A), 075.C-0689(A), 075.C-0689(B), 077.C-0295(D), 075.C-0689(B), 077.C-0295(D), 077.C-0295(C), 184.C-0815(F), 179.C-0197(A), 077.C-0138(A), 091.D-0122(A), 081.D-2002(A), 293.D-5036(A), and 083.D-0034(A).

%%%%%%%%%%%%%%%%%%%%%%%%%%%%%%%%%%%%%%%%%%%%%%%%%%

%%%%%%%%%%%%%%%%%%%% REFERENCES %%%%%%%%%%%%%%%%%%

% The best way to enter references is to use BibTeX:

\bibliographystyle{mnras}
%\bibliography{biblio} % if your bibtex file is called example.bib

\begin{thebibliography}{}
\makeatletter
\relax
\def\mn@urlcharsother{\let\do\@makeother \do\$\do\&\do\#\do\^\do\_\do\%\do\~}
\def\mn@doi{\begingroup\mn@urlcharsother \@ifnextchar [ {\mn@doi@}
  {\mn@doi@[]}}
\def\mn@doi@[#1]#2{\def\@tempa{#1}\ifx\@tempa\@empty \href
  {http://dx.doi.org/#2} {doi:#2}\else \href {http://dx.doi.org/#2} {#1}\fi
  \endgroup}
\def\mn@eprint#1#2{\mn@eprint@#1:#2::\@nil}
\def\mn@eprint@arXiv#1{\href {http://arxiv.org/abs/#1} {{\tt arXiv:#1}}}
\def\mn@eprint@dblp#1{\href {http://dblp.uni-trier.de/rec/bibtex/#1.xml}
  {dblp:#1}}
\def\mn@eprint@#1:#2:#3:#4\@nil{\def\@tempa {#1}\def\@tempb {#2}\def\@tempc
  {#3}\ifx \@tempc \@empty \let \@tempc \@tempb \let \@tempb \@tempa \fi \ifx
  \@tempb \@empty \def\@tempb {arXiv}\fi \@ifundefined
  {mn@eprint@\@tempb}{\@tempb:\@tempc}{\expandafter \expandafter \csname
  mn@eprint@\@tempb\endcsname \expandafter{\@tempc}}}

\bibitem[\protect\citeauthoryear{{Alexander}, {Clarke}  \&
  {Pringle}}{{Alexander} et~al.}{2006}]{Alexander2006}
{Alexander} R.~D.,  {Clarke} C.~J.,   {Pringle} J.~E.,  2006, \mn@doi [\mnras]
  {10.1111/j.1365-2966.2006.10293.x}, \href
  {http://adsabs.harvard.edu/abs/2006MNRAS.369..216A} {369, 216}

\bibitem[\protect\citeauthoryear{{Ballering}, {Rieke}, {Su}  \&
  {Montiel}}{{Ballering} et~al.}{2013}]{Ballering13}
{Ballering} N.~P.,  {Rieke} G.~H.,  {Su} K.~Y.~L.,   {Montiel} E.,  2013,
  \mn@doi [\apj] {10.1088/0004-637X/775/1/55}, \href
  {http://adsabs.harvard.edu/abs/2013ApJ...775...55B} {775, 55}

\bibitem[\protect\citeauthoryear{{Baraffe}, {Chabrier}, {Allard}  \&
  {Hauschildt}}{{Baraffe} et~al.}{1998}]{Baraffe1998}
{Baraffe} I.,  {Chabrier} G.,  {Allard} F.,   {Hauschildt} P.~H.,  1998, \aap,
  \href {http://adsabs.harvard.edu/abs/1998A%26A...337..403B} {337, 403}

\bibitem[\protect\citeauthoryear{{Barnes}, {Tobin}  \& {Pollard}}{{Barnes}
  et~al.}{2000}]{Barnes2000}
{Barnes} S.~I.,  {Tobin} W.,   {Pollard} K.~R.,  2000, \mn@doi [\pasa]
  {10.1071/AS00036}, \href {http://adsabs.harvard.edu/abs/2000PASA...17..241B}
  {17, 241}

\bibitem[\protect\citeauthoryear{{Bayo}, {Rodrigo}, {Barrado Y Navascu{\'e}s},
  {Solano}, {Guti{\'e}rrez}, {Morales-Calder{\'o}n}  \& {Allard}}{{Bayo}
  et~al.}{2008}]{Bayo2008}
{Bayo} A.,  {Rodrigo} C.,  {Barrado Y Navascu{\'e}s} D.,  {Solano} E.,
  {Guti{\'e}rrez} R.,  {Morales-Calder{\'o}n} M.,   {Allard} F.,  2008, \mn@doi
  [\aap] {10.1051/0004-6361:200810395}, \href
  {http://adsabs.harvard.edu/abs/2008A%26A...492..277B} {492, 277}

\bibitem[\protect\citeauthoryear{{Bertin}, {Lallement}, {Ferlet}  \&
  {Vidal-Madjar}}{{Bertin} et~al.}{1993}]{Bertin1993}
{Bertin} P.,  {Lallement} R.,  {Ferlet} R.,   {Vidal-Madjar} A.,  1993, \aap,
  \href {http://adsabs.harvard.edu/abs/1993A%26A...278..549B} {278, 549}

\bibitem[\protect\citeauthoryear{{Beust}, {Lagrange}, {Crawford}, {Goudard},
  {Spyromilio}  \& {Vidal-Madjar}}{{Beust} et~al.}{1998}]{Beust1998}
{Beust} H.,  {Lagrange} A.-M.,  {Crawford} I.~A.,  {Goudard} C.,  {Spyromilio}
  J.,   {Vidal-Madjar} A.,  1998, \aap, \href
  {http://adsabs.harvard.edu/abs/1998A%26A...338.1015B} {338, 1015}

\bibitem[\protect\citeauthoryear{{Bohlin}, {Savage}  \& {Drake}}{{Bohlin}
  et~al.}{1978}]{Bohlin78}
{Bohlin} R.~C.,  {Savage} B.~D.,   {Drake} J.~F.,  1978, \mn@doi [\apj]
  {10.1086/156357}, \href {http://adsabs.harvard.edu/abs/1978ApJ...224..132B}
  {224, 132}

\bibitem[\protect\citeauthoryear{{Brandeker}, {Liseau}, {Olofsson}  \&
  {Fridlund}}{{Brandeker} et~al.}{2004}]{Brandeker2004}
{Brandeker} A.,  {Liseau} R.,  {Olofsson} G.,   {Fridlund} M.,  2004, \mn@doi
  [\aap] {10.1051/0004-6361:20034326}, \href
  {http://adsabs.harvard.edu/abs/2004A%26A...413..681B} {413, 681}

\bibitem[\protect\citeauthoryear{{Brown}, {de Geus}  \& {de Zeeuw}}{{Brown}
  et~al.}{1994}]{Brown1994}
{Brown} A.~G.~A.,  {de Geus} E.~J.,   {de Zeeuw} P.~T.,  1994, \aap, \href
  {http://adsabs.harvard.edu/abs/1994A%26A...289..101B} {289, 101}

\bibitem[\protect\citeauthoryear{{Canovas} et~al.,}{{Canovas}
  et~al.}{2017}]{Canovas2017}
{Canovas} H.,  et~al., 2017, preprint, \href
  {http://adsabs.harvard.edu/abs/2017arXiv171009393C} {} (\mn@eprint {arXiv}
  {1710.09393})

\bibitem[\protect\citeauthoryear{{Castelli}, {Gratton}  \& {Kurucz}}{{Castelli}
  et~al.}{1997}]{Castelli1997}
{Castelli} F.,  {Gratton} R.~G.,   {Kurucz} R.~L.,  1997, \aap, \href
  {http://adsabs.harvard.edu/abs/1997A%26A...318..841C} {318, 841}

\bibitem[\protect\citeauthoryear{{Chen} \& {Jura}}{{Chen} \&
  {Jura}}{2003}]{Chen&Jura2003}
{Chen} C.~H.,  {Jura} M.,  2003, \mn@doi [\apj] {10.1086/344589}, \href
  {http://adsabs.harvard.edu/abs/2003ApJ...582..443C} {582, 443}

\bibitem[\protect\citeauthoryear{{Cleeves}, {{\"O}berg}, {Wilner}, {Huang},
  {Loomis}, {Andrews}  \& {Czekala}}{{Cleeves} et~al.}{2016}]{Cleeves2016}
{Cleeves} L.~I.,  {{\"O}berg} K.~I.,  {Wilner} D.~J.,  {Huang} J.,  {Loomis}
  R.~A.,  {Andrews} S.~M.,   {Czekala} I.,  2016, \mn@doi [\apj]
  {10.3847/0004-637X/832/2/110}, \href
  {http://adsabs.harvard.edu/abs/2016ApJ...832..110C} {832, 110}

\bibitem[\protect\citeauthoryear{{Dekker}, {D'Odorico}, {Kaufer}, {Delabre}  \&
  {Kotzlowski}}{{Dekker} et~al.}{2000}]{2000SPIE.4008..534D}
{Dekker} H.,  {D'Odorico} S.,  {Kaufer} A.,  {Delabre} B.,   {Kotzlowski} H.,
  2000, in {Iye} M.,  {Moorwood} A.~F.,  eds,  \procspie Vol. 4008, Optical and
  IR Telescope Instrumentation and Detectors. pp 534--545

\bibitem[\protect\citeauthoryear{{D{\'{\i}}az-Luis},
  {Garc{\'{\i}}a-Hern{\'a}ndez}, {Kameswara Rao}, {Manchado}  \&
  {Cataldo}}{{D{\'{\i}}az-Luis} et~al.}{2015}]{DiazLuis2015}
{D{\'{\i}}az-Luis} J.~J.,  {Garc{\'{\i}}a-Hern{\'a}ndez} D.~A.,  {Kameswara
  Rao} N.,  {Manchado} A.,   {Cataldo} F.,  2015, \mn@doi [\aap]
  {10.1051/0004-6361/201424710}, \href
  {http://adsabs.harvard.edu/abs/2015A%26A...573A..97D} {573, A97}

\bibitem[\protect\citeauthoryear{{Docobo} \& {Ling}}{{Docobo} \&
  {Ling}}{2007}]{DocoboLing2007}
{Docobo} J.~A.,  {Ling} J.~F.,  2007, \mn@doi [\aj] {10.1086/511070}, \href
  {http://adsabs.harvard.edu/abs/2007AJ....133.1209D} {133, 1209}

\bibitem[\protect\citeauthoryear{{Dommanget} \& {Nys}}{{Dommanget} \&
  {Nys}}{2002}]{Dommanget2002}
{Dommanget} J.,  {Nys} O.,  2002, VizieR Online Data Catalog, \href
  {http://adsabs.harvard.edu/abs/2002yCat.1274....0D} {1274}

\bibitem[\protect\citeauthoryear{{Eiroa} et~al.,}{{Eiroa}
  et~al.}{2016}]{Eiroa16}
{Eiroa} C.,  et~al., 2016, \mn@doi [\aap] {10.1051/0004-6361/201629514}, \href
  {http://adsabs.harvard.edu/abs/2016A%26A...594L...1E} {594, L1}

\bibitem[\protect\citeauthoryear{{Fedele}, {van den Ancker}, {Henning},
  {Jayawardhana}  \& {Oliveira}}{{Fedele} et~al.}{2010}]{Fedele10}
{Fedele} D.,  {van den Ancker} M.~E.,  {Henning} T.,  {Jayawardhana} R.,
  {Oliveira} J.~M.,  2010, \mn@doi [\aap] {10.1051/0004-6361/200912810}, \href
  {http://adsabs.harvard.edu/abs/2010A%26A...510A..72F} {510, A72}

\bibitem[\protect\citeauthoryear{{Fern{\'a}ndez}, {Brandeker}  \&
  {Wu}}{{Fern{\'a}ndez} et~al.}{2006}]{Fernandez2006}
{Fern{\'a}ndez} R.,  {Brandeker} A.,   {Wu} Y.,  2006, \mn@doi [\apj]
  {10.1086/500788}, \href {http://adsabs.harvard.edu/abs/2006ApJ...643..509F}
  {643, 509}

\bibitem[\protect\citeauthoryear{{Figueira}}{{Figueira}}{2013}]{Figueira2013}
{Figueira} P.,  2013, in {Chavez} M.,  {Bertone} E.,  {Vega} O.,   {De la Luz}
  V.,  eds,  Astronomical Society of the Pacific Conference Series Vol. 472,
  New Quests in Stellar Astrophysics III: A Panchromatic View of Solar-Like
  Stars, With and Without Planets. p.~137

\bibitem[\protect\citeauthoryear{{Gaia Collaboration} et~al.,}{{Gaia
  Collaboration} et~al.}{2016}]{Gaia2016}
{Gaia Collaboration} et~al., 2016, \mn@doi [\aap]
  {10.1051/0004-6361/201629512}, \href
  {http://adsabs.harvard.edu/abs/2016A%26A...595A...2G} {595, A2}

\bibitem[\protect\citeauthoryear{{Gontcharov}}{{Gontcharov}}{2012}]{Gontcharov12}
{Gontcharov} G.~A.,  2012, \mn@doi [Astronomy Letters]
  {10.1134/S1063773712110035}, \href
  {http://adsabs.harvard.edu/abs/2012AstL...38..694G} {38, 694}

\bibitem[\protect\citeauthoryear{{Gray} \& {Corbally}}{{Gray} \&
  {Corbally}}{2009}]{GrayCorbally2009}
{Gray} R.~O.,  {Corbally} J. C.,  2009, {Stellar Spectral Classification}

\bibitem[\protect\citeauthoryear{{Greaves} et~al.,}{{Greaves}
  et~al.}{2016}]{Greaves2016}
{Greaves} J.~S.,  et~al., 2016, \mn@doi [\mnras] {10.1093/mnras/stw1569}, \href
  {http://adsabs.harvard.edu/abs/2016MNRAS.461.3910G} {461, 3910}

\bibitem[\protect\citeauthoryear{{Hales} et~al.,}{{Hales}
  et~al.}{2014}]{Hales2014}
{Hales} A.~S.,  et~al., 2014, \mn@doi [\aj] {10.1088/0004-6256/148/3/47}, \href
  {http://adsabs.harvard.edu/abs/2014AJ....148...47H} {148, 47}

\bibitem[\protect\citeauthoryear{{Hales}, {Barlow}, {Crawford}  \&
  {Casassus}}{{Hales} et~al.}{2017}]{Hales2017}
{Hales} A.~S.,  {Barlow} M.~J.,  {Crawford} I.~A.,   {Casassus} S.,  2017,
  \mn@doi [\mnras] {10.1093/mnras/stw3274}, \href
  {http://adsabs.harvard.edu/abs/2017MNRAS.466.3582H} {466, 3582}

\bibitem[\protect\citeauthoryear{{Hauck} \& {Jaschek}}{{Hauck} \&
  {Jaschek}}{2000}]{HauckJaschek2000}
{Hauck} B.,  {Jaschek} C.,  2000, \aap, \href
  {http://adsabs.harvard.edu/abs/2000A%26A...354..157H} {354, 157}

\bibitem[\protect\citeauthoryear{{Herbig}}{{Herbig}}{1995}]{Herbig1995}
{Herbig} G.~H.,  1995, \mn@doi [\araa] {10.1146/annurev.aa.33.090195.000315},
  \href {http://adsabs.harvard.edu/abs/1995ARA%26A..33...19H} {33, 19}

\bibitem[\protect\citeauthoryear{{Hern{\'a}ndez}, {Brice{\~n}o}, {Calvet},
  {Hartmann}, {Muzerolle}  \& {Quintero}}{{Hern{\'a}ndez}
  et~al.}{2006}]{Hernandez2006}
{Hern{\'a}ndez} J.,  {Brice{\~n}o} C.,  {Calvet} N.,  {Hartmann} L.,
  {Muzerolle} J.,   {Quintero} A.,  2006, \mn@doi [\apj] {10.1086/507942},
  \href {http://adsabs.harvard.edu/abs/2006ApJ...652..472H} {652, 472}

\bibitem[\protect\citeauthoryear{{Hern{\'a}ndez} et~al.,}{{Hern{\'a}ndez}
  et~al.}{2007}]{Hernandez07}
{Hern{\'a}ndez} J.,  et~al., 2007, \mn@doi [\apj] {10.1086/513735}, \href
  {http://adsabs.harvard.edu/abs/2007ApJ...662.1067H} {662, 1067}

\bibitem[\protect\citeauthoryear{{Higuchi} et~al.,}{{Higuchi}
  et~al.}{2017}]{Higuchi2017}
{Higuchi} A.~E.,  et~al., 2017, \mn@doi [\apjl] {10.3847/2041-8213/aa67f4},
  \href {http://adsabs.harvard.edu/abs/2017ApJ...839L..14H} {839, L14}

\bibitem[\protect\citeauthoryear{{Houck} et~al.,}{{Houck}
  et~al.}{2004}]{Houck2004}
{Houck} J.~R.,  et~al., 2004, \mn@doi [\apjs] {10.1086/423134}, \href
  {http://adsabs.harvard.edu/abs/2004ApJS..154...18H} {154, 18}

\bibitem[\protect\citeauthoryear{{Howarth}, {Price}, {Crawford}  \&
  {Hawkins}}{{Howarth} et~al.}{2002}]{Howarth2002}
{Howarth} I.~D.,  {Price} R.~J.,  {Crawford} I.~A.,   {Hawkins} I.,  2002,
  \mn@doi [\mnras] {10.1046/j.1365-8711.2002.05519.x}, \href
  {http://adsabs.harvard.edu/abs/2002MNRAS.335..267H} {335, 267}

\bibitem[\protect\citeauthoryear{{Hughes}, {Duchene}  \& {Matthews}}{{Hughes}
  et~al.}{2018}]{Hughes2018}
{Hughes} A.~M.,  {Duchene} G.,   {Matthews} B.,  2018, preprint, \href
  {http://adsabs.harvard.edu/abs/2018arXiv180204313H} {} (\mn@eprint {arXiv}
  {1802.04313})

\bibitem[\protect\citeauthoryear{{Jaschek} \& {Andrillat}}{{Jaschek} \&
  {Andrillat}}{1998}]{Jaschek1998}
{Jaschek} C.,  {Andrillat} Y.,  1998, \mn@doi [\aaps] {10.1051/aas:1998101},
  \href {http://adsabs.harvard.edu/abs/1998A%26AS..130..507J} {130, 507}

\bibitem[\protect\citeauthoryear{{Jaschek}, {Jaschek}  \&
  {Andrillat}}{{Jaschek} et~al.}{1988}]{Jaschek1988}
{Jaschek} M.,  {Jaschek} C.,   {Andrillat} Y.,  1988, \aaps, \href
  {http://adsabs.harvard.edu/abs/1988A%26AS...72..505J} {72, 505}

\bibitem[\protect\citeauthoryear{{Jaschek}, {Jaschek}  \&
  {Andrillat}}{{Jaschek} et~al.}{1991}]{Jaschek1991}
{Jaschek} M.,  {Jaschek} C.,   {Andrillat} Y.,  1991, \aap, \href
  {http://adsabs.harvard.edu/abs/1991A%26A...250..127J} {250, 127}

\bibitem[\protect\citeauthoryear{{Jura}, {Zuckerman}, {Becklin}  \&
  {Smith}}{{Jura} et~al.}{1993}]{Jura1993}
{Jura} M.,  {Zuckerman} B.,  {Becklin} E.~E.,   {Smith} R.~C.,  1993, \mn@doi
  [\apjl] {10.1086/187110}, \href
  {http://adsabs.harvard.edu/abs/1993ApJ...418L..37J} {418, L37}

\bibitem[\protect\citeauthoryear{{Kasper}, {Apai}, {Wagner}  \&
  {Robberto}}{{Kasper} et~al.}{2015}]{Kasper2015}
{Kasper} M.,  {Apai} D.,  {Wagner} K.,   {Robberto} M.,  2015, \mn@doi [\apjl]
  {10.1088/2041-8205/812/2/L33}, \href
  {http://adsabs.harvard.edu/abs/2015ApJ...812L..33K} {812, L33}

\bibitem[\protect\citeauthoryear{{Kaufer}, {Stahl}, {Tubbesing},
  {N{\o}rregaard}, {Avila}, {Francois}, {Pasquini}  \& {Pizzella}}{{Kaufer}
  et~al.}{1999}]{1999Msngr..95....8K}
{Kaufer} A.,  {Stahl} O.,  {Tubbesing} S.,  {N{\o}rregaard} P.,  {Avila} G.,
  {Francois} P.,  {Pasquini} L.,   {Pizzella} A.,  1999, The Messenger, \href
  {http://adsabs.harvard.edu/abs/1999Msngr..95....8K} {95, 8}

\bibitem[\protect\citeauthoryear{{Kausch} et~al.,}{{Kausch}
  et~al.}{2015}]{Kausch2015}
{Kausch} W.,  et~al., 2015, \mn@doi [\aap] {10.1051/0004-6361/201423909}, \href
  {http://adsabs.harvard.edu/abs/2015A%26A...576A..78K} {576, A78}

\bibitem[\protect\citeauthoryear{{Kennedy}, {Marino}, {Matr{\'a}}, {Pani{\'c}},
  {Wilner}, {Wyatt}  \& {Yelverton}}{{Kennedy} et~al.}{2018}]{Kennedy2018}
{Kennedy} G.~M.,  {Marino} S.,  {Matr{\'a}} L.,  {Pani{\'c}} O.,  {Wilner} D.,
  {Wyatt} M.~C.,   {Yelverton} B.,  2018, \mn@doi [\mnras]
  {10.1093/mnras/sty135}, \href
  {http://adsabs.harvard.edu/abs/2018MNRAS.tmp..137K} {}

\bibitem[\protect\citeauthoryear{{Kenyon}, {Najita}  \& {Bromley}}{{Kenyon}
  et~al.}{2016}]{Kenyon2016}
{Kenyon} S.~J.,  {Najita} J.~R.,   {Bromley} B.~C.,  2016, \mn@doi [\apj]
  {10.3847/0004-637X/831/1/8}, \href
  {http://adsabs.harvard.edu/abs/2016ApJ...831....8K} {831, 8}

\bibitem[\protect\citeauthoryear{{Kharchenko}}{{Kharchenko}}{2001}]{Kharchenko2001}
{Kharchenko} N.~V.,  2001, Kinematika i Fizika Nebesnykh Tel, \href
  {http://adsabs.harvard.edu/abs/2001KFNT...17..409K} {17, 409}

\bibitem[\protect\citeauthoryear{{Kiefer}, {Lecavelier des Etangs}  \&
  {Vidal-Madjar}}{{Kiefer} et~al.}{2014a}]{Kiefer2014twoAtype}
{Kiefer} F.,  {Lecavelier des Etangs} A.,   {Vidal-Madjar} A.,  2014a, in
  {Ballet} J.,  {Martins} F.,  {Bournaud} F.,  {Monier} R.,   {Reyl{\'e}} C.,
  eds, SF2A-2014: Proceedings of the Annual meeting of the French Society of
  Astronomy and Astrophysics. pp 39--43

\bibitem[\protect\citeauthoryear{{Kiefer}, {Lecavelier des Etangs}, {Boissier},
  {Vidal-Madjar}, {Beust}, {Lagrange}, {H{\'e}brard}  \& {Ferlet}}{{Kiefer}
  et~al.}{2014b}]{Kiefer2014BetaPicNature}
{Kiefer} F.,  {Lecavelier des Etangs} A.,  {Boissier} J.,  {Vidal-Madjar} A.,
  {Beust} H.,  {Lagrange} A.-M.,  {H{\'e}brard} G.,   {Ferlet} R.,  2014b,
  \mn@doi [\nat] {10.1038/nature13849}, \href
  {http://adsabs.harvard.edu/abs/2014Natur.514..462K} {514, 462}

\bibitem[\protect\citeauthoryear{{Kiefer}, {Lecavelier des Etangs}, {Augereau},
  {Vidal-Madjar}, {Lagrange}  \& {Beust}}{{Kiefer}
  et~al.}{2014c}]{Kiefer2014HD172555}
{Kiefer} F.,  {Lecavelier des Etangs} A.,  {Augereau} J.-C.,  {Vidal-Madjar}
  A.,  {Lagrange} A.-M.,   {Beust} H.,  2014c, \mn@doi [\aap]
  {10.1051/0004-6361/201323128}, \href
  {http://adsabs.harvard.edu/abs/2014A%26A...561L..10K} {561, L10}

\bibitem[\protect\citeauthoryear{{K{\'o}sp{\'a}l} \&
  {Mo{\'o}r}}{{K{\'o}sp{\'a}l} \& {Mo{\'o}r}}{2016}]{Kospal2016}
{K{\'o}sp{\'a}l} {\'A}.,  {Mo{\'o}r} A.,  2016, in {Kastner} J.~H.,  {Stelzer}
  B.,   {Metchev} S.~A.,  eds,  IAU Symposium Vol. 314, Young Stars {\&}
  Planets Near the Sun. pp 183--188 (\mn@eprint {arXiv} {1510.03618}),
  \mn@doi{10.1017/S1743921315006614}

\bibitem[\protect\citeauthoryear{{K{\'o}sp{\'a}l} et~al.,}{{K{\'o}sp{\'a}l}
  et~al.}{2013}]{Kospal2013}
{K{\'o}sp{\'a}l} {\'A}.,  et~al., 2013, \mn@doi [\apj]
  {10.1088/0004-637X/776/2/77}, \href
  {http://adsabs.harvard.edu/abs/2013ApJ...776...77K} {776, 77}

\bibitem[\protect\citeauthoryear{{Kral}, {Clarke}  \& {Wyatt}}{{Kral}
  et~al.}{2017a}]{Kral2017review}
{Kral} Q.,  {Clarke} C.,   {Wyatt} M.,  2017a, preprint, \href
  {http://adsabs.harvard.edu/abs/2017arXiv170308560K} {} (\mn@eprint {arXiv}
  {1703.08560})

\bibitem[\protect\citeauthoryear{{Kral}, {Matr{\`a}}, {Wyatt}  \&
  {Kennedy}}{{Kral} et~al.}{2017b}]{Kral2017}
{Kral} Q.,  {Matr{\`a}} L.,  {Wyatt} M.~C.,   {Kennedy} G.~M.,  2017b, \mn@doi
  [\mnras] {10.1093/mnras/stx730}, \href
  {http://adsabs.harvard.edu/abs/2017MNRAS.469..521K} {469, 521}

\bibitem[\protect\citeauthoryear{{Kuschnig}, {Paunzen}  \& {Weiss}}{{Kuschnig}
  et~al.}{1994}]{Kuschnig1994}
{Kuschnig} R.,  {Paunzen} E.,   {Weiss} W.~W.,  1994, Information Bulletin on
  Variable Stars, \href {http://adsabs.harvard.edu/abs/1994IBVS.4070....1K}
  {4070}

\bibitem[\protect\citeauthoryear{{Lyra} \& {Kuchner}}{{Lyra} \&
  {Kuchner}}{2013}]{LyraKuchner2013}
{Lyra} W.,  {Kuchner} M.,  2013, \mn@doi [\nat] {10.1038/nature12281}, \href
  {http://adsabs.harvard.edu/abs/2013Natur.499..184L} {499, 184}

\bibitem[\protect\citeauthoryear{{Marino} et~al.,}{{Marino}
  et~al.}{2016}]{Marino2016}
{Marino} S.,  et~al., 2016, \mn@doi [\mnras] {10.1093/mnras/stw1216}, \href
  {http://adsabs.harvard.edu/abs/2016MNRAS.460.2933M} {460, 2933}

\bibitem[\protect\citeauthoryear{{Marino} et~al.,}{{Marino}
  et~al.}{2017}]{Marino2017}
{Marino} S.,  et~al., 2017, \mn@doi [\mnras] {10.1093/mnras/stw2867}, \href
  {http://adsabs.harvard.edu/abs/2017MNRAS.465.2595M} {465, 2595}

\bibitem[\protect\citeauthoryear{{Mashonkina}, {Shimanski{\u i}}  \&
  {Sakhibullin}}{{Mashonkina} et~al.}{2000}]{Mashonkina2000}
{Mashonkina} L.~I.,  {Shimanski{\u i}} V.~V.,   {Sakhibullin} N.~A.,  2000,
  \mn@doi [Astronomy Reports] {10.1134/1.1327637}, \href
  {http://adsabs.harvard.edu/abs/2000ARep...44..790M} {44, 790}

\bibitem[\protect\citeauthoryear{{Matthews}, {Krivov}, {Wyatt}, {Bryden}  \&
  {Eiroa}}{{Matthews} et~al.}{2014}]{Matthews2014}
{Matthews} B.~C.,  {Krivov} A.~V.,  {Wyatt} M.~C.,  {Bryden} G.,   {Eiroa} C.,
  2014, \mn@doi [Protostars and Planets VI]
  {10.2458/azu_uapress_9780816531240-ch023}, \href
  {http://adsabs.harvard.edu/abs/2014prpl.conf..521M} {pp 521--544}

\bibitem[\protect\citeauthoryear{{Mayor} et~al.,}{{Mayor}
  et~al.}{2003}]{2003Msngr.114...20M}
{Mayor} M.,  et~al., 2003, The Messenger, \href
  {http://adsabs.harvard.edu/abs/2003Msngr.114...20M} {114, 20}

\bibitem[\protect\citeauthoryear{{Milli} et~al.,}{{Milli}
  et~al.}{2017}]{Milli2017}
{Milli} J.,  et~al., 2017, \mn@doi [\aap] {10.1051/0004-6361/201527838}, \href
  {http://adsabs.harvard.edu/abs/2017A%26A...599A.108M} {599, A108}

\bibitem[\protect\citeauthoryear{{Mittal}, {Chen}, {Jang-Condell}, {Manoj},
  {Sargent}, {Watson}  \& {Lisse}}{{Mittal} et~al.}{2015}]{Mittal2015}
{Mittal} T.,  {Chen} C.~H.,  {Jang-Condell} H.,  {Manoj} P.,  {Sargent} B.~A.,
  {Watson} D.~M.,   {Lisse} C.~M.,  2015, \mn@doi [\apj]
  {10.1088/0004-637X/798/2/87}, \href
  {http://adsabs.harvard.edu/abs/2015ApJ...798...87M} {798, 87}

\bibitem[\protect\citeauthoryear{{Montgomery} \& {Welsh}}{{Montgomery} \&
  {Welsh}}{2012}]{Montgomery2012}
{Montgomery} S.~L.,  {Welsh} B.~Y.,  2012, \mn@doi [\pasp] {10.1086/668293},
  \href {http://adsabs.harvard.edu/abs/2012PASP..124.1042M} {124, 1042}

\bibitem[\protect\citeauthoryear{{Montgomery} \& {Welsh}}{{Montgomery} \&
  {Welsh}}{2017}]{Montgomery2017}
{Montgomery} S.~L.,  {Welsh} B.~Y.,  2017, \mn@doi [\mnras]
  {10.1093/mnrasl/slx016}, \href
  {http://adsabs.harvard.edu/abs/2017MNRAS.468L..55M} {468, L55}

\bibitem[\protect\citeauthoryear{{Mo{\'o}r} et~al.,}{{Mo{\'o}r}
  et~al.}{2011}]{Moor2011}
{Mo{\'o}r} A.,  et~al., 2011, \mn@doi [\apjl] {10.1088/2041-8205/740/1/L7},
  \href {http://adsabs.harvard.edu/abs/2011ApJ...740L...7M} {740, L7}

\bibitem[\protect\citeauthoryear{{Mo{\'o}r} et~al.,}{{Mo{\'o}r}
  et~al.}{2015}]{Moor2015}
{Mo{\'o}r} A.,  et~al., 2015, \mn@doi [\mnras] {10.1093/mnras/stu2442}, \href
  {http://adsabs.harvard.edu/abs/2015MNRAS.447..577M} {447, 577}

\bibitem[\protect\citeauthoryear{{Morales}, {Bryden}, {Werner}  \&
  {Stapelfeldt}}{{Morales} et~al.}{2016}]{Morales2016}
{Morales} F.~Y.,  {Bryden} G.,  {Werner} M.~W.,   {Stapelfeldt} K.~R.,  2016,
  \mn@doi [\apj] {10.3847/0004-637X/831/1/97}, \href
  {http://adsabs.harvard.edu/abs/2016ApJ...831...97M} {831, 97}

\bibitem[\protect\citeauthoryear{{Morton}}{{Morton}}{1991}]{Morton1991}
{Morton} D.~C.,  1991, \mn@doi [\apjs] {10.1086/191601}, \href
  {http://adsabs.harvard.edu/abs/1991ApJS...77..119M} {77, 119}

\bibitem[\protect\citeauthoryear{{Nesvorn{\'y}}, {Vokrouhlick{\'y}}, {Dones},
  {Levison}, {Kaib}  \& {Morbidelli}}{{Nesvorn{\'y}}
  et~al.}{2017}]{Nesvorny2017}
{Nesvorn{\'y}} D.,  {Vokrouhlick{\'y}} D.,  {Dones} L.,  {Levison} H.~F.,
  {Kaib} N.,   {Morbidelli} A.,  2017, \mn@doi [\apj]
  {10.3847/1538-4357/aa7cf6}, \href
  {http://adsabs.harvard.edu/abs/2017ApJ...845...27N} {845, 27}

\bibitem[\protect\citeauthoryear{{Olofsson}, {Juh{\'a}sz}, {Henning},
  {Mutschke}, {Tamanai}, {Mo{\'o}r}  \& {{\'A}brah{\'a}m}}{{Olofsson}
  et~al.}{2012}]{Olofsson2012}
{Olofsson} J.,  {Juh{\'a}sz} A.,  {Henning} T.,  {Mutschke} H.,  {Tamanai} A.,
  {Mo{\'o}r} A.,   {{\'A}brah{\'a}m} P.,  2012, \mn@doi [\aap]
  {10.1051/0004-6361/201118735}, \href
  {http://adsabs.harvard.edu/abs/2012A%26A...542A..90O} {542, A90}

\bibitem[\protect\citeauthoryear{{Plez}}{{Plez}}{2013}]{Plez2013}
{Plez} B.,  2013, in {Cambresy} L.,  {Martins} F.,  {Nuss} E.,   {Palacios} A.,
   eds, SF2A-2013: Proceedings of the Annual meeting of the French Society of
  Astronomy and Astrophysics. pp 141--146

\bibitem[\protect\citeauthoryear{{Pontoppidan}, {Salyk}, {Bergin}, {Brittain},
  {Marty}, {Mousis}  \& {{\"O}berg}}{{Pontoppidan}
  et~al.}{2014}]{Pontoppidan14}
{Pontoppidan} K.~M.,  {Salyk} C.,  {Bergin} E.~A.,  {Brittain} S.,  {Marty} B.,
   {Mousis} O.,   {{\"O}berg} K.~I.,  2014, \mn@doi [Protostars and Planets VI]
  {10.2458/azu_uapress_9780816531240-ch016}, \href
  {http://adsabs.harvard.edu/abs/2014prpl.conf..363P} {pp 363--385}

\bibitem[\protect\citeauthoryear{{Przybilla}, {Nieva}  \& {Butler}}{{Przybilla}
  et~al.}{2011}]{Przybilla2011}
{Przybilla} N.,  {Nieva} M.-F.,   {Butler} K.,  2011, in Journal of Physics
  Conference Series. p. 012015 (\mn@eprint {arXiv} {1111.1445}),
  \mn@doi{10.1088/1742-6596/328/1/012015}

\bibitem[\protect\citeauthoryear{{Rebollido} et~al.,}{{Rebollido}
  et~al.}{2018}]{Rebollido2018}
{Rebollido} I.,  et~al., 2018, preprint, \href
  {http://adsabs.harvard.edu/abs/2018arXiv180107951R} {} (\mn@eprint {arXiv}
  {1801.07951})

\bibitem[\protect\citeauthoryear{{Redfield}}{{Redfield}}{2007}]{Redfield2007}
{Redfield} S.,  2007, \mn@doi [\apjl] {10.1086/512237}, \href
  {http://adsabs.harvard.edu/abs/2007ApJ...656L..97R} {656, L97}

\bibitem[\protect\citeauthoryear{{Redfield} \& {Linsky}}{{Redfield} \&
  {Linsky}}{2008}]{Redfield08}
{Redfield} S.,  {Linsky} J.~L.,  2008, \mn@doi [\apj] {10.1086/524002}, \href
  {http://adsabs.harvard.edu/abs/2008ApJ...673..283R} {673, 283}

\bibitem[\protect\citeauthoryear{{Rhee}, {Song}, {Zuckerman}  \&
  {McElwain}}{{Rhee} et~al.}{2007}]{Rhee07}
{Rhee} J.~H.,  {Song} I.,  {Zuckerman} B.,   {McElwain} M.,  2007, \mn@doi
  [\apj] {10.1086/509912}, \href
  {http://adsabs.harvard.edu/abs/2007ApJ...660.1556R} {660, 1556}

\bibitem[\protect\citeauthoryear{{Riviere-Marichalar}
  et~al.,}{{Riviere-Marichalar} et~al.}{2012}]{Riviere-Marichalar2012}
{Riviere-Marichalar} P.,  et~al., 2012, \mn@doi [\aap]
  {10.1051/0004-6361/201219745}, \href
  {http://adsabs.harvard.edu/abs/2012A%26A...546L...8R} {546, L8}

\bibitem[\protect\citeauthoryear{{Roberge}, {Feldman}, {Weinberger}, {Deleuil}
  \& {Bouret}}{{Roberge} et~al.}{2006}]{Roberge2006}
{Roberge} A.,  {Feldman} P.~D.,  {Weinberger} A.~J.,  {Deleuil} M.,   {Bouret}
  J.-C.,  2006, \mn@doi [\nat] {10.1038/nature04832}, \href
  {http://adsabs.harvard.edu/abs/2006Natur.441..724R} {441, 724}

\bibitem[\protect\citeauthoryear{{Roberge}, {Welsh}, {Kamp}, {Weinberger}  \&
  {Grady}}{{Roberge} et~al.}{2014}]{Roberge2014}
{Roberge} A.,  {Welsh} B.~Y.,  {Kamp} I.,  {Weinberger} A.~J.,   {Grady} C.~A.,
   2014, \mn@doi [\apjl] {10.1088/2041-8205/796/1/L11}, \href
  {http://adsabs.harvard.edu/abs/2014ApJ...796L..11R} {796, L11}

\bibitem[\protect\citeauthoryear{{Savage} \& {Sembach}}{{Savage} \&
  {Sembach}}{1991}]{Savage1991}
{Savage} B.~D.,  {Sembach} K.~R.,  1991, \mn@doi [\apj] {10.1086/170498}, \href
  {http://adsabs.harvard.edu/abs/1991ApJ...379..245S} {379, 245}

\bibitem[\protect\citeauthoryear{{Sbordone}, {Bonifacio}, {Castelli}  \&
  {Kurucz}}{{Sbordone} et~al.}{2004}]{Sbordone04}
{Sbordone} L.,  {Bonifacio} P.,  {Castelli} F.,   {Kurucz} R.~L.,  2004,
  Memorie della Societa Astronomica Italiana Supplementi, \href
  {http://adsabs.harvard.edu/abs/2004MSAIS...5...93S} {5, 93}

\bibitem[\protect\citeauthoryear{{Siess}, {Dufour}  \& {Forestini}}{{Siess}
  et~al.}{2000}]{Siess2000}
{Siess} L.,  {Dufour} E.,   {Forestini} M.,  2000, \aap, \href
  {http://adsabs.harvard.edu/abs/2000A%26A...358..593S} {358, 593}

\bibitem[\protect\citeauthoryear{{Sitnova}, {Ryabchikova}, {Alexeeva}  \&
  {Mashonkina}}{{Sitnova} et~al.}{2017}]{Sitnova2017}
{Sitnova} T.,  {Ryabchikova} T.,  {Alexeeva} S.,   {Mashonkina} L.,  2017,
  preprint, \href {http://adsabs.harvard.edu/abs/2017arXiv171006726S} {}
  (\mn@eprint {arXiv} {1710.06726})

\bibitem[\protect\citeauthoryear{{Smette} et~al.,}{{Smette}
  et~al.}{2015}]{Smette2015}
{Smette} A.,  et~al., 2015, \mn@doi [\aap] {10.1051/0004-6361/201423932}, \href
  {http://adsabs.harvard.edu/abs/2015A%26A...576A..77S} {576, A77}

\bibitem[\protect\citeauthoryear{{Th{\'e}bault}}{{Th{\'e}bault}}{2009}]{Thebault2009}
{Th{\'e}bault} P.,  2009, \mn@doi [\aap] {10.1051/0004-6361/200912396}, \href
  {http://adsabs.harvard.edu/abs/2009A%26A...505.1269T} {505, 1269}

\bibitem[\protect\citeauthoryear{{Th{\'e}bault} \& {Beust}}{{Th{\'e}bault} \&
  {Beust}}{2001}]{Thebault2001}
{Th{\'e}bault} P.,  {Beust} H.,  2001, \mn@doi [\aap]
  {10.1051/0004-6361:20010983}, \href
  {http://adsabs.harvard.edu/abs/2001A%26A...376..621T} {376, 621}

\bibitem[\protect\citeauthoryear{{Thi} et~al.,}{{Thi} et~al.}{2013}]{Thi2013}
{Thi} W.~F.,  et~al., 2013, \mn@doi [\aap] {10.1051/0004-6361/201221002}, \href
  {http://adsabs.harvard.edu/abs/2013A%26A...557A.111T} {557, A111}

\bibitem[\protect\citeauthoryear{{Welsh} \& {Montgomery}}{{Welsh} \&
  {Montgomery}}{2013}]{Welsh2013}
{Welsh} B.~Y.,  {Montgomery} S.,  2013, \mn@doi [\pasp] {10.1086/671757}, \href
  {http://adsabs.harvard.edu/abs/2013PASP..125..759W} {125, 759}

\bibitem[\protect\citeauthoryear{{Welsh} \& {Montgomery}}{{Welsh} \&
  {Montgomery}}{2015}]{WelshMontg2015}
{Welsh} B.~Y.,  {Montgomery} S.~L.,  2015, \mn@doi [Advances in Astronomy]
  {10.1155/2015/980323}, \href
  {http://adsabs.harvard.edu/abs/2015AdAst2015E..26W} {2015, 980323}

\bibitem[\protect\citeauthoryear{{Welsh} \& {Montgomery}}{{Welsh} \&
  {Montgomery}}{2018}]{Welsh2018}
{Welsh} B.~Y.,  {Montgomery} S.~L.,  2018, \mn@doi [\mnras]
  {10.1093/mnras/stx2800}, \href
  {http://adsabs.harvard.edu/abs/2018MNRAS.474.1515W} {474, 1515}

\bibitem[\protect\citeauthoryear{{Wenger} et~al.,}{{Wenger}
  et~al.}{2000}]{Wenger2000}
{Wenger} M.,  et~al., 2000, \mn@doi [\aaps] {10.1051/aas:2000332}, \href
  {http://adsabs.harvard.edu/abs/2000A%26AS..143....9W} {143, 9}

\bibitem[\protect\citeauthoryear{{Werner} et~al.,}{{Werner}
  et~al.}{2004}]{Werner2004}
{Werner} M.~W.,  et~al., 2004, \mn@doi [\apjs] {10.1086/422992}, \href
  {http://adsabs.harvard.edu/abs/2004ApJS..154....1W} {154, 1}

\bibitem[\protect\citeauthoryear{{Williams} \& {Best}}{{Williams} \&
  {Best}}{2014}]{Williams14}
{Williams} J.~P.,  {Best} W.~M.~J.,  2014, \mn@doi [\apj]
  {10.1088/0004-637X/788/1/59}, \href
  {http://adsabs.harvard.edu/abs/2014ApJ...788...59W} {788, 59}

\bibitem[\protect\citeauthoryear{{Williams} \& {Cieza}}{{Williams} \&
  {Cieza}}{2011}]{WilliamsCieza2011}
{Williams} J.~P.,  {Cieza} L.~A.,  2011, \mn@doi [\araa]
  {10.1146/annurev-astro-081710-102548}, \href
  {http://adsabs.harvard.edu/abs/2011ARA%26A..49...67W} {49, 67}

\bibitem[\protect\citeauthoryear{{Wilson}}{{Wilson}}{1953}]{Wilson1953}
{Wilson} R.~E.,  1953, Carnegie Institute Washington D.C.~Publication, \href
  {http://adsabs.harvard.edu/abs/1953GCRV..C......0W} {}

\bibitem[\protect\citeauthoryear{{Wyatt}}{{Wyatt}}{2008}]{Wyatt2008}
{Wyatt} M.~C.,  2008, \mn@doi [\araa] {10.1146/annurev.astro.45.051806.110525},
  \href {http://adsabs.harvard.edu/abs/2008ARA%26A..46..339W} {46, 339}

\bibitem[\protect\citeauthoryear{{Wyatt}, {Pani{\'c}}, {Kennedy}  \&
  {Matr{\`a}}}{{Wyatt} et~al.}{2015}]{Wyatt2015}
{Wyatt} M.~C.,  {Pani{\'c}} O.,  {Kennedy} G.~M.,   {Matr{\`a}} L.,  2015,
  \mn@doi [\apss] {10.1007/s10509-015-2315-6}, \href
  {http://adsabs.harvard.edu/abs/2015Ap%26SS.357..103W} {357, 103}

\bibitem[\protect\citeauthoryear{{Zorec} \& {Royer}}{{Zorec} \&
  {Royer}}{2012}]{Zorec2012}
{Zorec} J.,  {Royer} F.,  2012, \mn@doi [\aap] {10.1051/0004-6361/201117691},
  \href {http://adsabs.harvard.edu/abs/2012A%26A...537A.120Z} {537, A120}

\bibitem[\protect\citeauthoryear{{van Leeuwen}}{{van
  Leeuwen}}{2007}]{VanLeeuwen2007}
{van Leeuwen} F.,  2007, \mn@doi [\aap] {10.1051/0004-6361:20078357}, \href
  {http://adsabs.harvard.edu/abs/2007A%26A...474..653V} {474, 653}

\makeatother
\end{thebibliography}

% Alternatively you could enter them by hand, like this:
% This method is tedious and prone to error if you have lots of references
%\begin{thebibliography}{99}
%\bibitem[\protect\citeauthoryear{Author}{2012}]{Author2012}
%Author A.~N., 2013, Journal of Improbable Astronomy, 1, 1
%\bibitem[\protect\citeauthoryear{Others}{2013}]{Others2013}
%Others S., 2012, Journal of Interesting Stuff, 17, 198
%\end{thebibliography}

%%%%%%%%%%%%%%%%%%%%%%%%%%%%%%%%%%%%%%%%%%%%%%%%%%

%%%%%%%%%%%%%%%%% APPENDICES %%%%%%%%%%%%%%%%%%%%%

\appendix

\section{Photospheric line fits}

%If you want to present additional material which would interrupt the flow of the main paper,
%it can be placed in an Appendix which appears after the list of references.
%\clearpage

%\bf{Models}
In this appendix we provide the Figures illustrating the Kurucz models that best reproduce each one of the lines of the reference spectrum for each star in the sample. In all cases the reference spectrum is displayed in black and the best fitting model in red. Please note that the parameters are allowed to vary between different lines for the same object (as these models do not include non-LTE effects).

\begin{figure*}
	\includegraphics[width=0.495\textwidth]{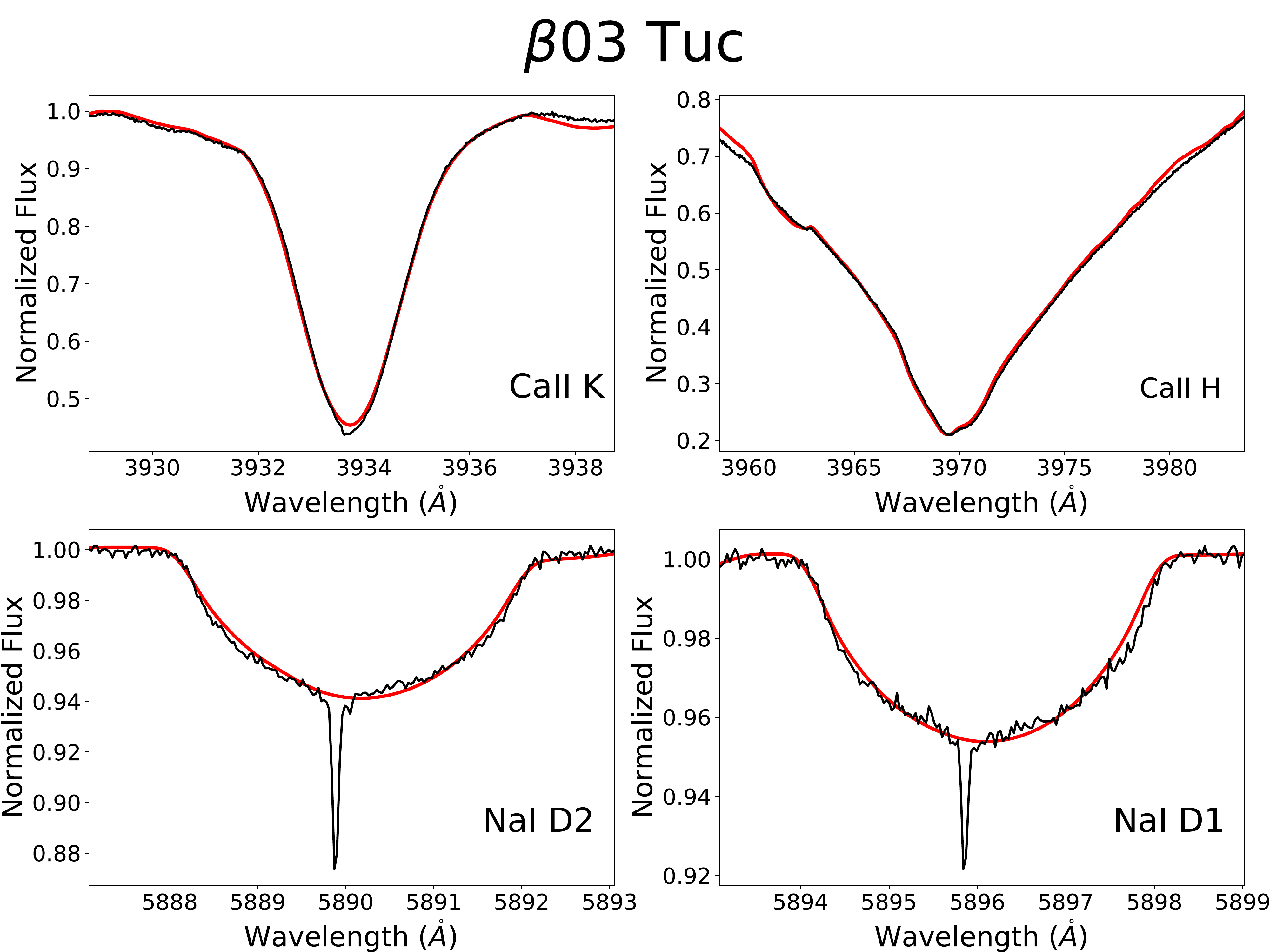}
	\includegraphics[width=0.495\textwidth]{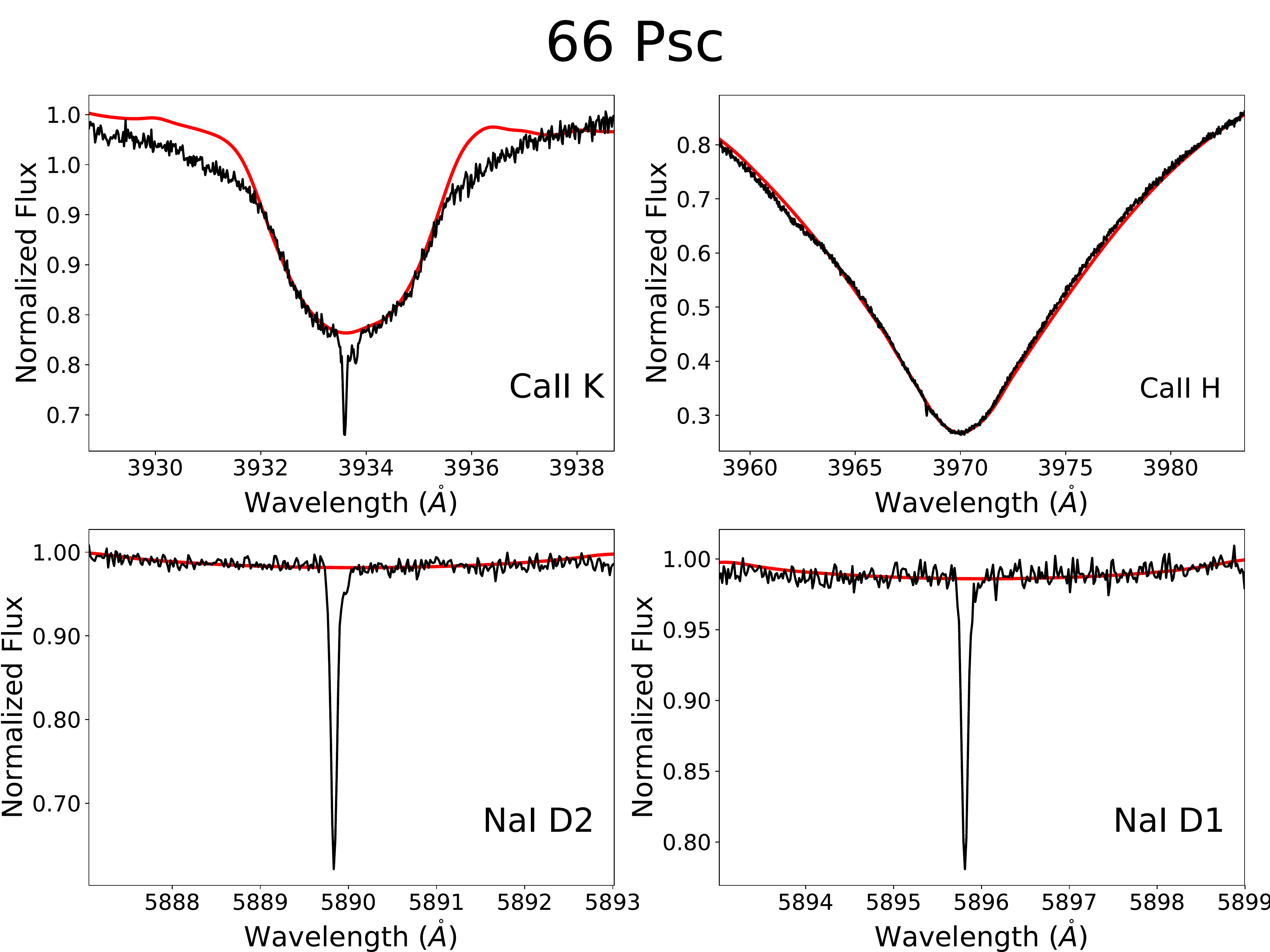}
\end{figure*}

\begin{figure*}
	\includegraphics[width=0.495\textwidth]{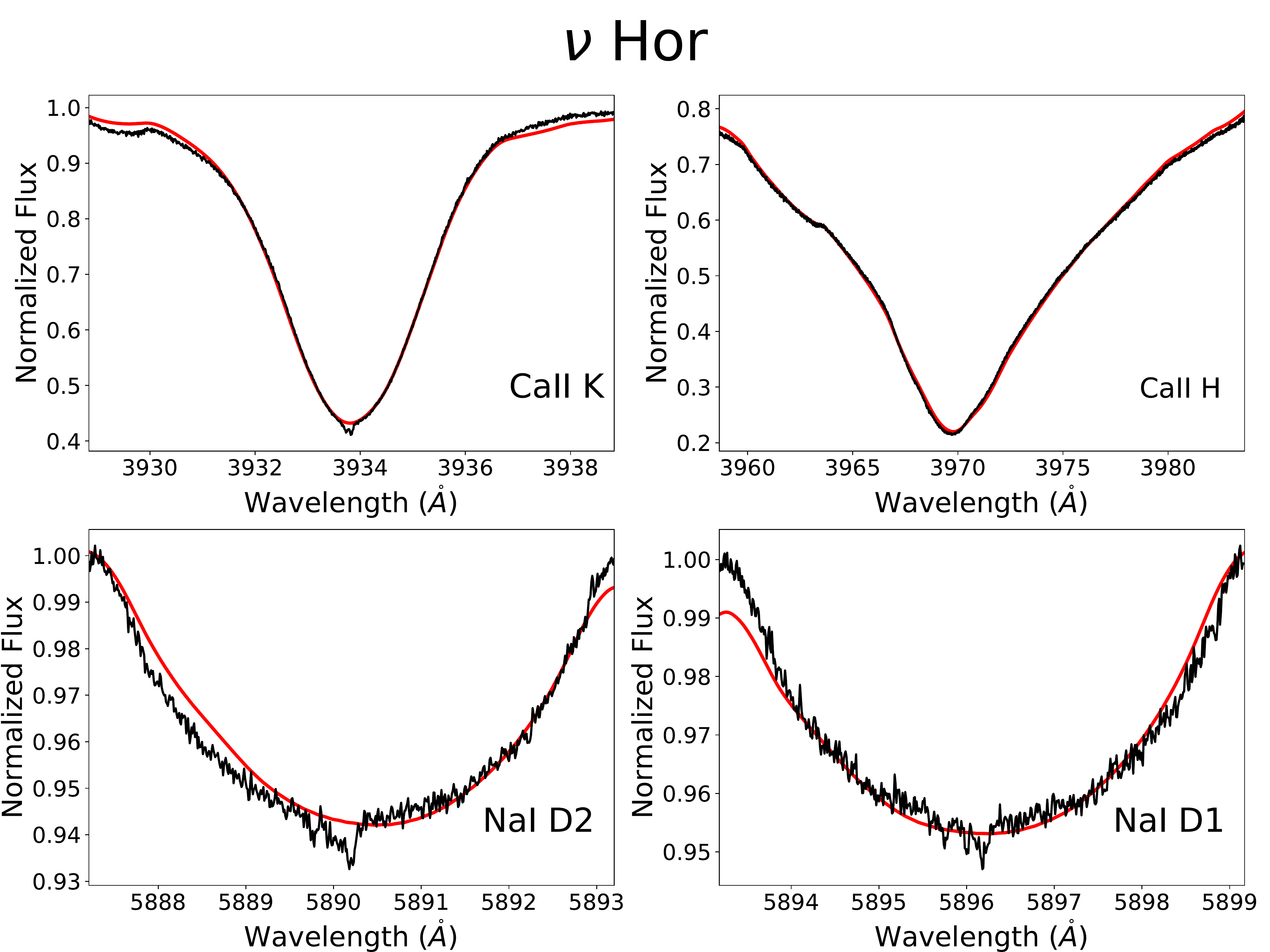}
	\includegraphics[width=0.495\textwidth]{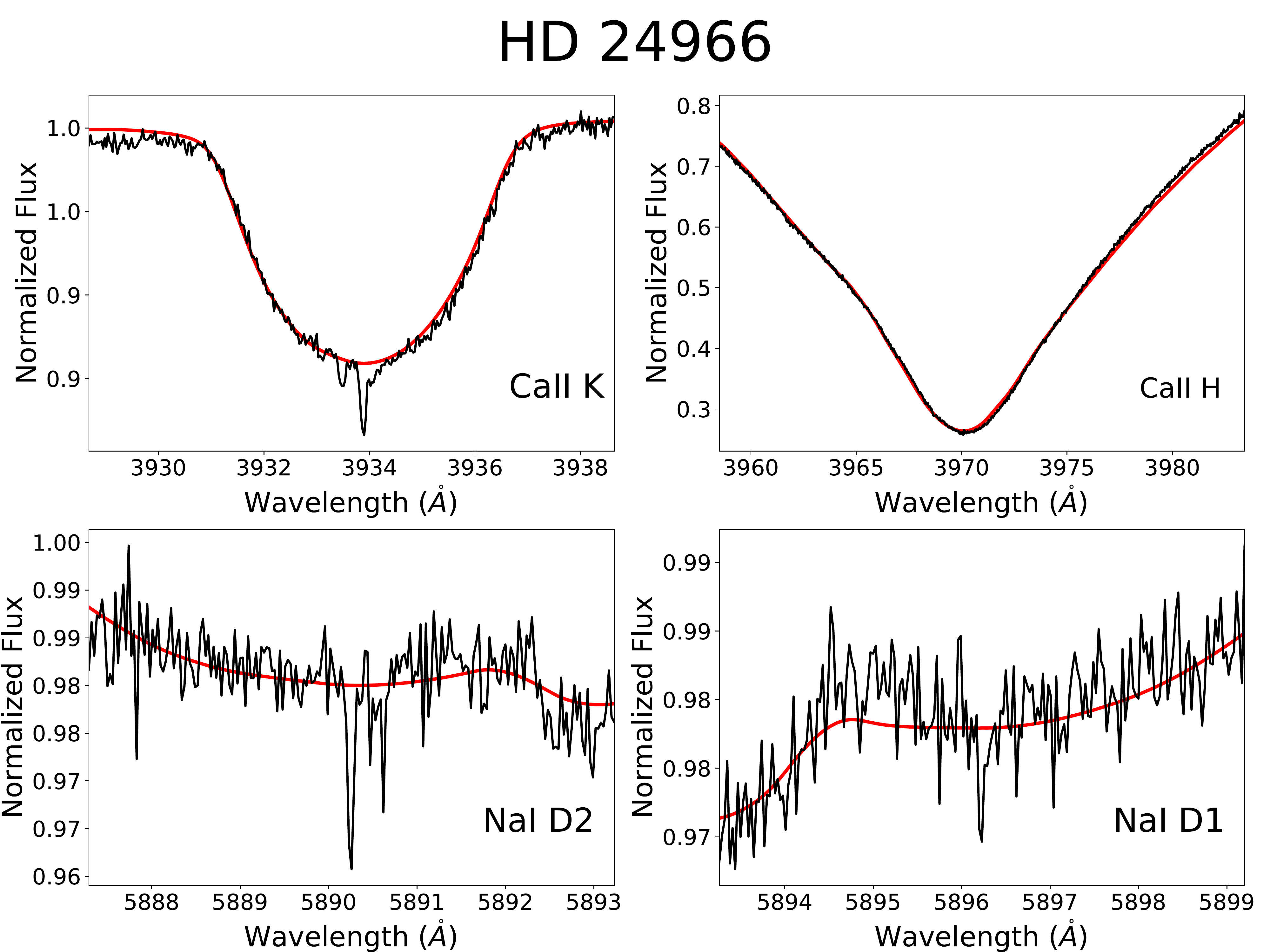}
\end{figure*}

\begin{figure*}
	\includegraphics[width=0.495\textwidth]{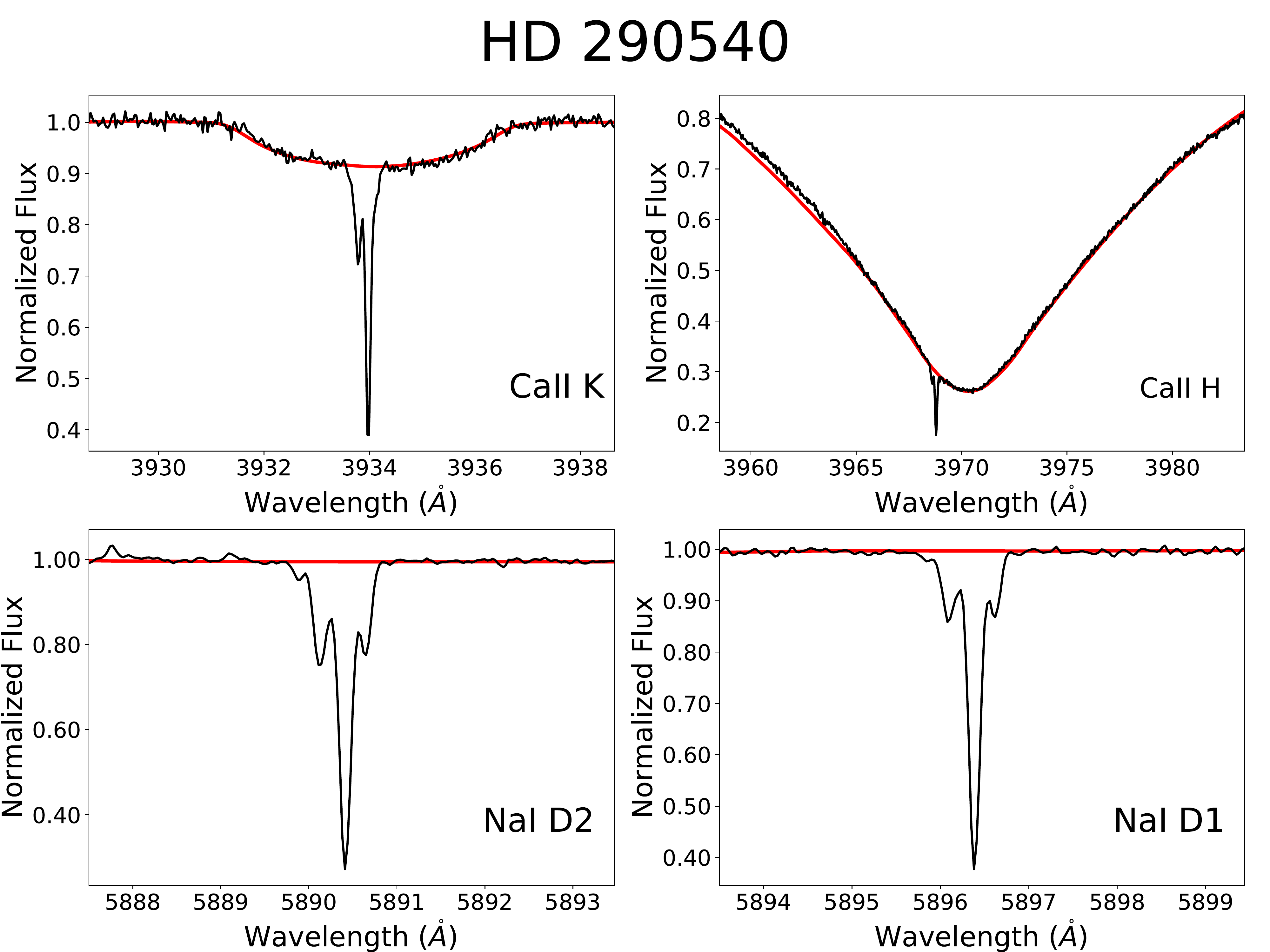}
	\includegraphics[width=0.495\textwidth]{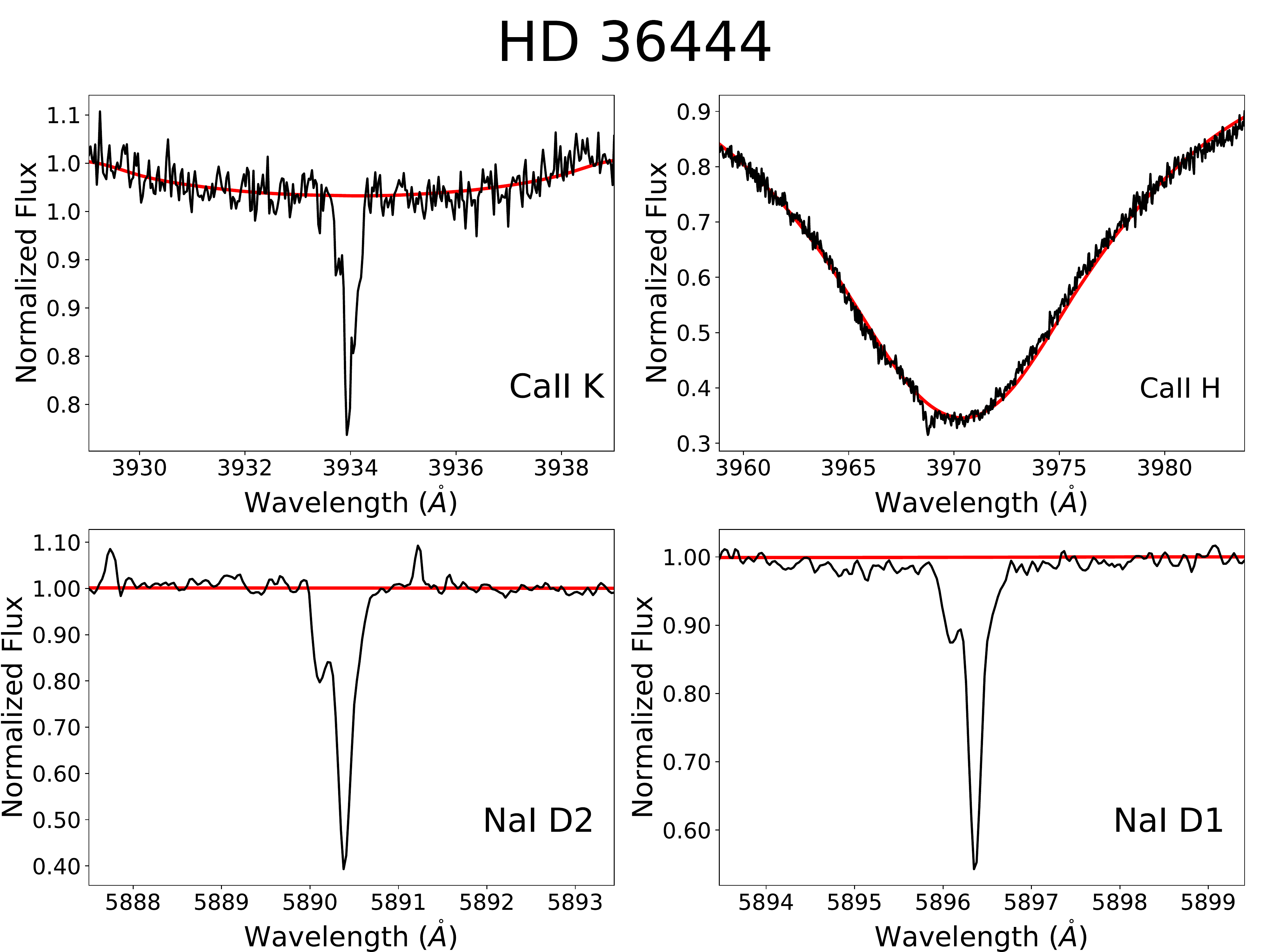}
	    \caption{Best-fit models for $\beta$03\,Tuc's, 66\,Psc's, $\nu$ Hor's, HD\,24966's, HD\,290540's and HD\,36444's Ca\,{\sc ii} H\&K and Na\,{\sc i} D1\&D2 lines. Median from the real spectra in black, synthetic spectrum in red.}
       \label{fig:mod1}
\end{figure*}

\begin{figure*}
	\includegraphics[width=0.495\textwidth]{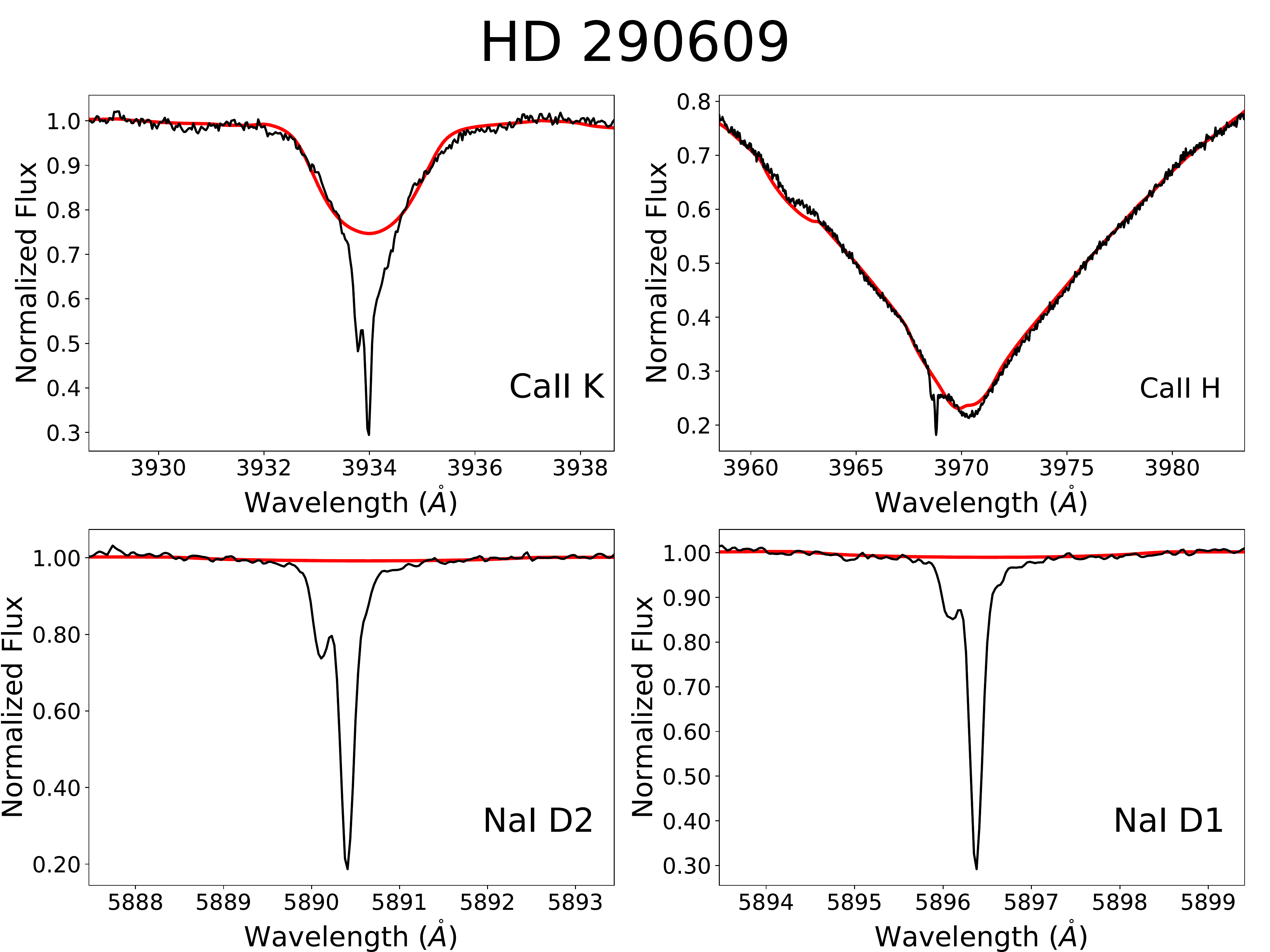}
	\includegraphics[width=0.495\textwidth]{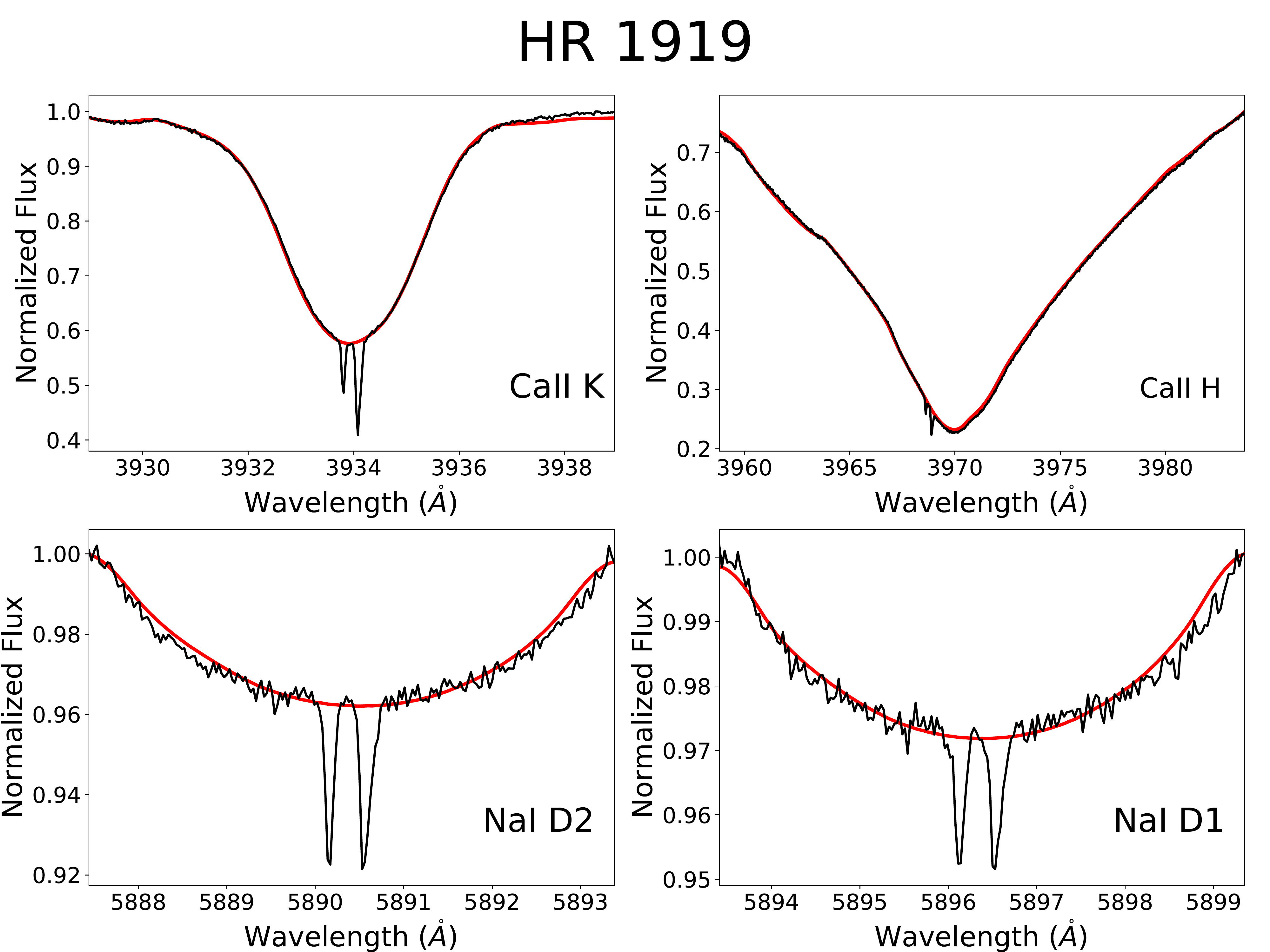}
\end{figure*}

\begin{figure*}
\includegraphics[width=0.495\textwidth]{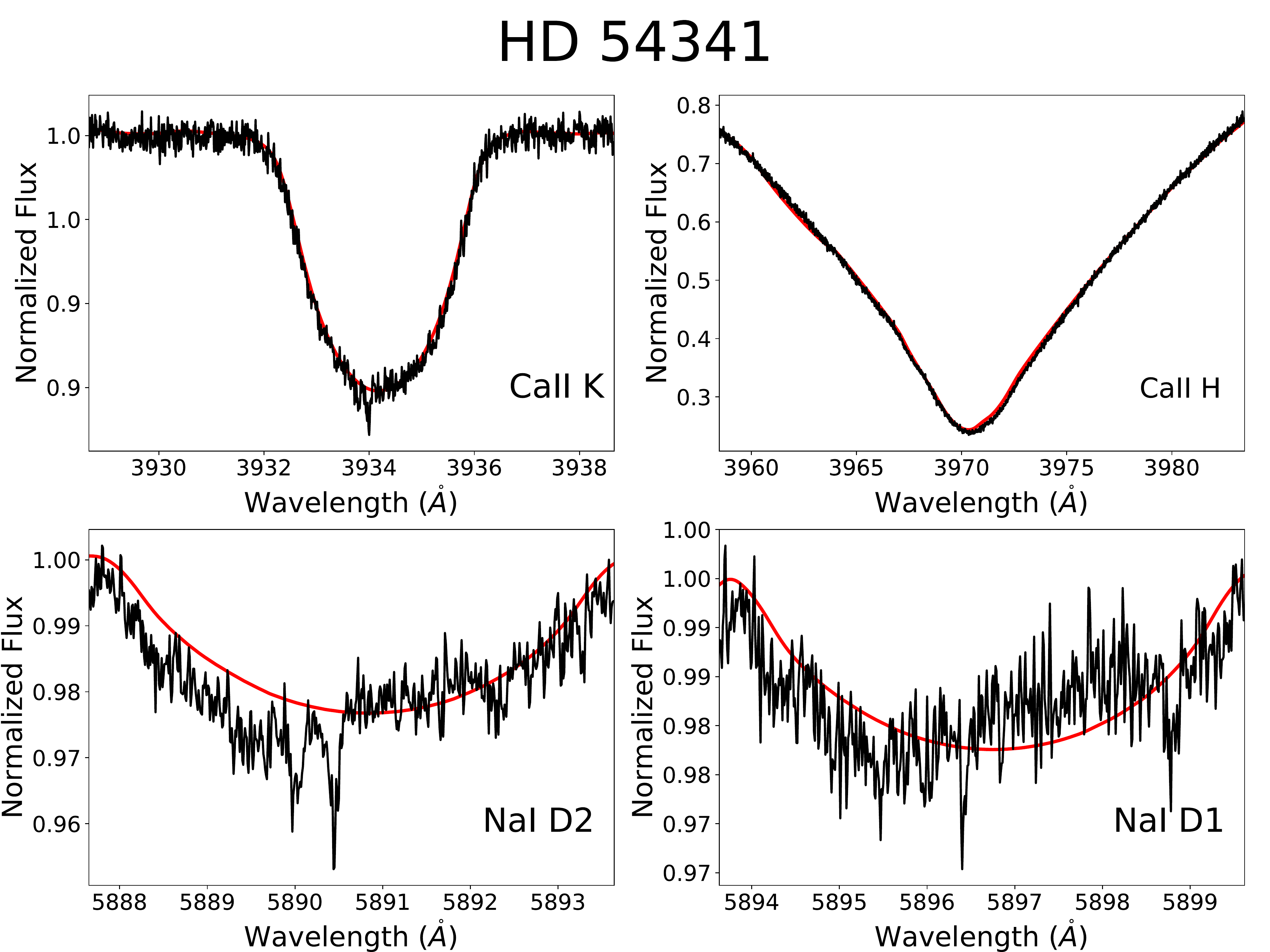}
	\includegraphics[width=0.495\textwidth]{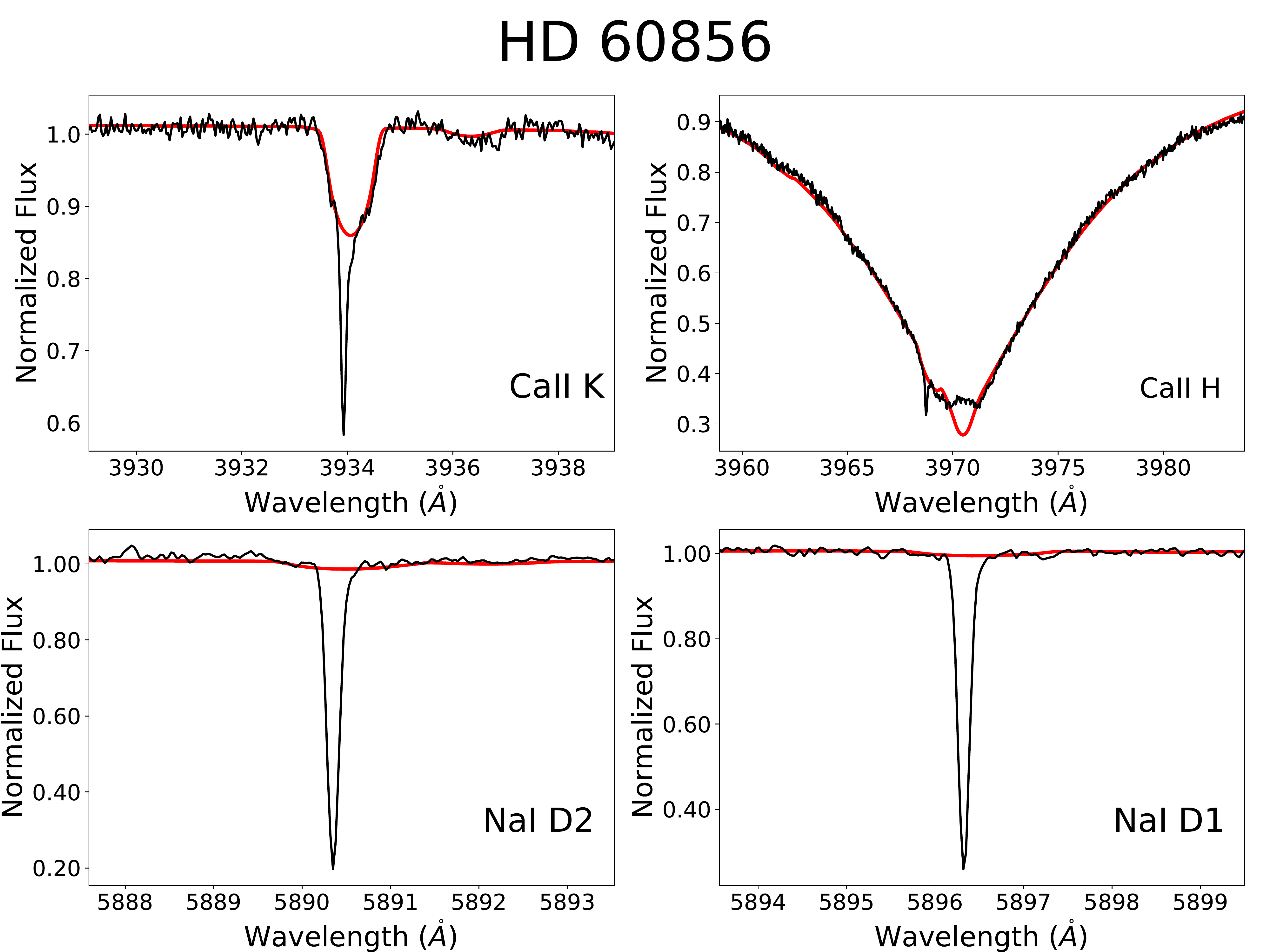}
\end{figure*}

\begin{figure*}
	\includegraphics[width=0.495\textwidth]{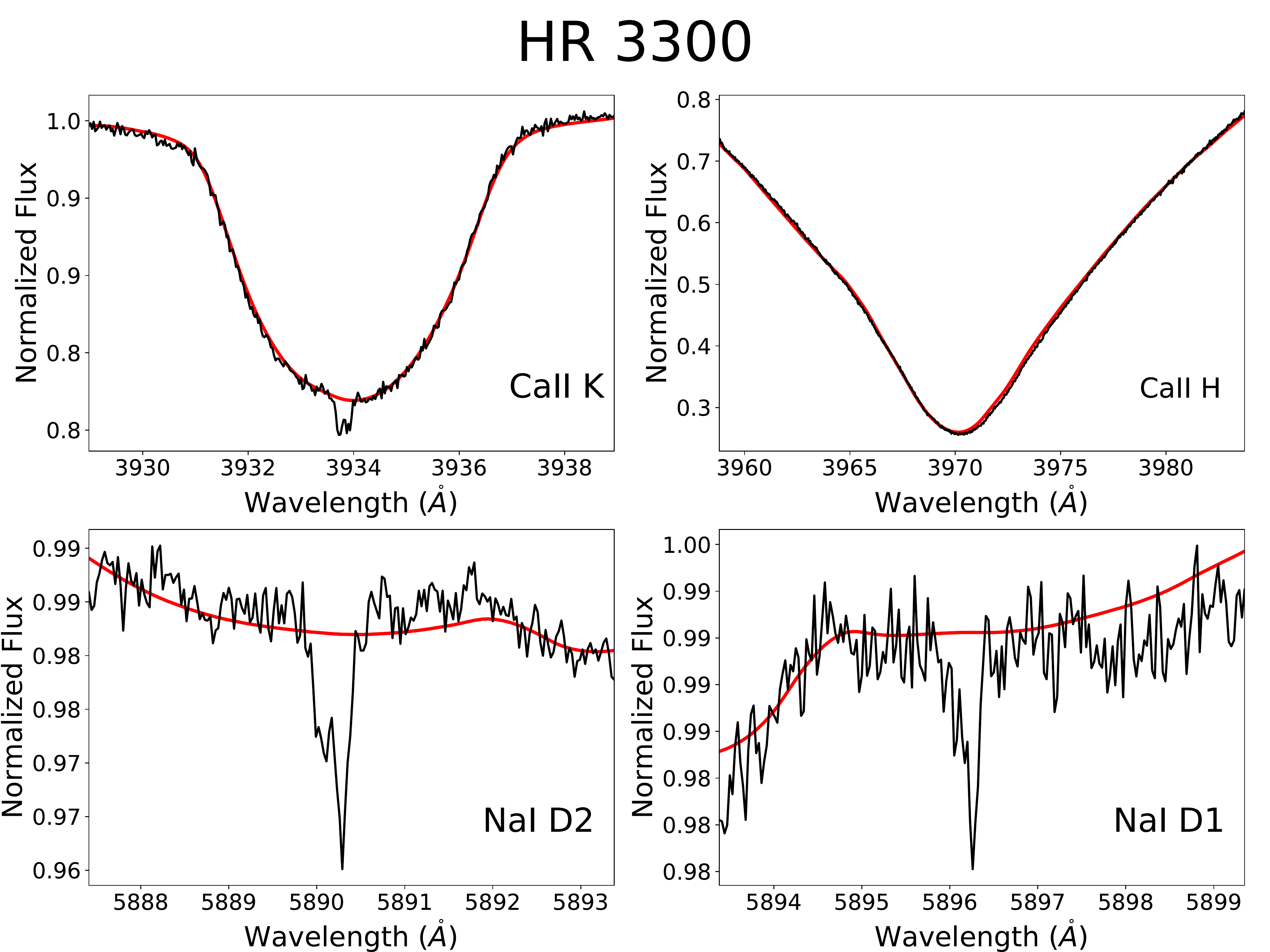}
	\includegraphics[width=0.495\textwidth]{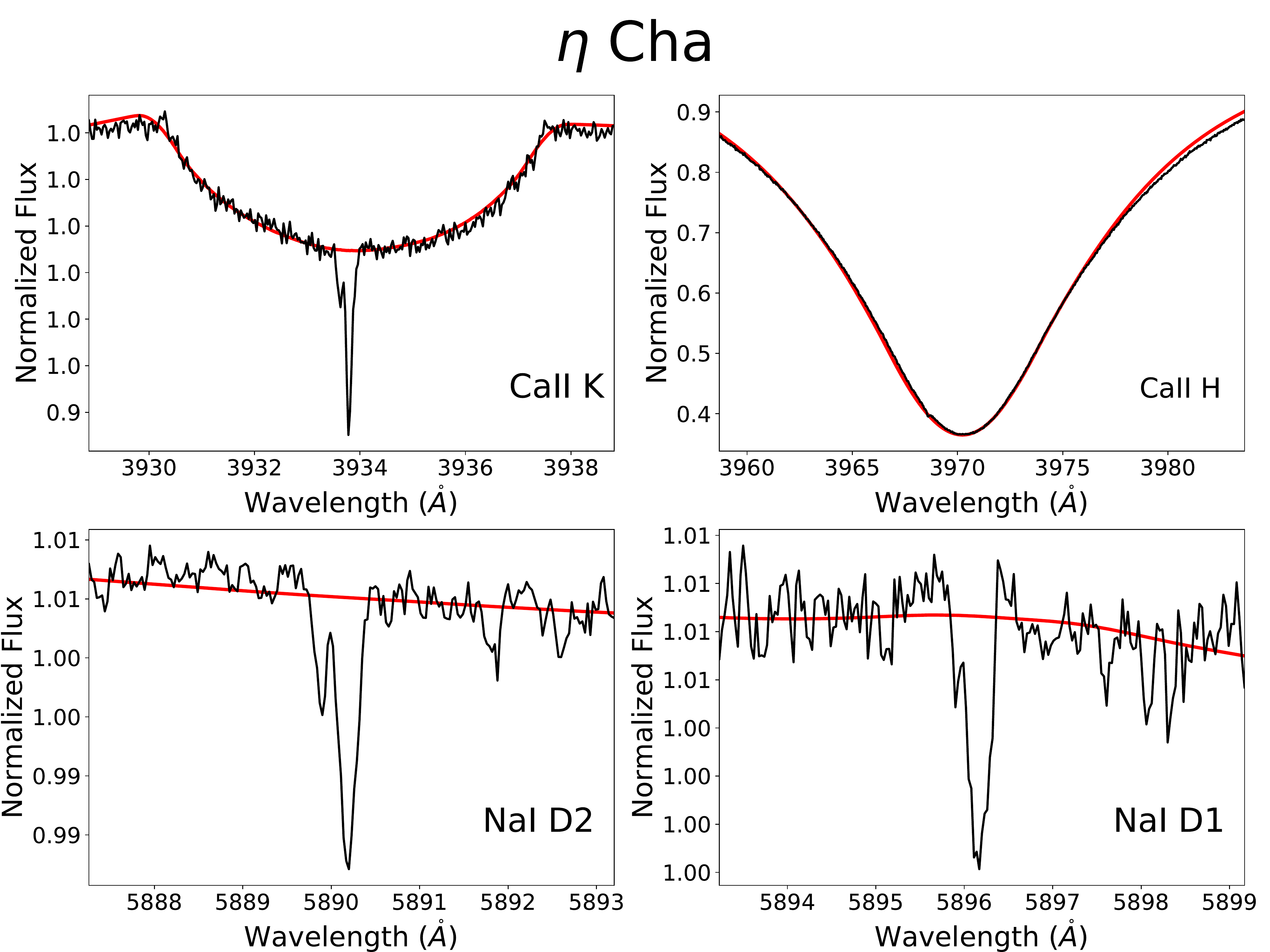}
	\caption{Best-fit models for HD\,290609's, HR\,1919's, HD\,54341's, HD\,60856's, HR 3300's and $\eta$ Cha's Ca\,{\sc ii} H\&K and Na\,{\sc i} D1\&D2 lines. Median from the real spectra in black, synthetic spectrum in red.}
       \label{fig:mod2}
\end{figure*}

\begin{figure*}
	\includegraphics[width=0.495\textwidth]{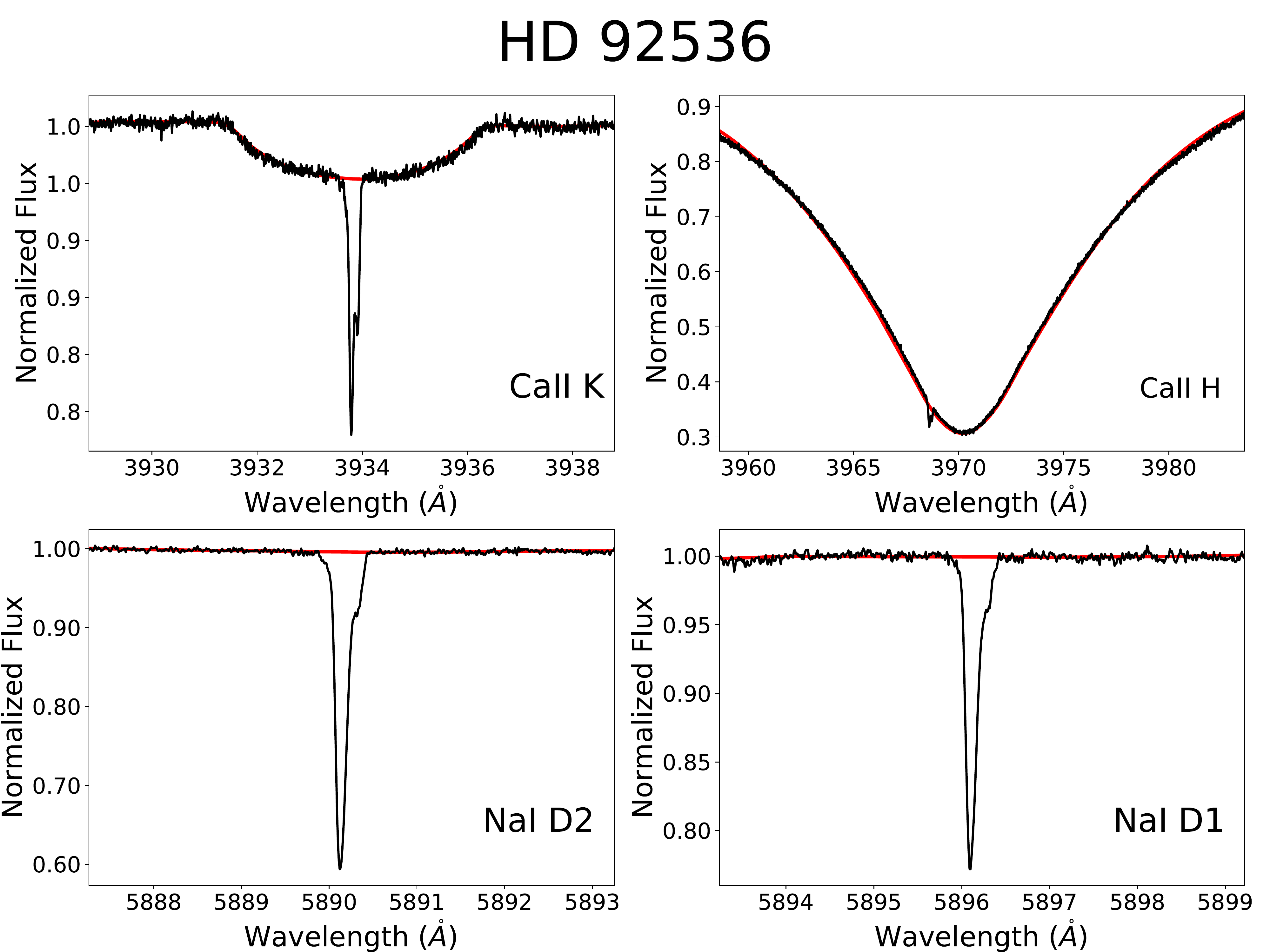}
	\includegraphics[width=0.495\textwidth]{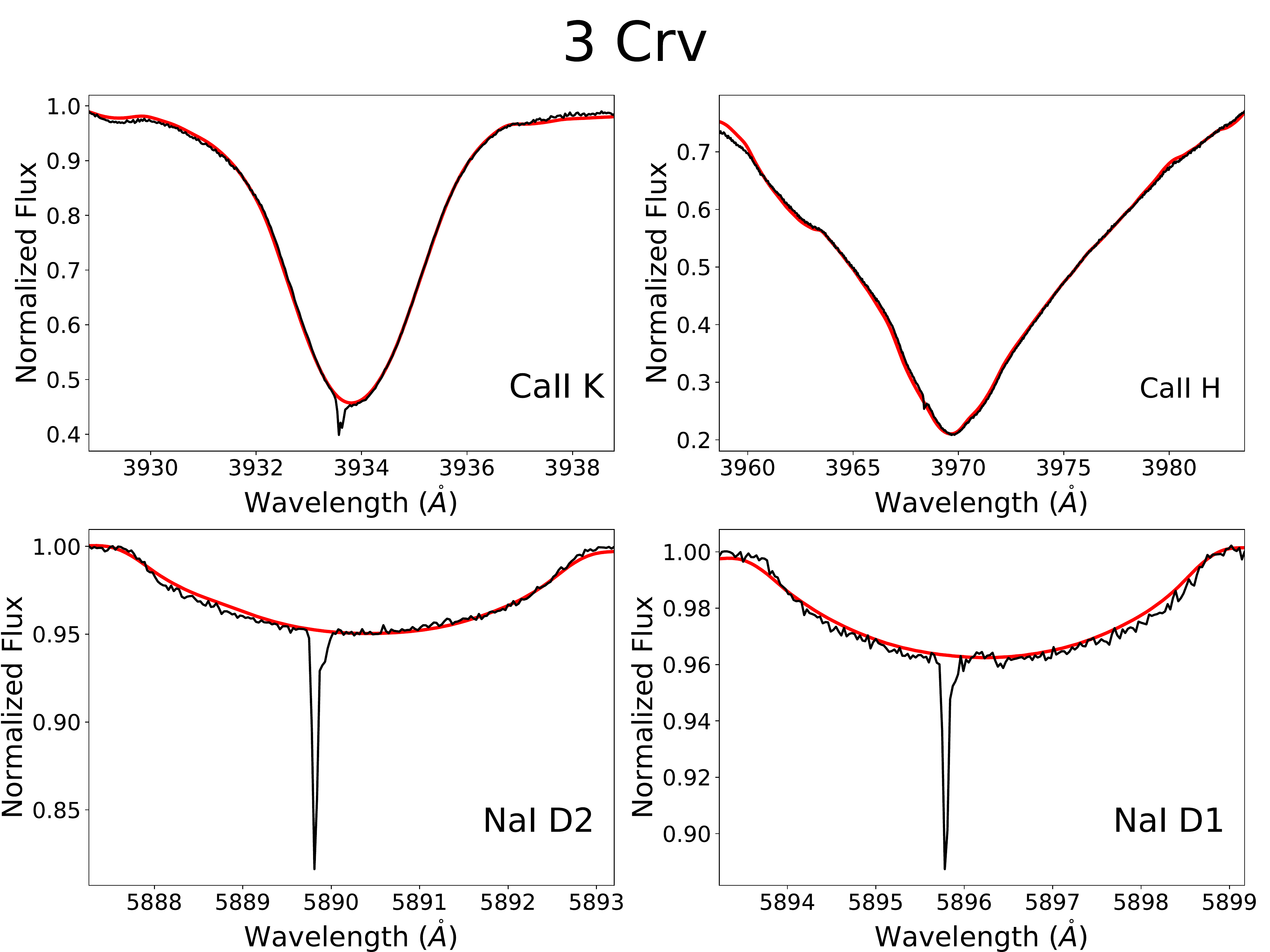}
\end{figure*}

\begin{figure*}
	\includegraphics[width=0.495\textwidth, valign=t]{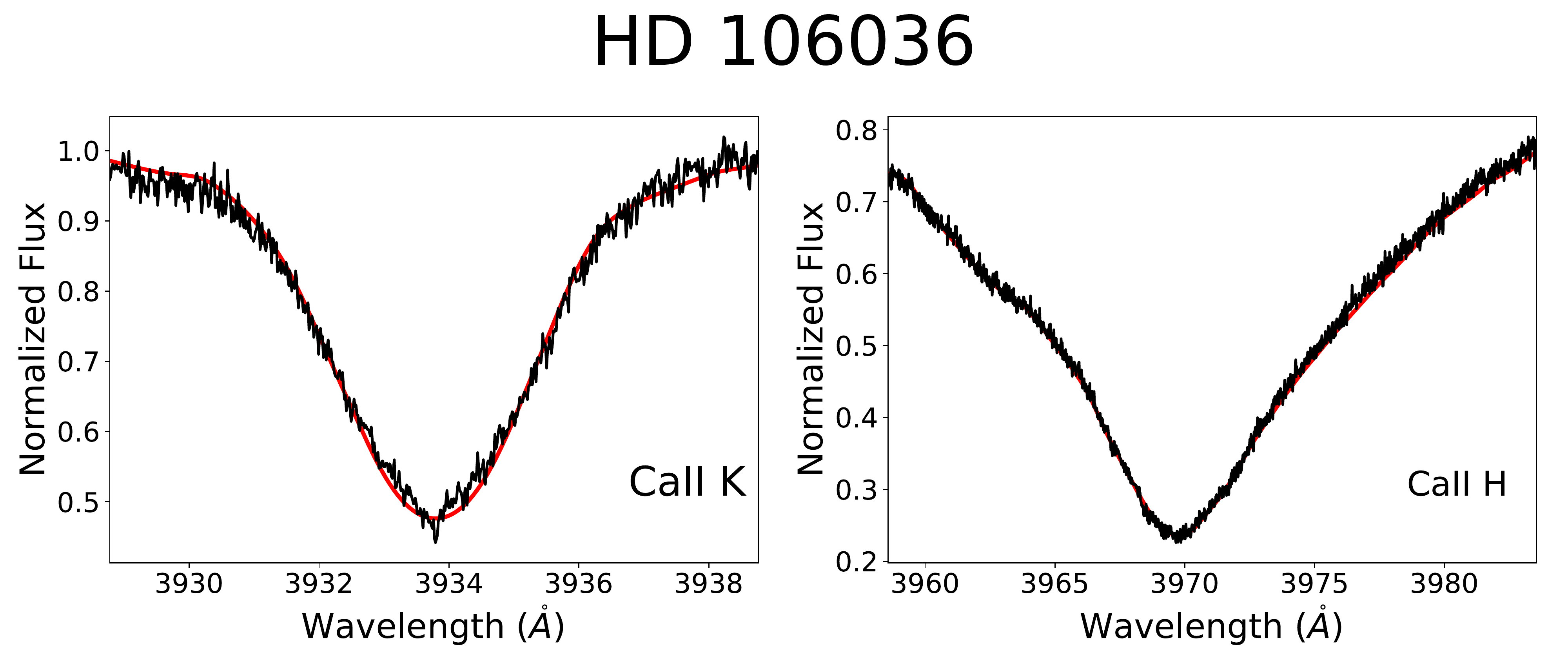}
	\includegraphics[width=0.495\textwidth, valign=t]{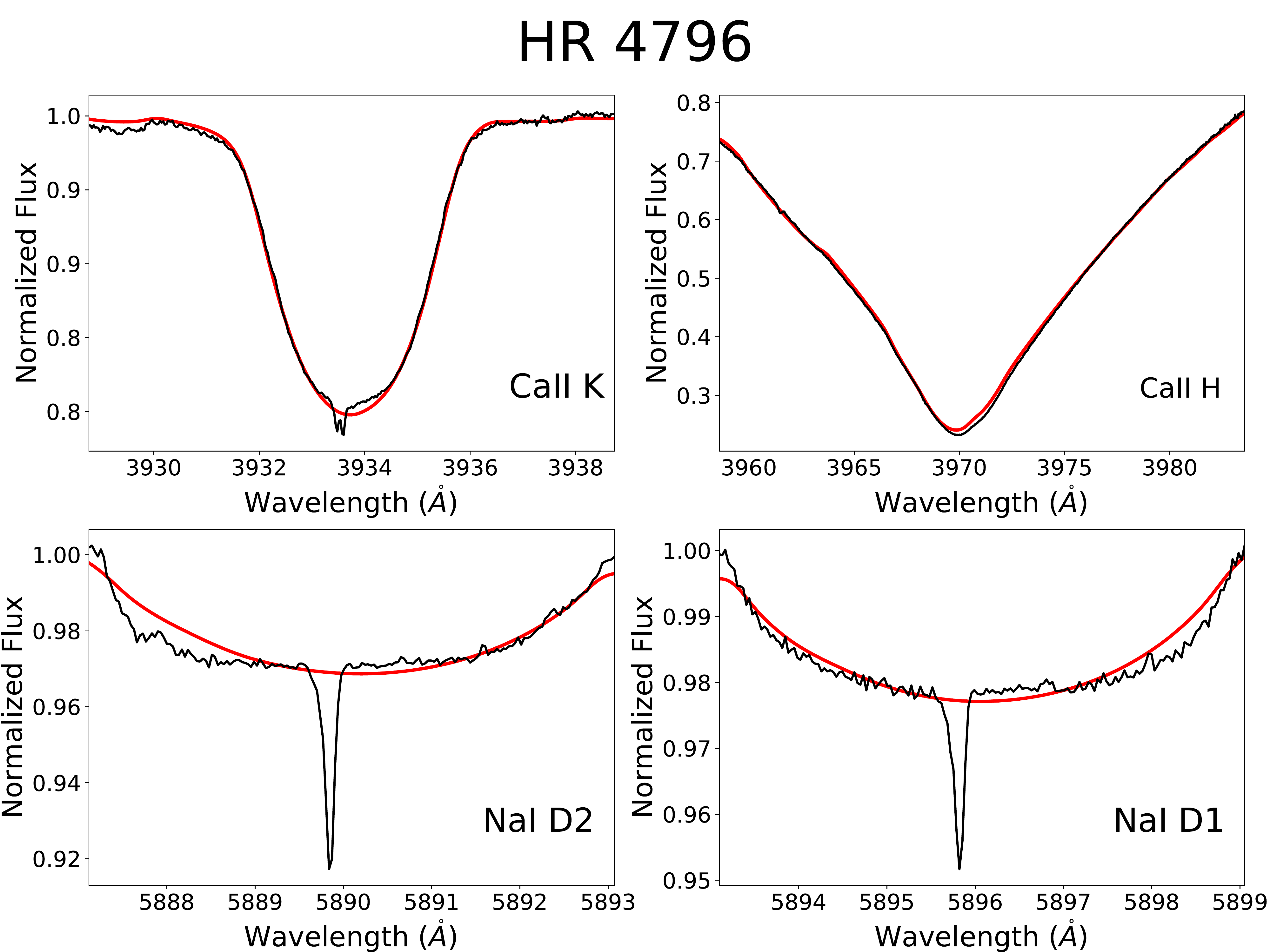}
\end{figure*}

\begin{figure*}
	\includegraphics[width=0.495\textwidth]{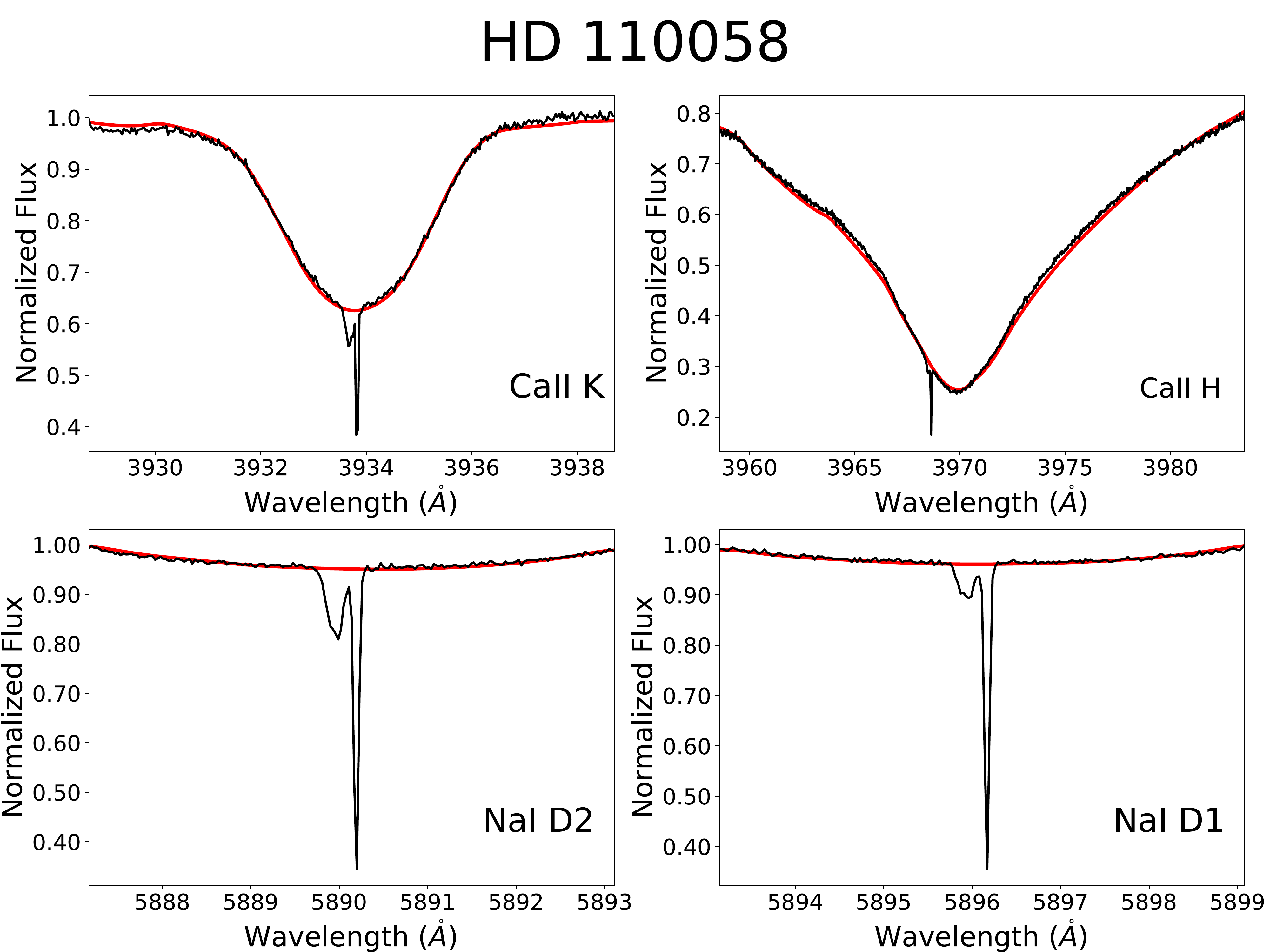}
	\includegraphics[width=0.495\textwidth]{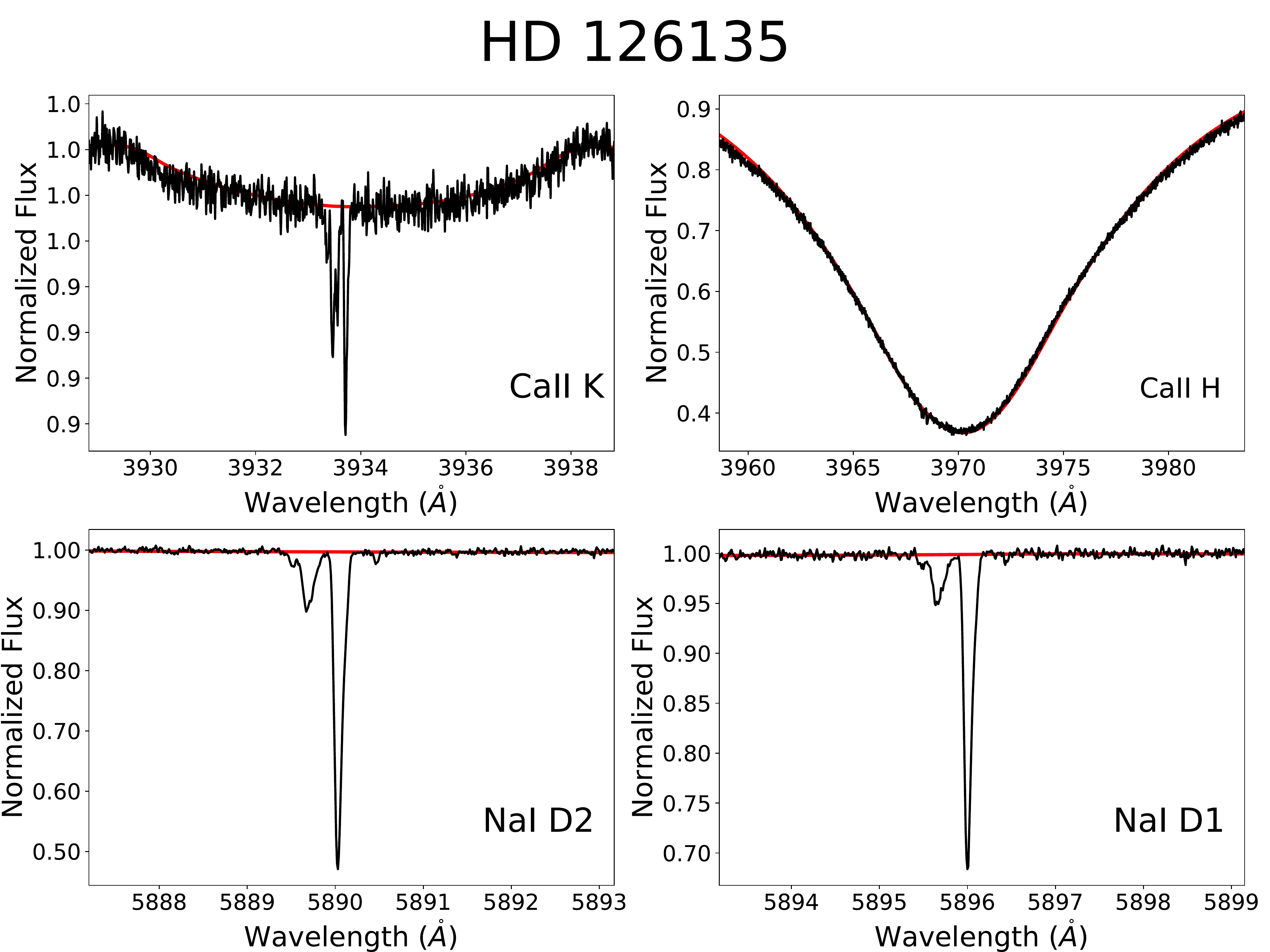}
	 \caption{Best-fit models for HD\,92536's, 3 Crv's, HD\,106036's, HR\,4796's, HD\,110058's and HD\,126235's Ca\,{\sc ii} H\&K and Na\,{\sc i} D1\&D2 lines. Median from the real spectra in black, synthetic spectrum in red.}
       \label{fig:mod3}
\end{figure*}

\begin{figure*}
	\includegraphics[width=0.495\textwidth]{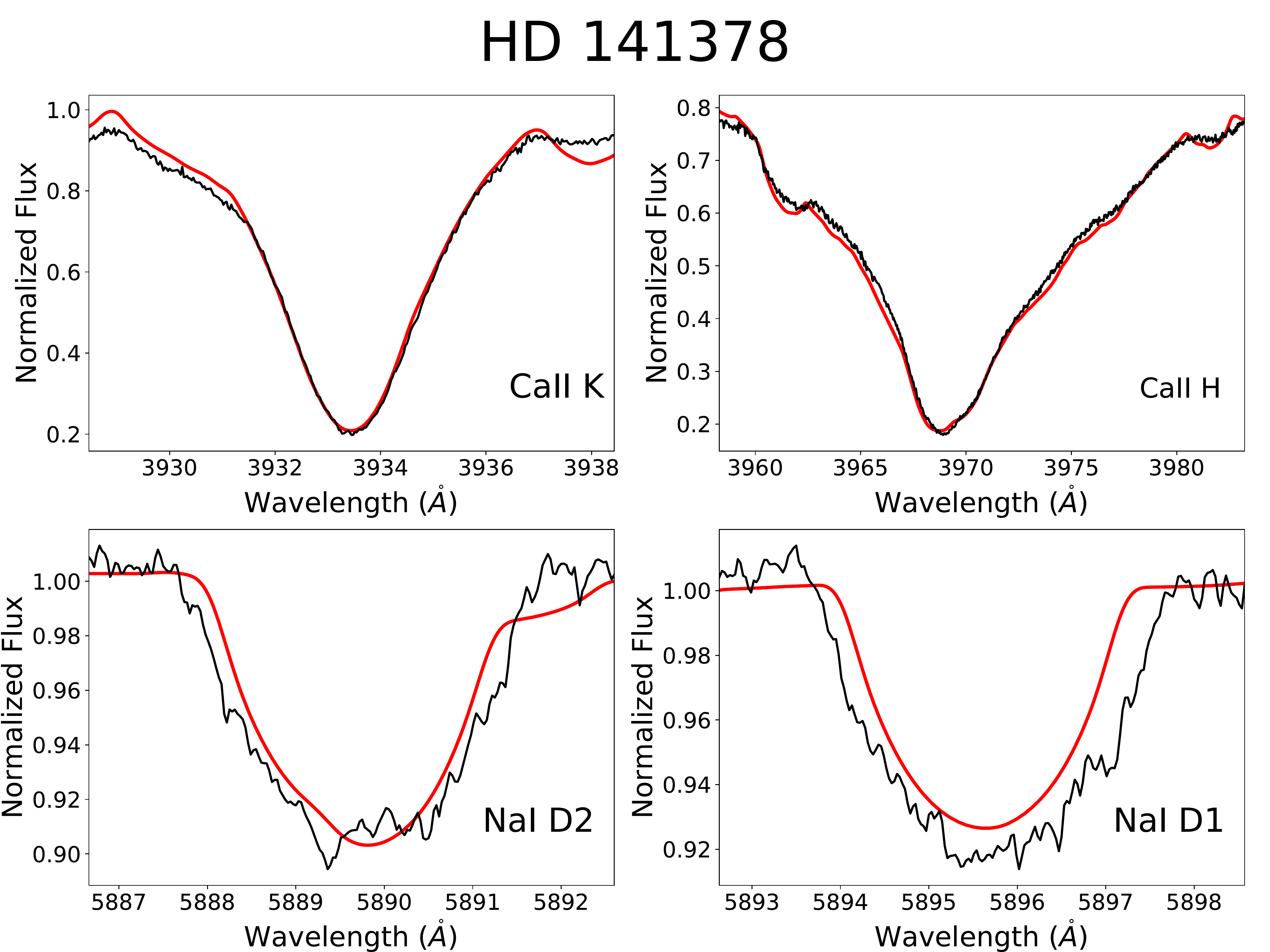}
	\includegraphics[width=0.495\textwidth]{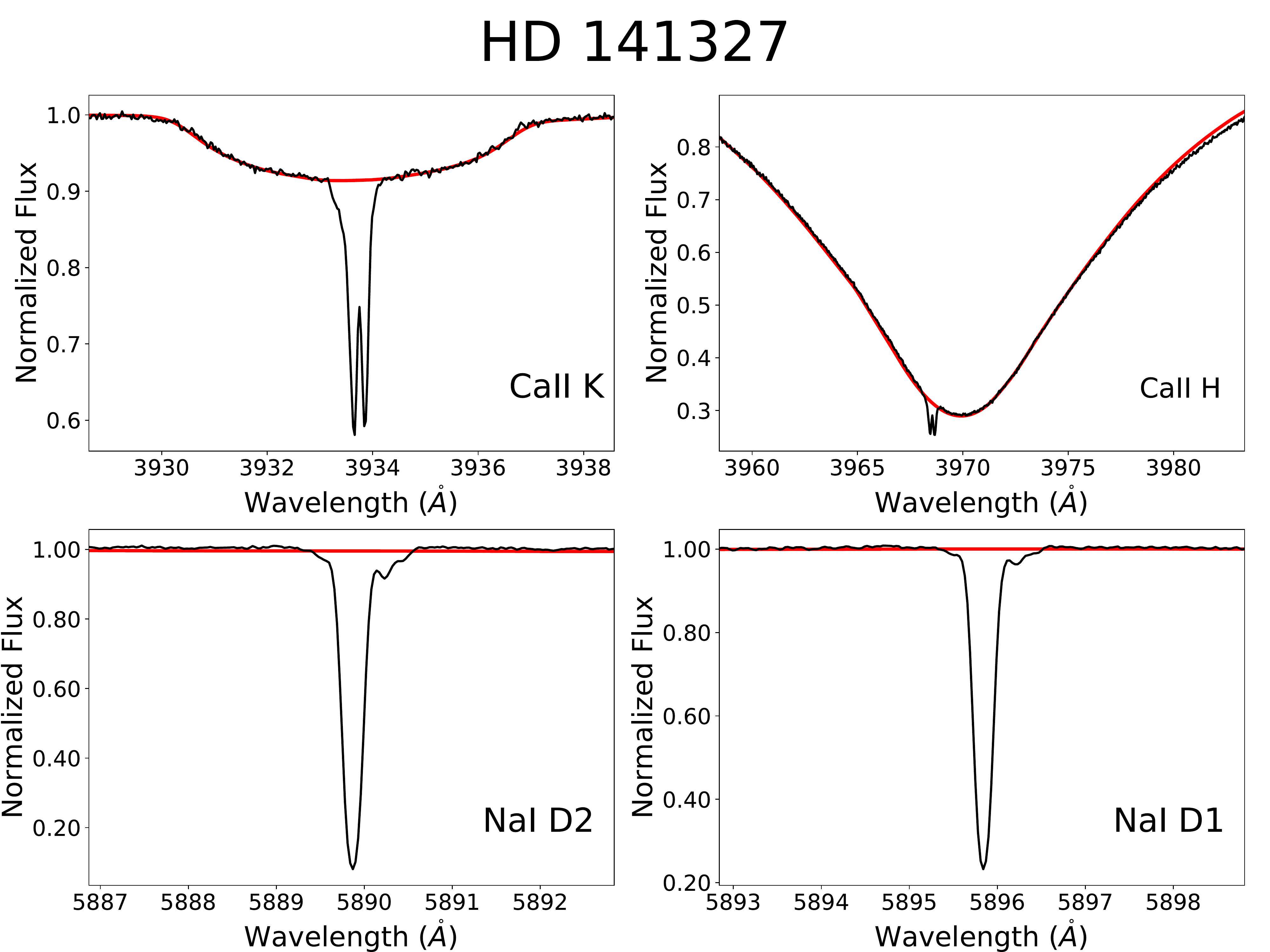}
\end{figure*}

\begin{figure*}
	\includegraphics[width=0.495\textwidth]{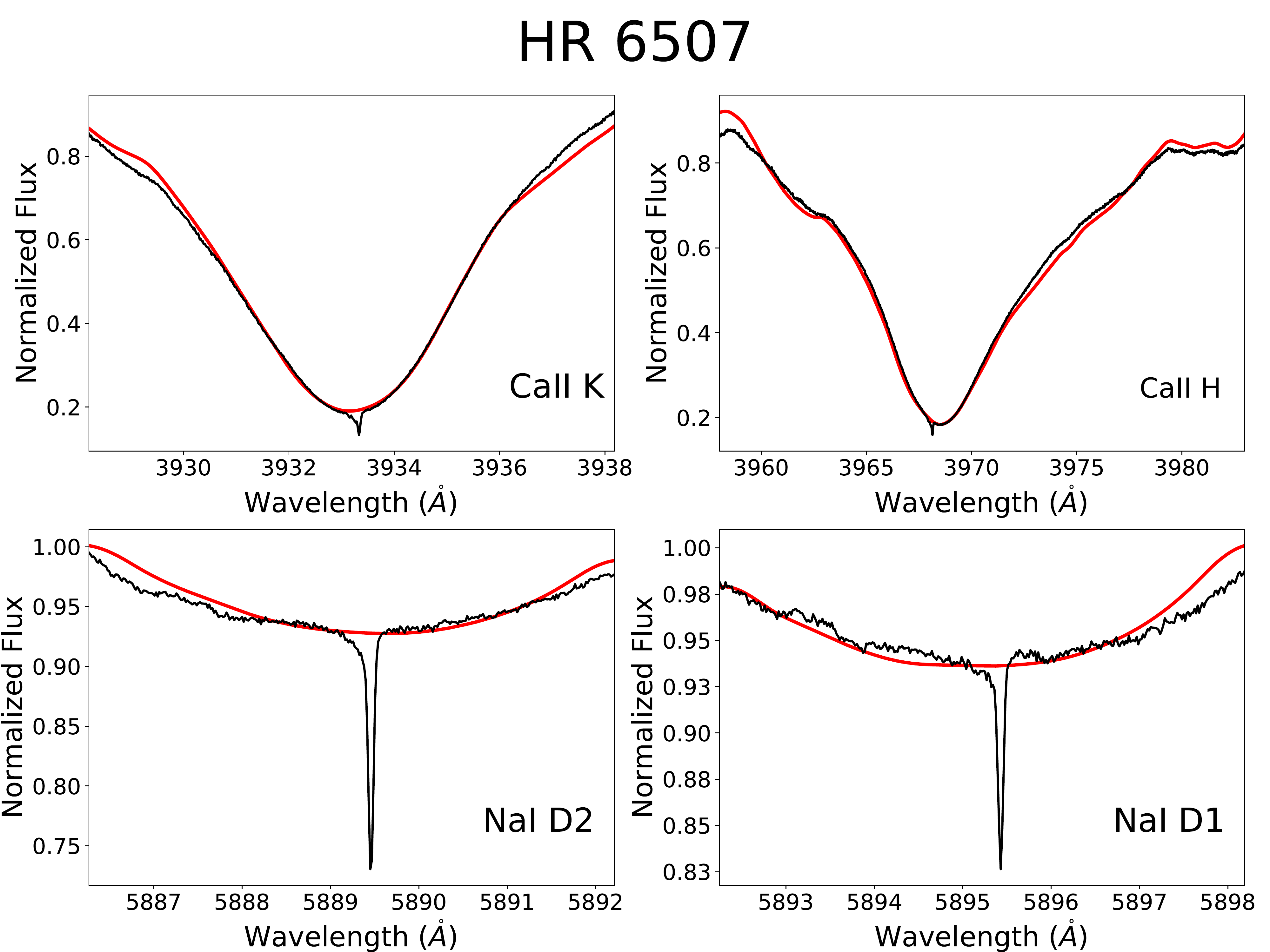}
	\includegraphics[width=0.495\textwidth]{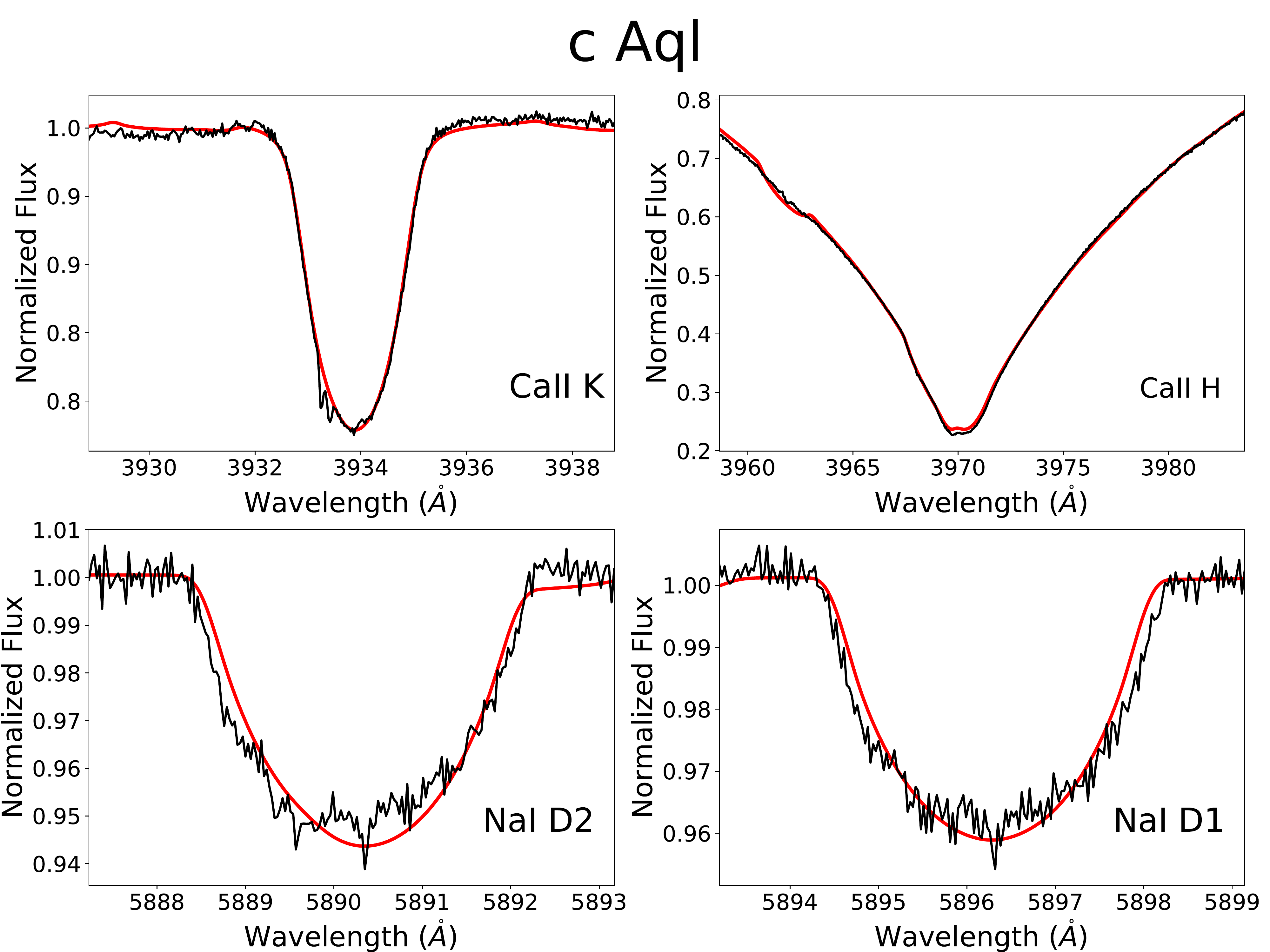}
	\caption{Best-fit models for HD\,141378's, HD\,141327's, HR\,6507 and c Aql's Ca\,{\sc ii} H\&K and Na\,{\sc i} D1\&D2 lines. Median from the real spectra in black, synthetic spectrum in red.}
       \label{fig:mod4}
\end{figure*}

\section{Residual components}

In this appendix we provide the Figures showing every isolated component and the multi-gaussian fit obtained to reproduce  and characterize those features.
%\clearpage

\begin{figure*}  %-> funciona bien con pdf
\includegraphics[page=1, width=0.32\textwidth]{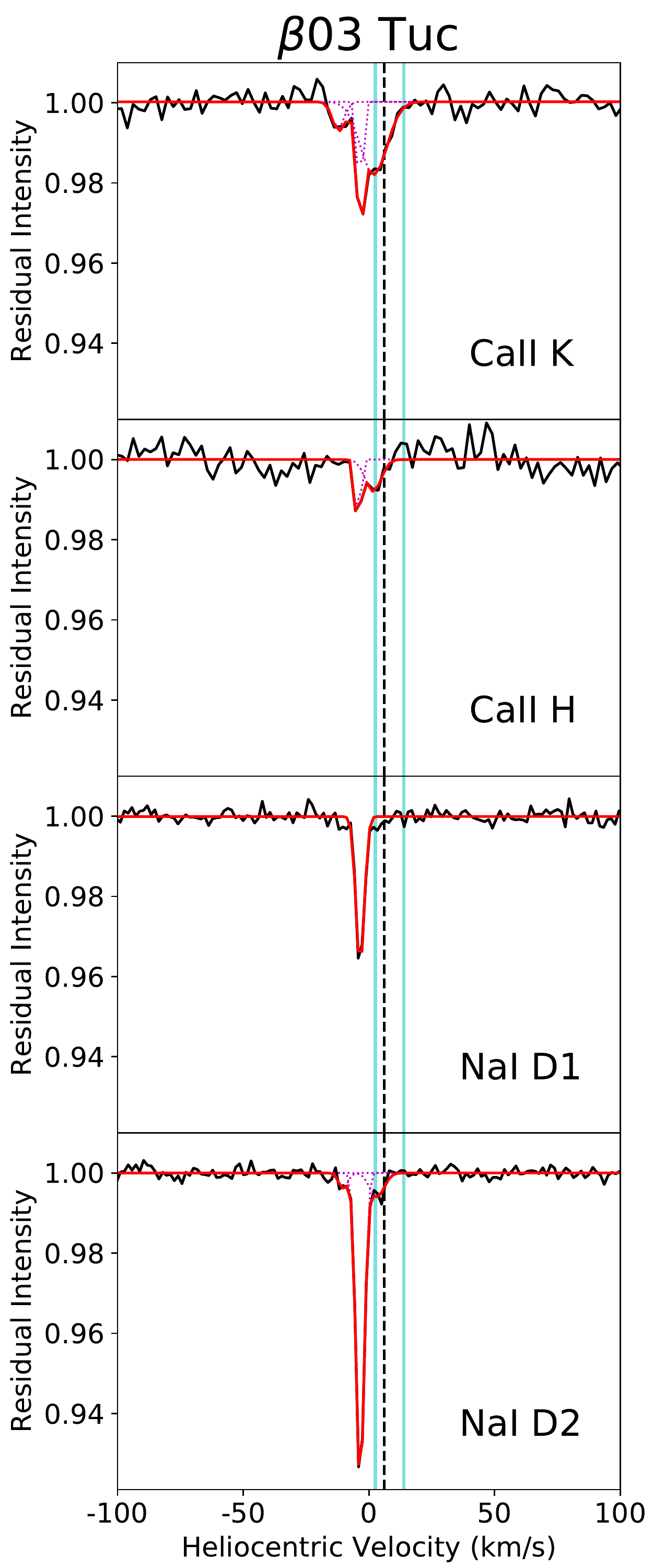}
\includegraphics[page=2, width=0.32\textwidth]{median_density.pdf}
\includegraphics[page=3, width=0.32\textwidth]{median_density.pdf}
\caption{Absorption profiles of the Ca\,{\sc ii} H \& K and Na\,{\sc i} D1 \& D2 lines for $\beta$03\,Tuc, 66\,Psc and $\nu$ Hor. Photospheric absorptions has been subtracted and the remaining extra components have been modelled by gaussian profiles. Individual gaussian fits are shown in dotted magenta lines and the combined profile is shown in red. Dashed black line marks the estimated radial velocity of the star and cyan lines mark the velocity of the traversing clouds in the line of sight with their respective errors as their line widths. }
\label{fig:abs1}
\end{figure*}

\begin{figure*}  %-> funciona bien con pdf
\includegraphics[page=4, width=0.32\textwidth]{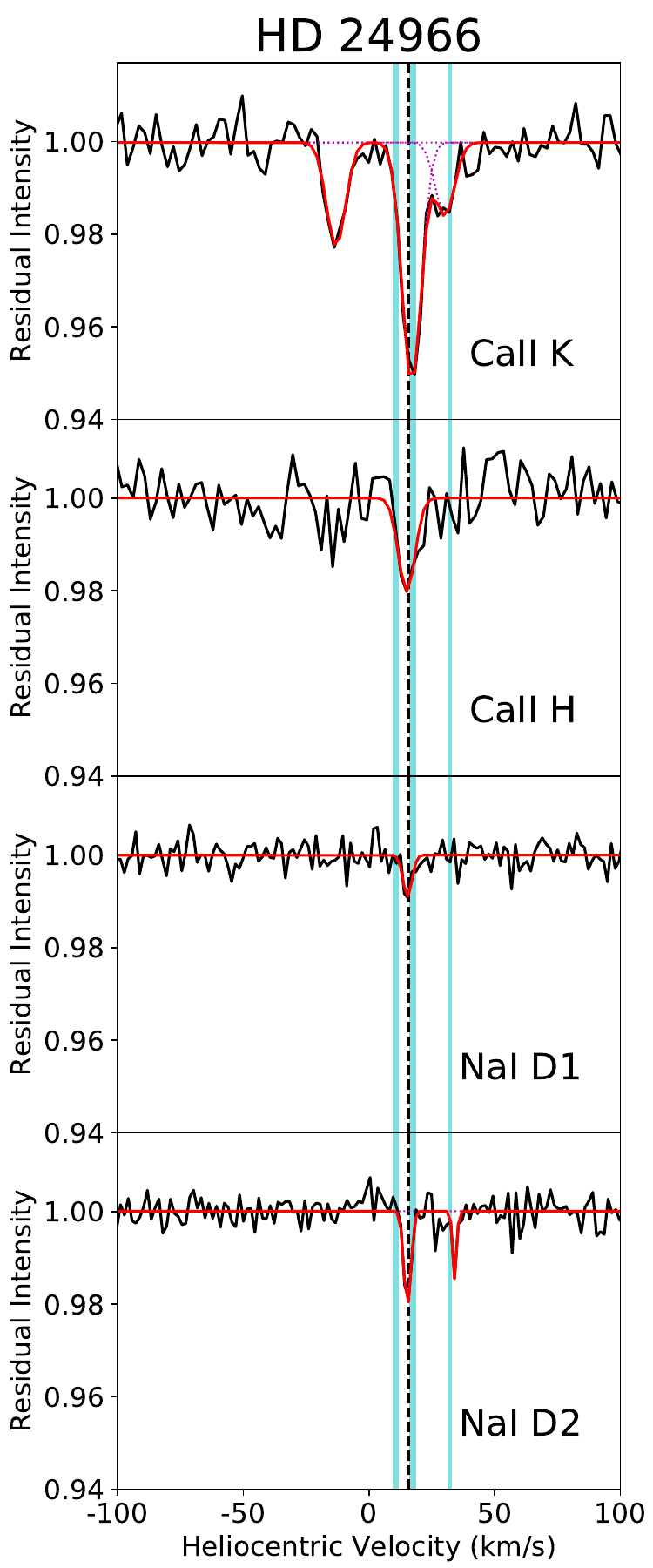}
\includegraphics[page=5, width=0.32\textwidth]{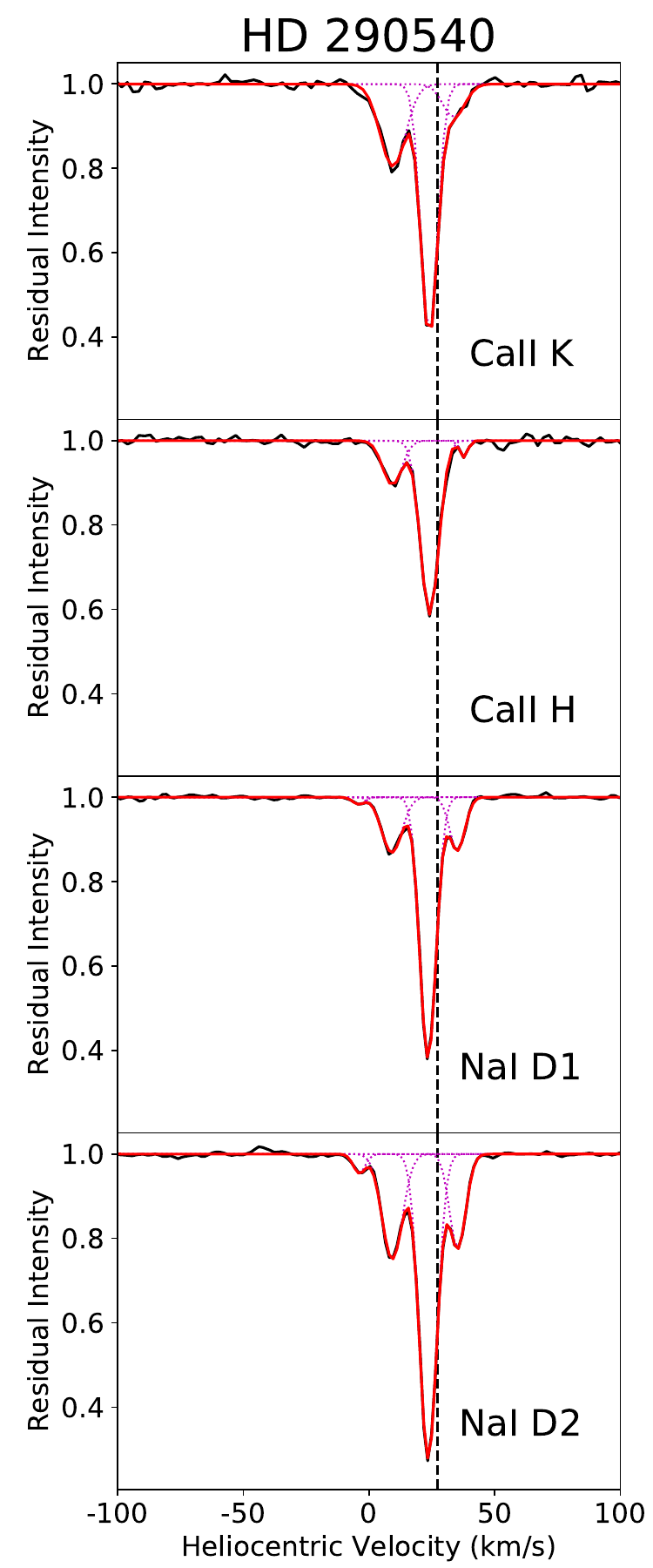}
\includegraphics[page=6, width=0.32\textwidth]{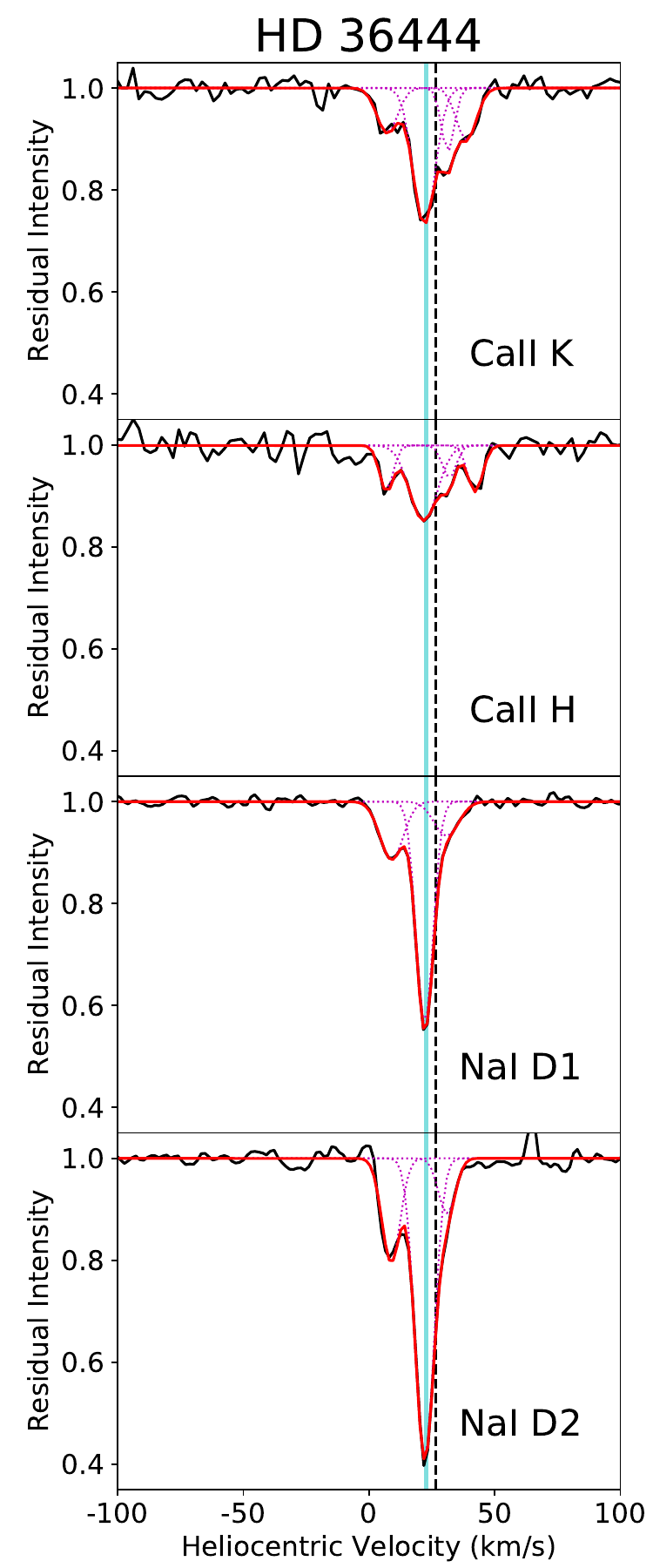}
\caption{Absorption profiles of the Ca\,{\sc ii} H \& K and Na\,{\sc i} D1 \& D2 lines for HD\,24966, HD\,290540 and HD\,36444. Photospheric absorptions has been subtracted and the remaining extra components have been modelled by gaussian profiles. Individual gaussian fits are shown in dotted magenta lines and the combined profile is shown in red. Dashed black line marks the estimated radial velocity of the star and cyan lines mark the velocity of the traversing clouds in the line of sight with their respective errors as their line widths. }
\label{fig:abs2}
\end{figure*}

\begin{figure*}  %-> funciona bien con pdf
\includegraphics[page=7, width=0.32\textwidth]{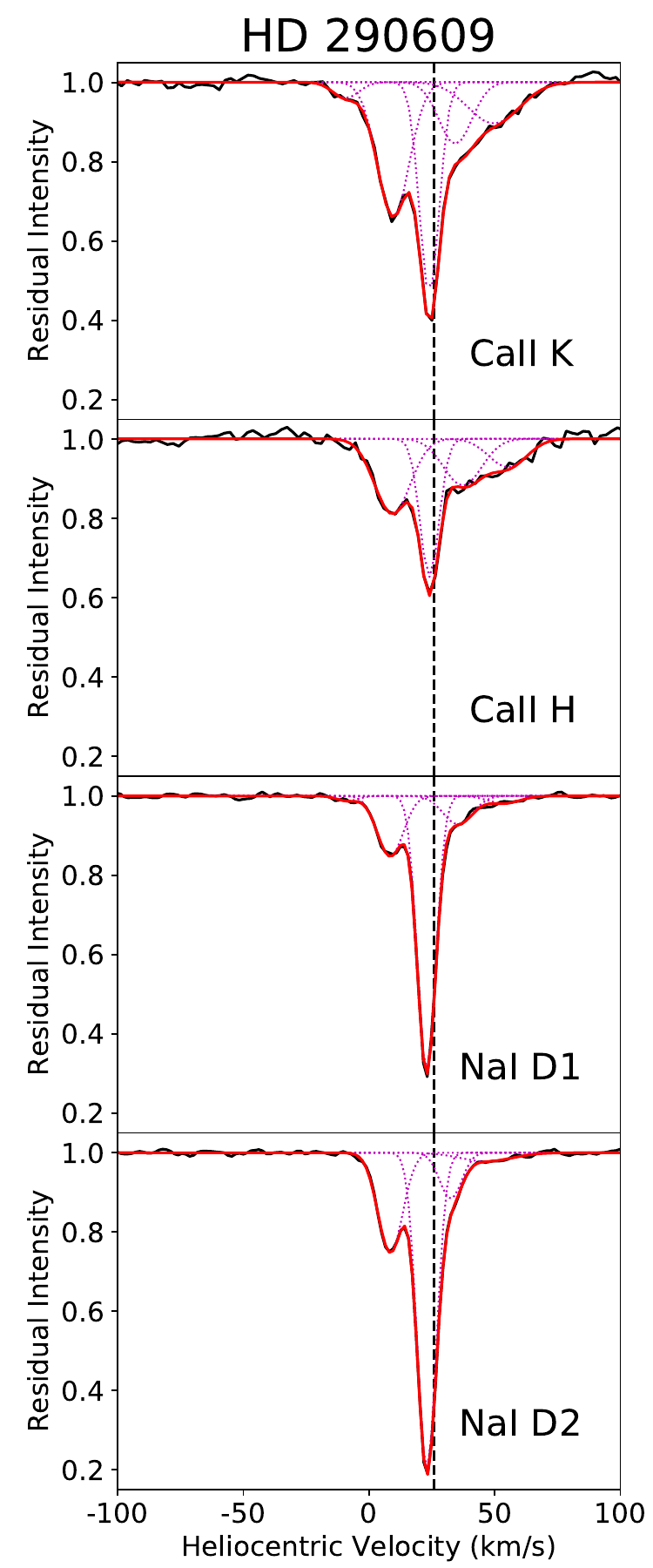}
\includegraphics[page=8, width=0.32\textwidth]{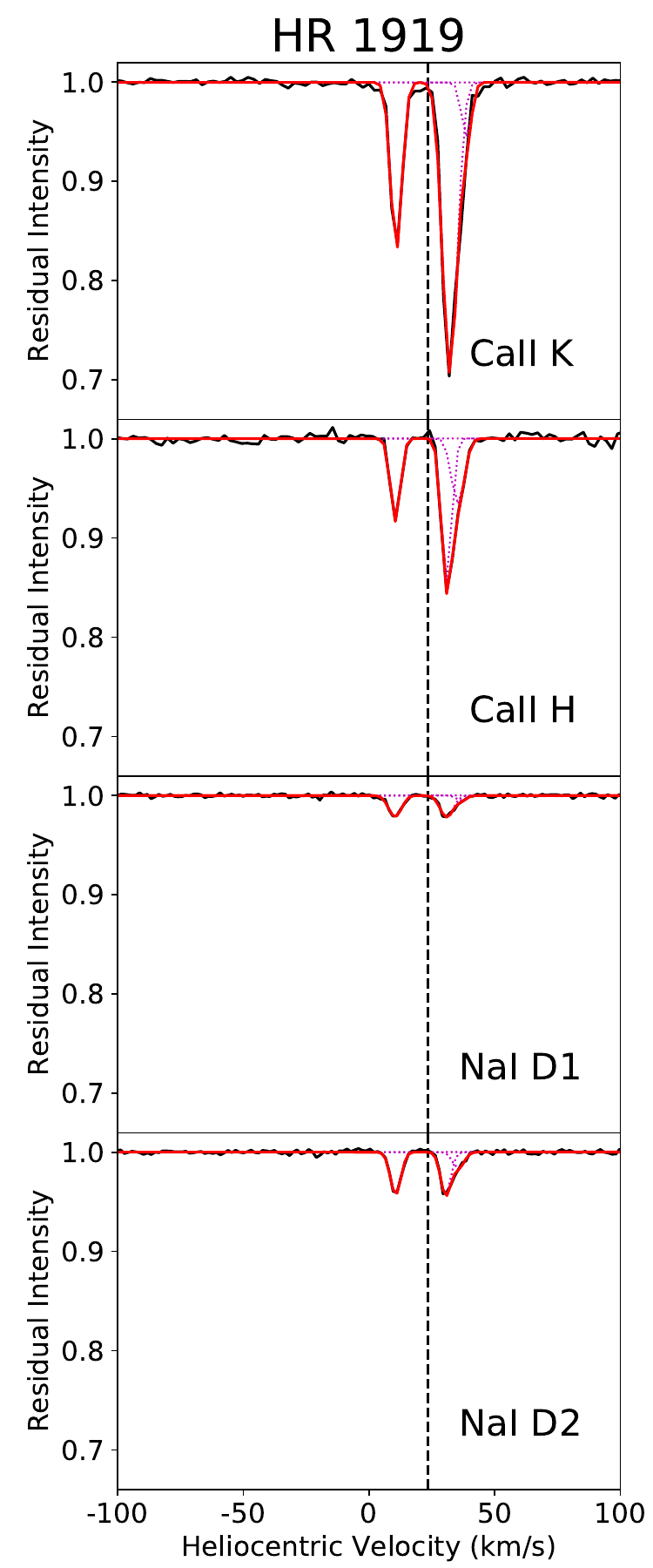}
\includegraphics[page=9, width=0.32\textwidth]{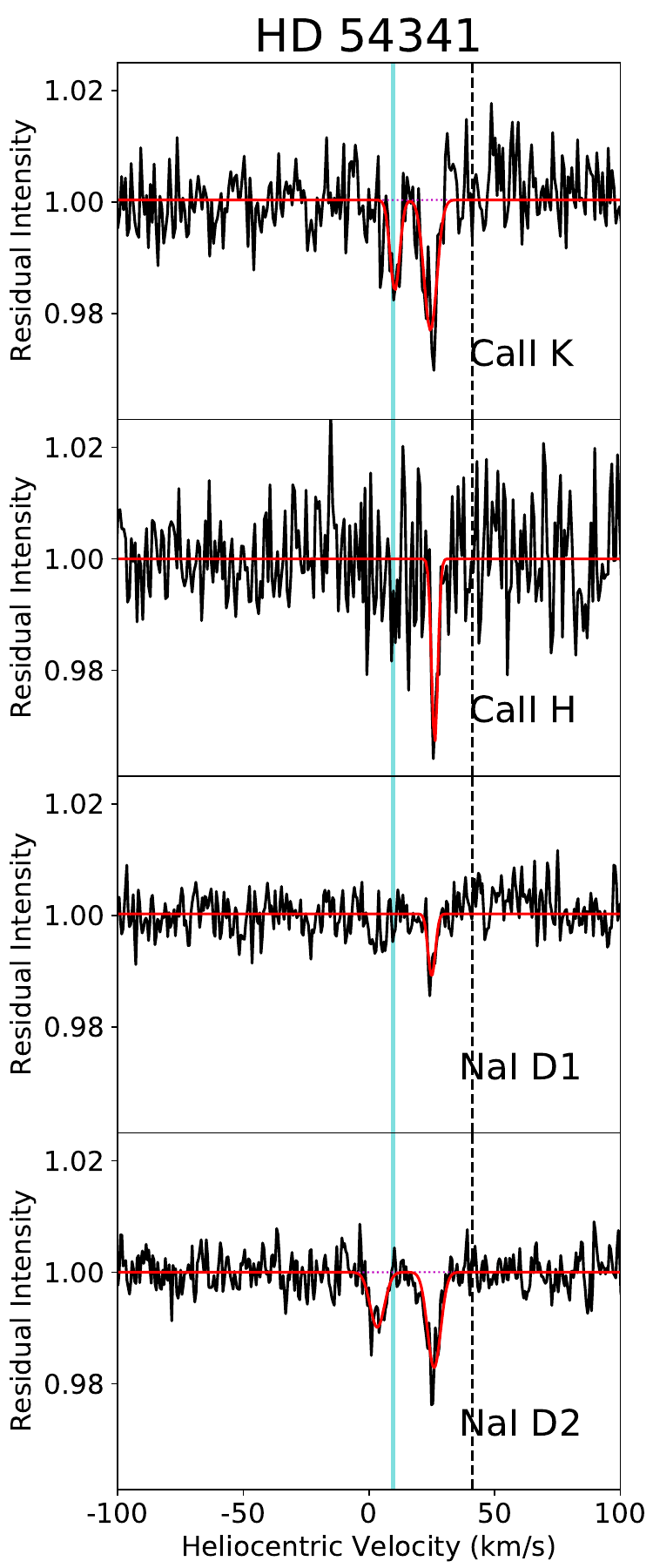}
\caption{Absorption profiles of the Ca\,{\sc ii} H \& K and Na\,{\sc i} D1 \& D2 lines for HD\,290609, HR\,1919 and HD\,54341. Photospheric absorptions has been subtracted and the remaining extra components have been modelled by gaussian profiles. Individual gaussian fits are shown in dotted magenta lines and the combined profile is shown in red. Dashed black line marks the estimated radial velocity of the star and cyan lines mark the velocity of the traversing clouds in the line of sight with their respective errors as their line widths.}
\label{fig:abs3}
\end{figure*}

\begin{figure*}  %-> funciona bien con pdf
\includegraphics[page=10, width=0.32\textwidth]{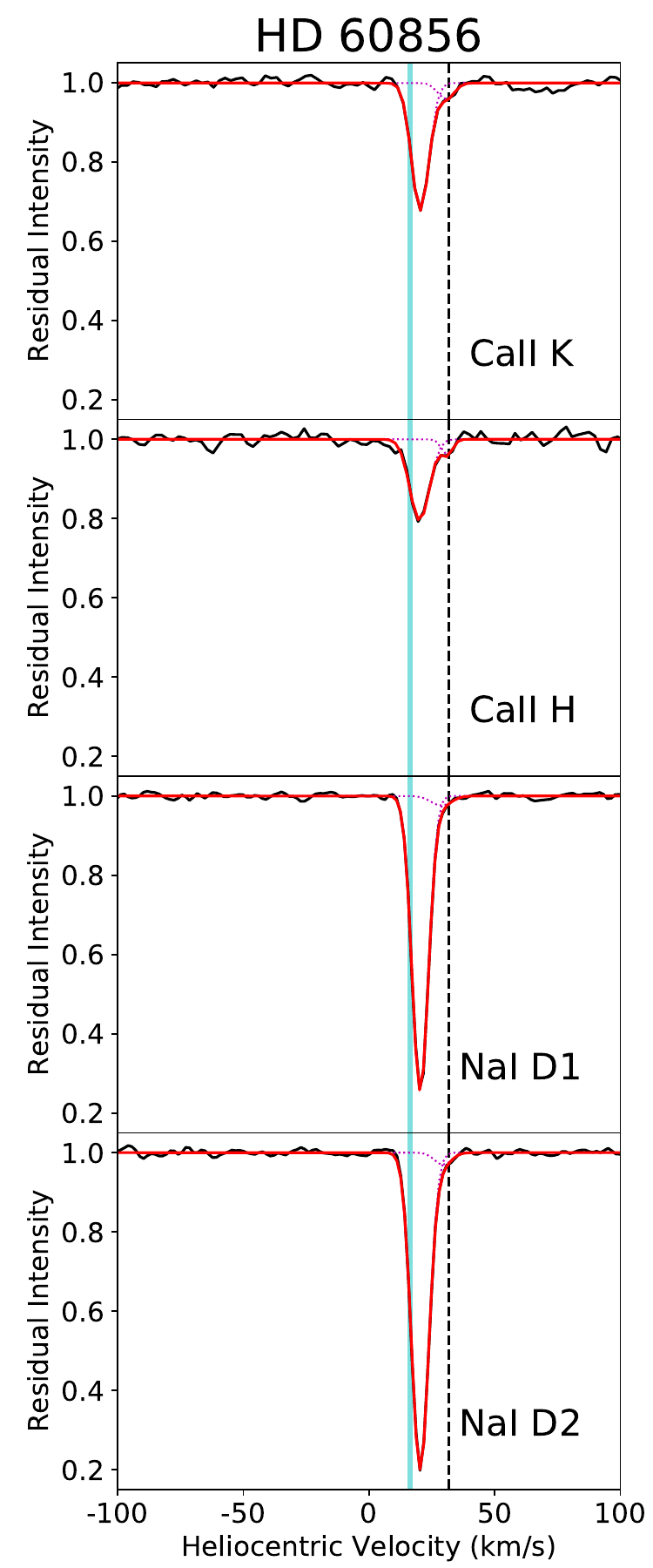}
\includegraphics[page=11, width=0.32\textwidth]{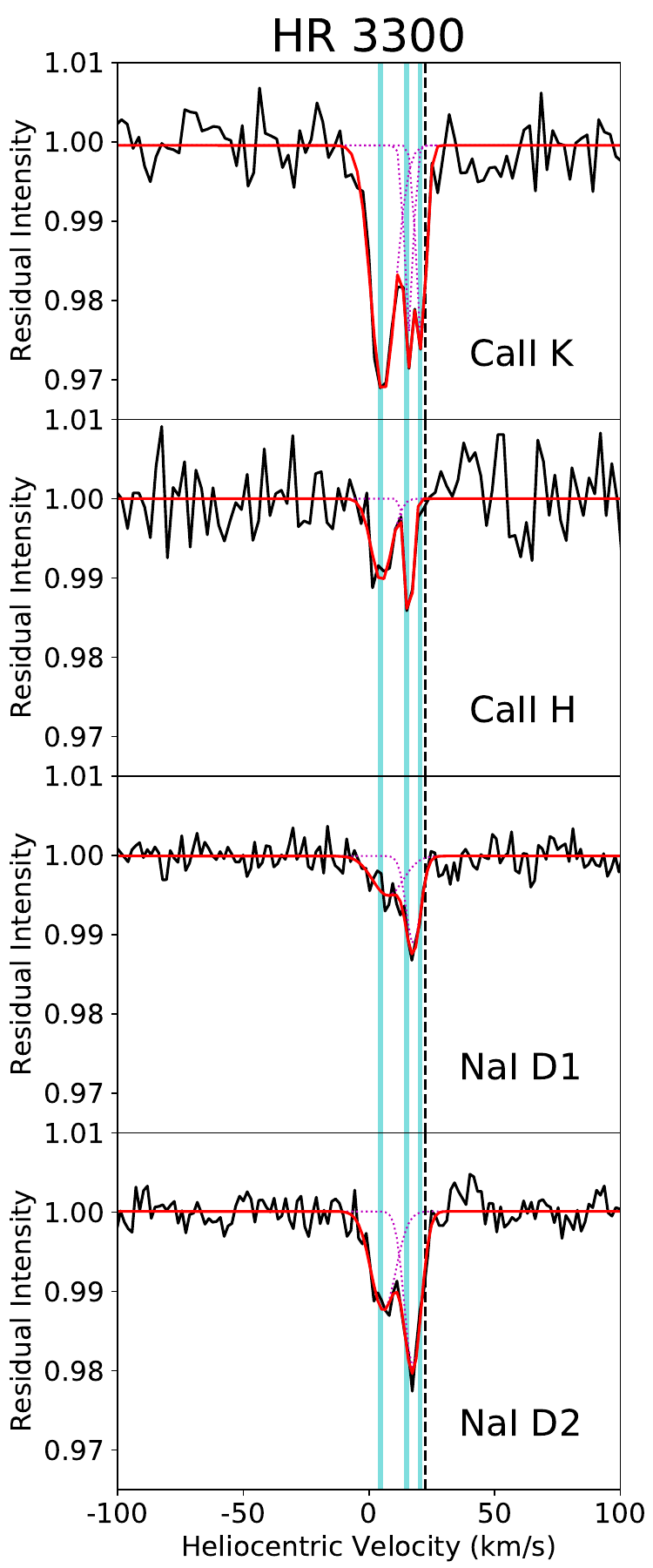}
\includegraphics[page=12, width=0.32\textwidth]{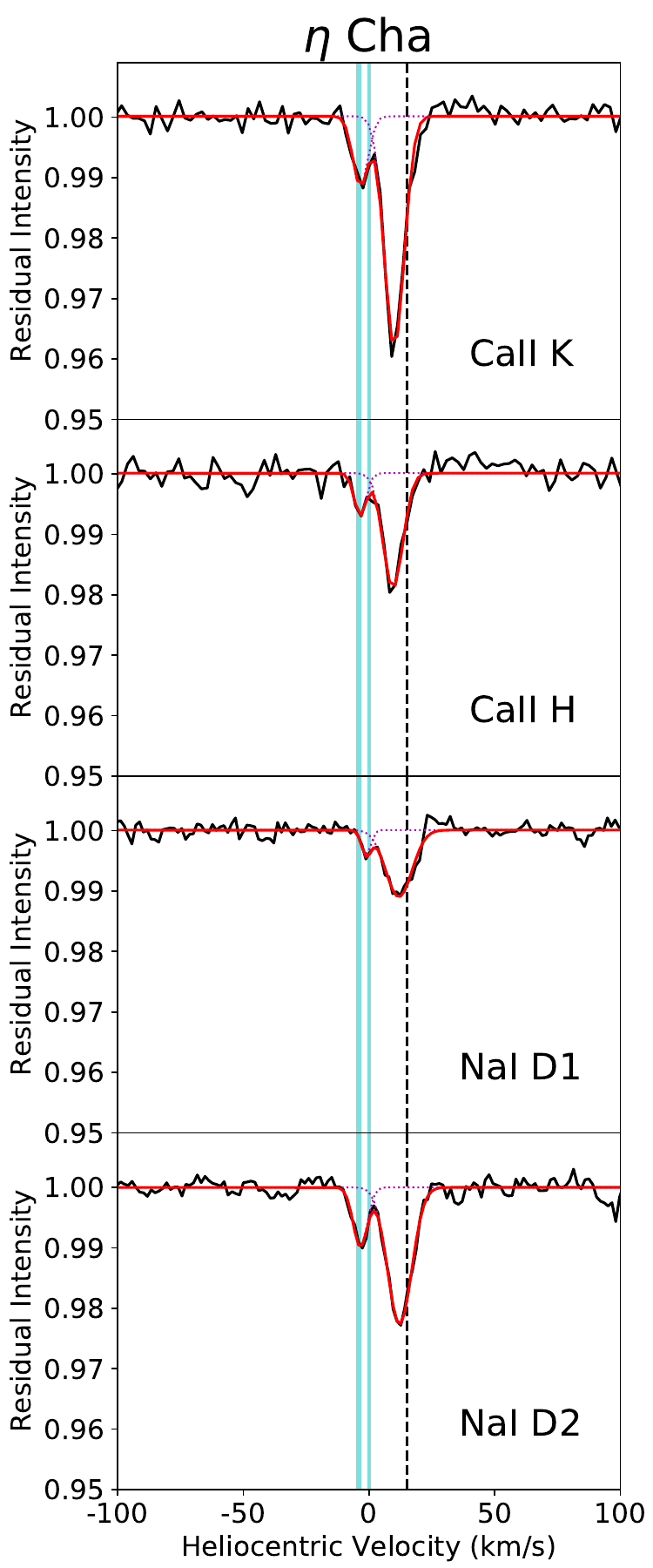}
\caption{Absorption profiles of the Ca\,{\sc ii} H \& K and Na\,{\sc i} D1 \& D2 lines for HD\,60856, HR 3300 and $\eta$ Cha. Photospheric absorptions has been subtracted and the remaining extra components have been modelled by gaussian profiles. Individual gaussian fits are shown in dotted magenta lines and the combined profile is shown in red. Dashed black line marks the estimated radial velocity of the star and cyan lines mark the velocity of the traversing clouds in the line of sight with their respective errors as their line widths. }
\label{fig:abs4}
\end{figure*}

\begin{figure*}  %-> funciona bien con pdf
\includegraphics[page=13, width=0.32\textwidth, valign=t]{median_density.pdf}
\includegraphics[page=14, width=0.32\textwidth, valign=t]{median_density.pdf}
\includegraphics[page=1, width=0.32\textwidth, valign=t]{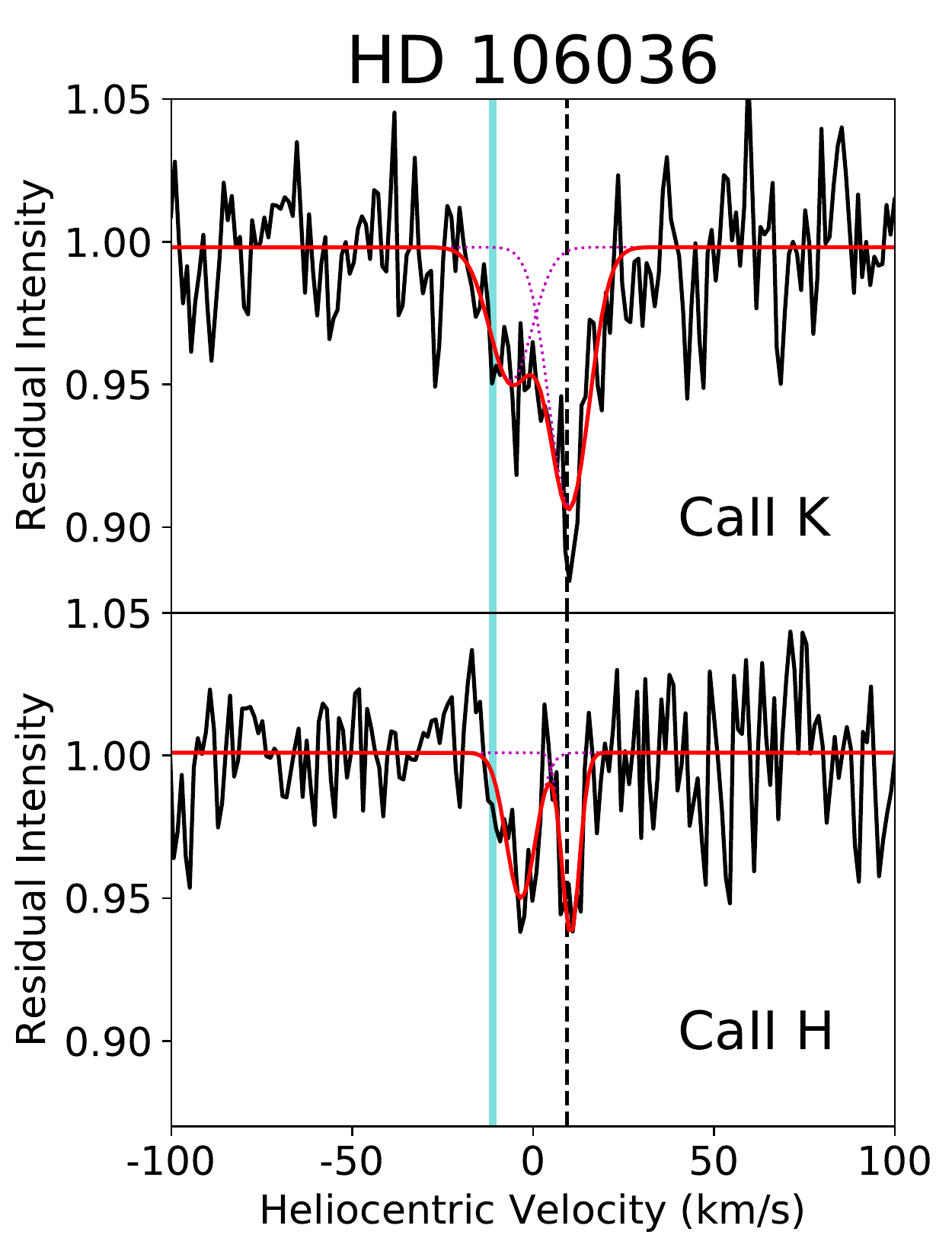}
\caption{Absorption profiles of the Ca\,{\sc ii} H \& K and Na\,{\sc i} D1 \& D2 lines for HD\,92536, 3 Crv and HD\,106036. Photospheric absorptions has been subtracted and the remaining extra components have been modelled by gaussian profiles. Individual gaussian fits are shown in dotted magenta lines and the combined profile is shown in red. Dashed black line marks the estimated radial velocity of the star and cyan lines mark the velocity of the traversing clouds in the line of sight with their respective errors as their line widths.}
\label{fig:abs5}
\end{figure*}

\begin{figure*}  %-> funciona bien con pdf
\includegraphics[page=15, width=0.32\textwidth, valign=t]{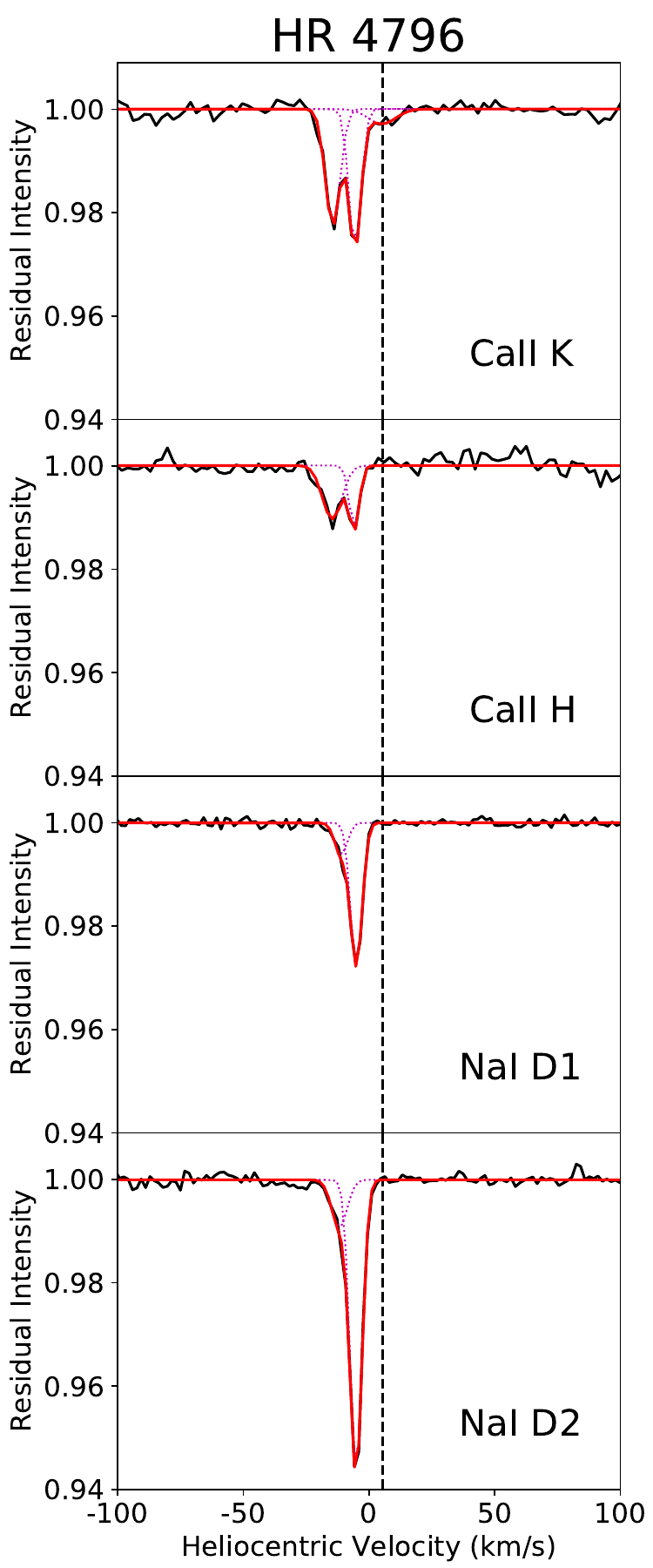}
\includegraphics[page=16, width=0.32\textwidth, valign=t]{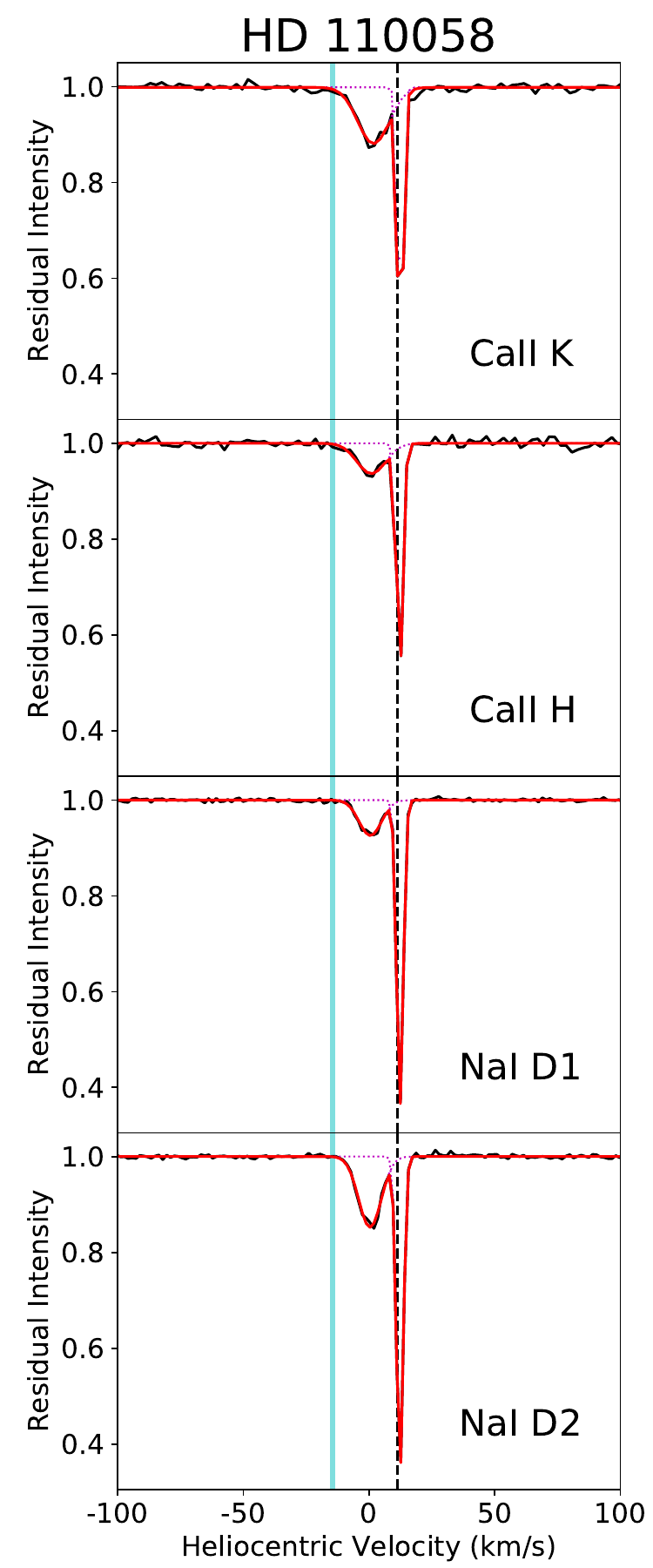}
\includegraphics[page=2, width=0.32\textwidth, valign=t]{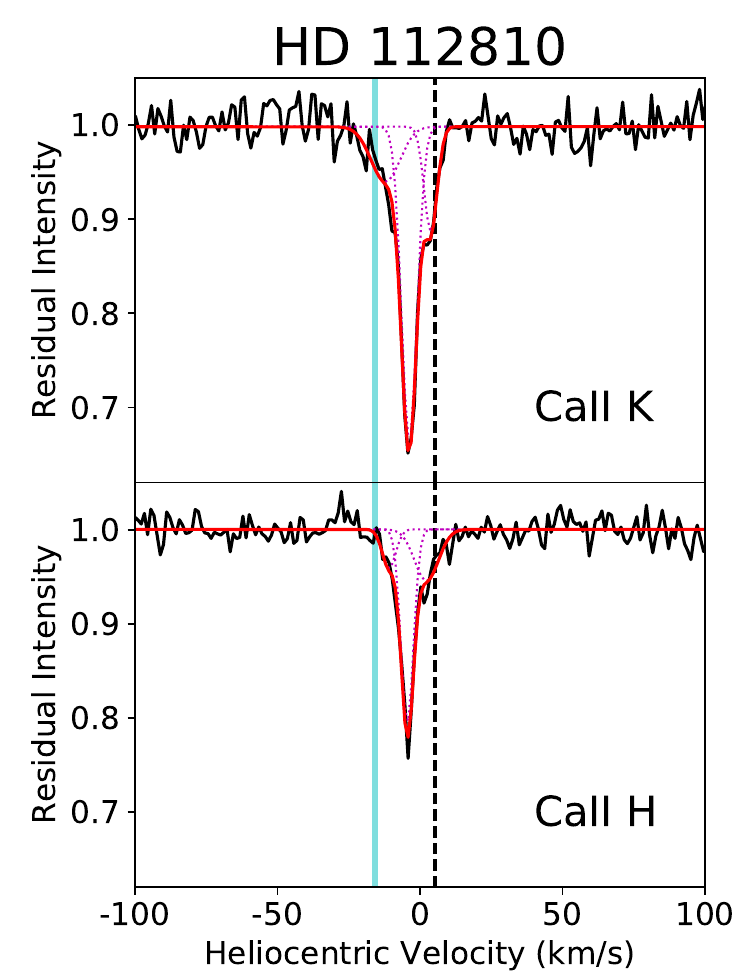}
\caption{Absorption profiles of the Ca\,{\sc ii} H \& K and Na\,{\sc i} D1 \& D2 lines for HR\,4796, HD\,110058 and HD\,112810. Photospheric absorptions has been subtracted and the remaining extra components have been modelled by gaussian profiles. Individual gaussian fits are shown in dotted magenta lines and the combined profile is shown in red. Dashed black line marks the estimated radial velocity of the star and cyan lines mark the velocity of the traversing clouds in the line of sight with their respective errors as their line widths. }
\label{fig:abs6}
\end{figure*}

\begin{figure*}  %-> funciona bien con pdf
\includegraphics[page=17, width=0.32\textwidth]{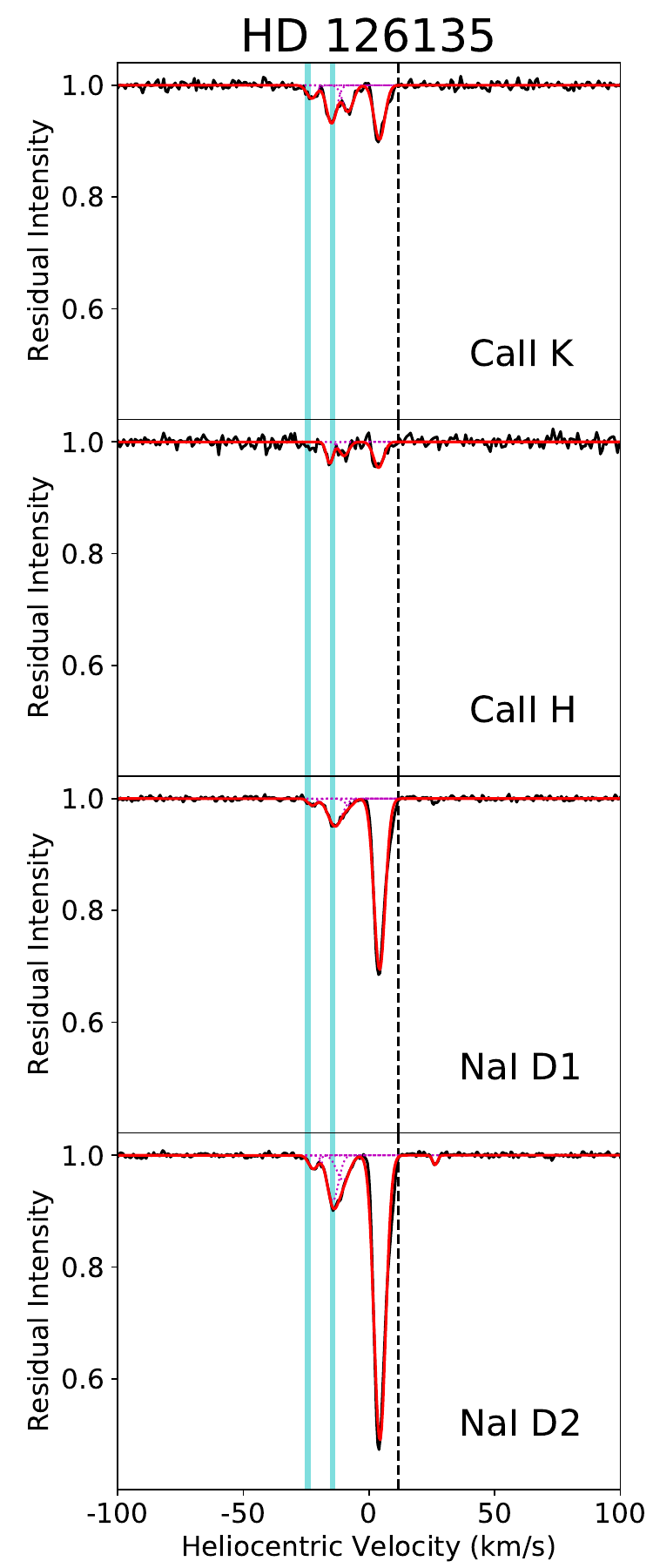}
\includegraphics[page=18, width=0.32\textwidth]{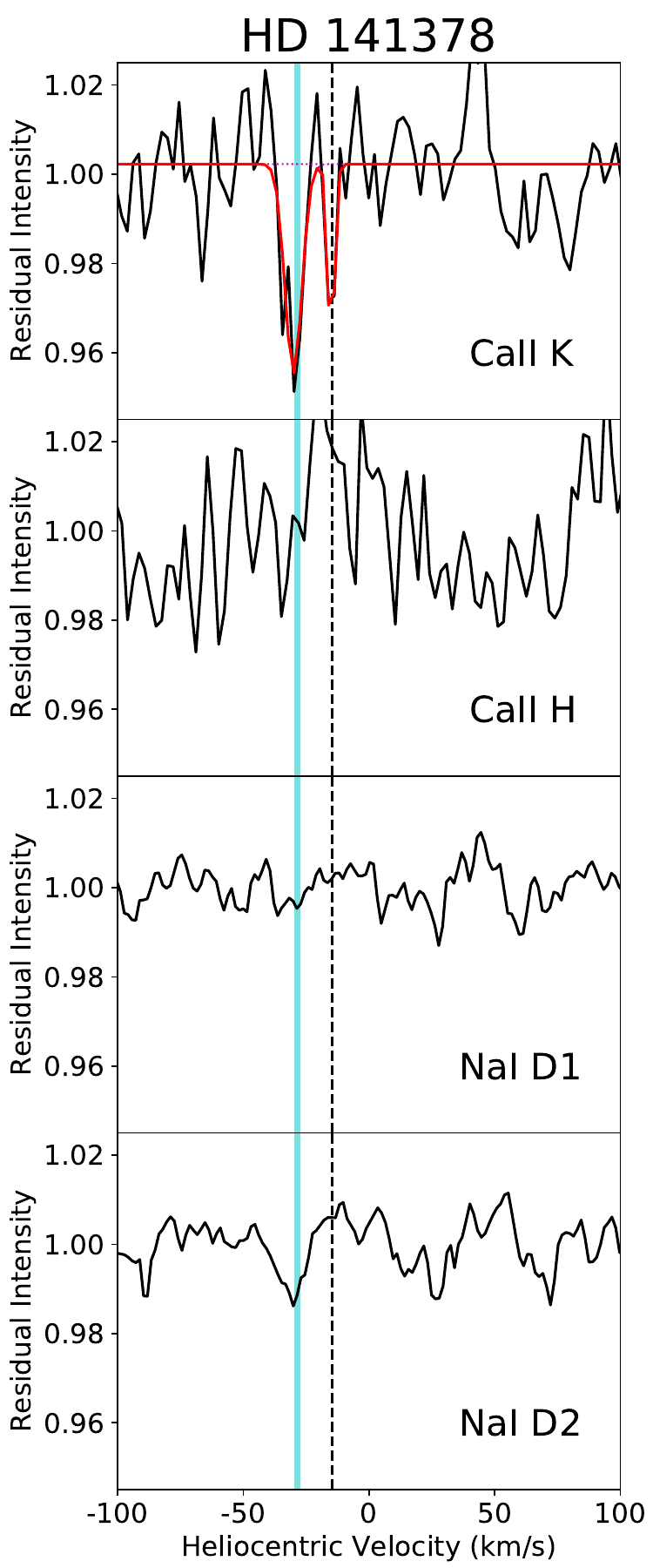}
\includegraphics[page=19, width=0.32\textwidth]{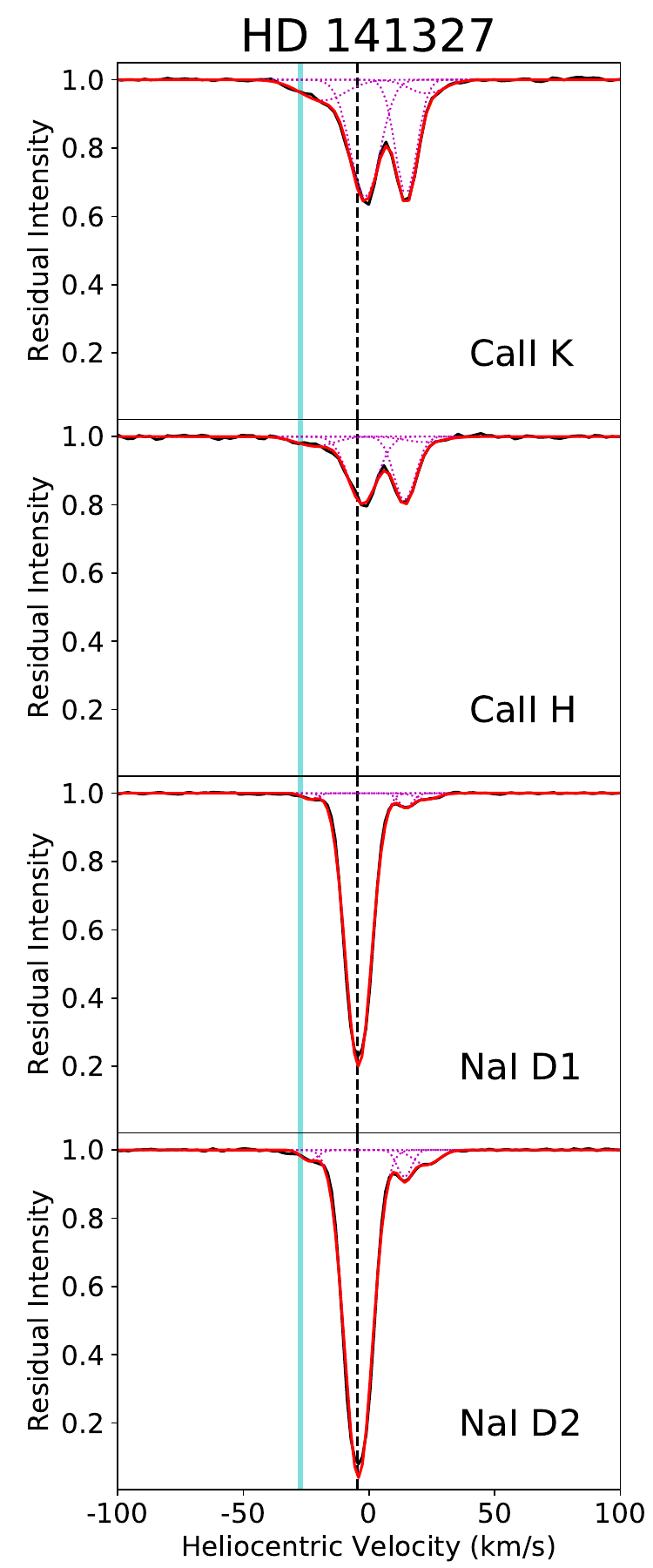}
\caption{Absorption profiles of the Ca\,{\sc ii} H \& K and Na\,{\sc i} D1 \& D2 lines for HD\,126135, HD\,141378 and HD\,141327. Photospheric absorptions has been subtracted and the remaining extra components have been modelled by gaussian profiles. Individual gaussian fits are shown in dotted magenta lines and the combined profile is shown in red. Dashed black line marks the estimated radial velocity of the star and cyan lines mark the velocity of the traversing clouds in the line of sight with their respective errors as their line widths. }
\label{fig:abs7}
\end{figure*}

\begin{figure*}  %-> funciona bien con pdf
\includegraphics[page=20, width=0.32\textwidth]{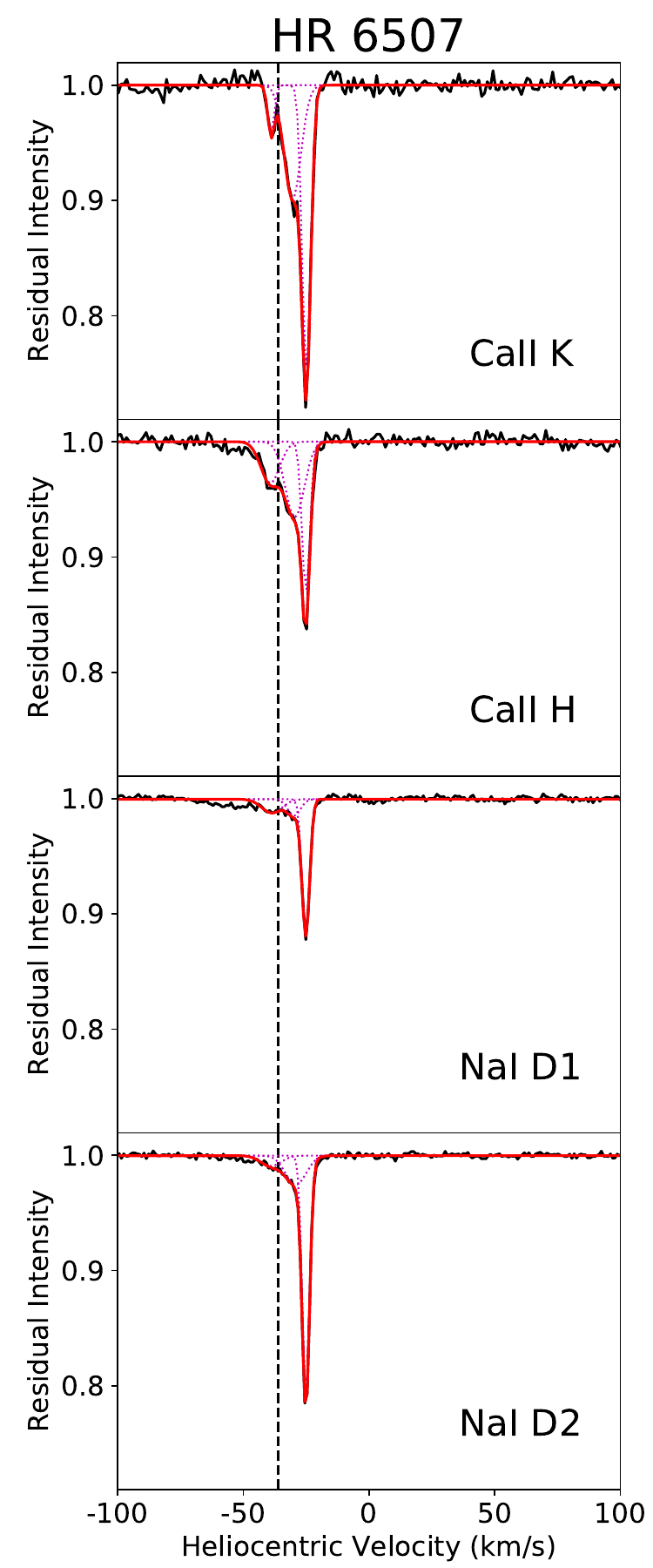}
\includegraphics[page=21, width=0.32\textwidth]{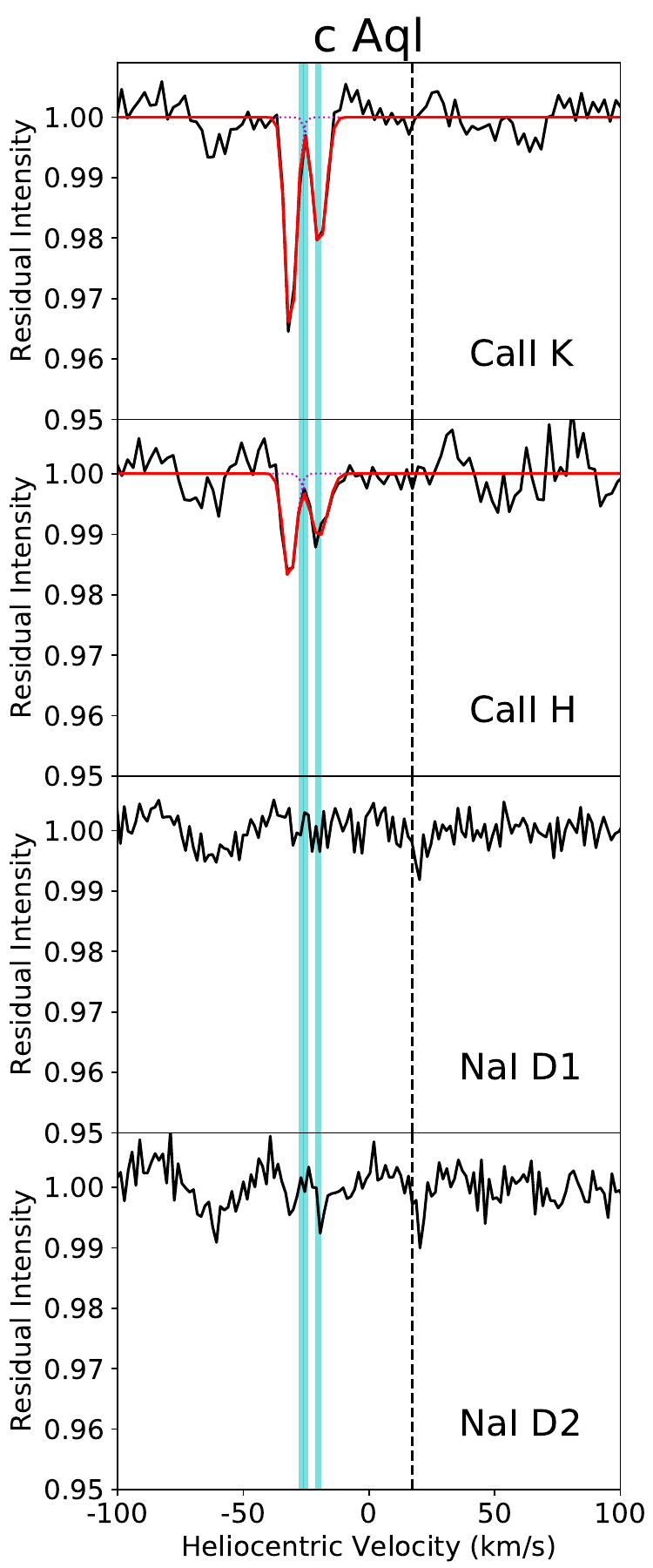}
\caption{Absorption profiles of the Ca\,{\sc ii} H \& K and Na\,{\sc i} D1 \& D2 lines for HR\,6507 and c Aql. Photospheric absorptions has been subtracted and the remaining extra components have been modelled by gaussian profiles. Individual gaussian fits are shown in dotted magenta lines and the combined profile is shown in red. Dashed black line marks the estimated radial velocity of the star and cyan lines mark the velocity of the traversing clouds in the line of sight with their respective errors as their line widths.
\vspace{250px} }
\label{fig:abs8}
\end{figure*}

\section{Neighbouring stars}
In this appendix we show the analysis performed with neighbouring stars looking for features of similar characteristics to those detected in each ``science" target (paying particular attention to the velocity). The detection of the same feature in several stars does automatically support an ISM origin of the feature.
%\clearpage

\begin{figure*}  %-> funciona bien con pdf
\includegraphics[width=0.49\textwidth]{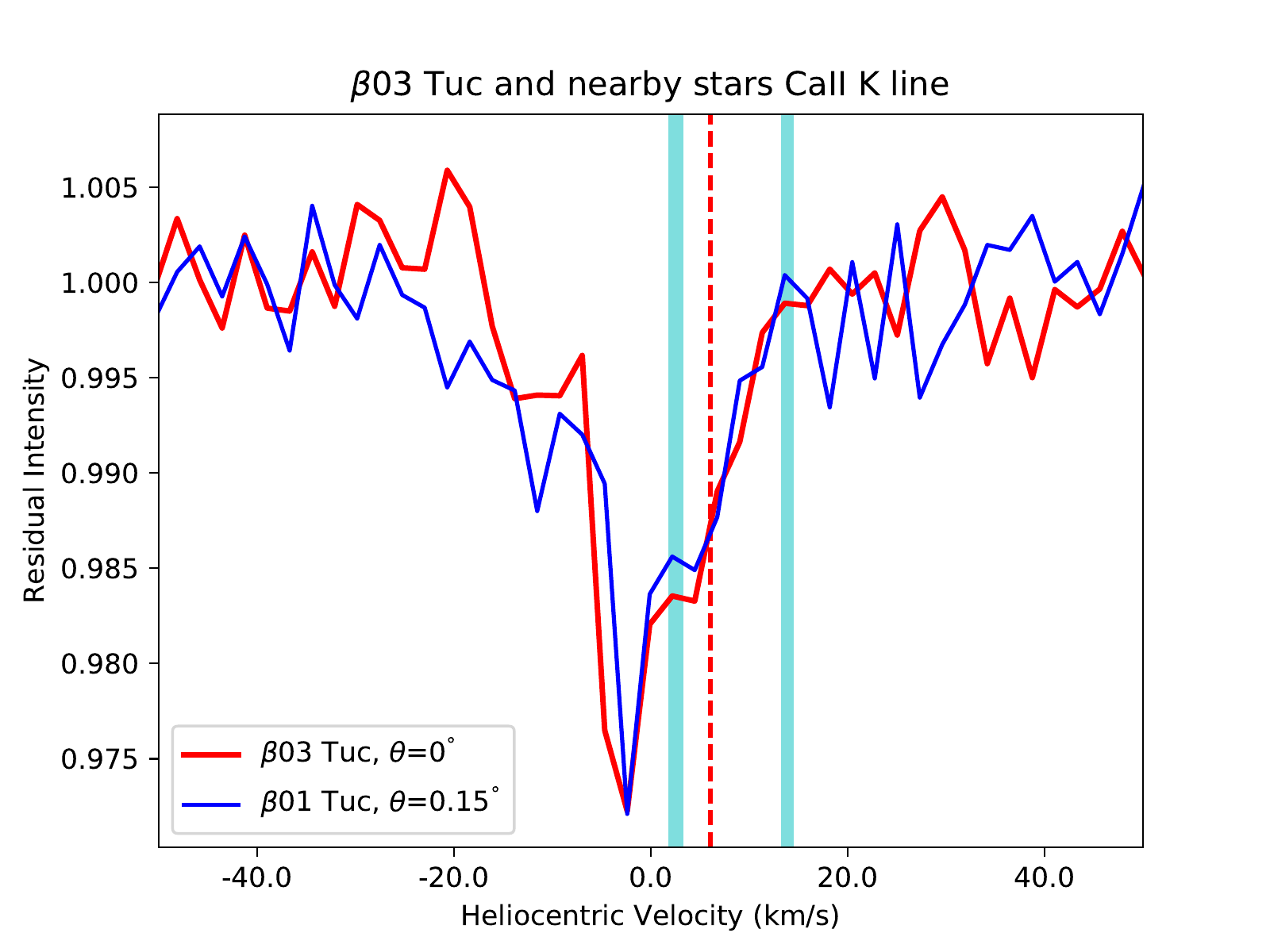}
\includegraphics[width=0.49\textwidth]{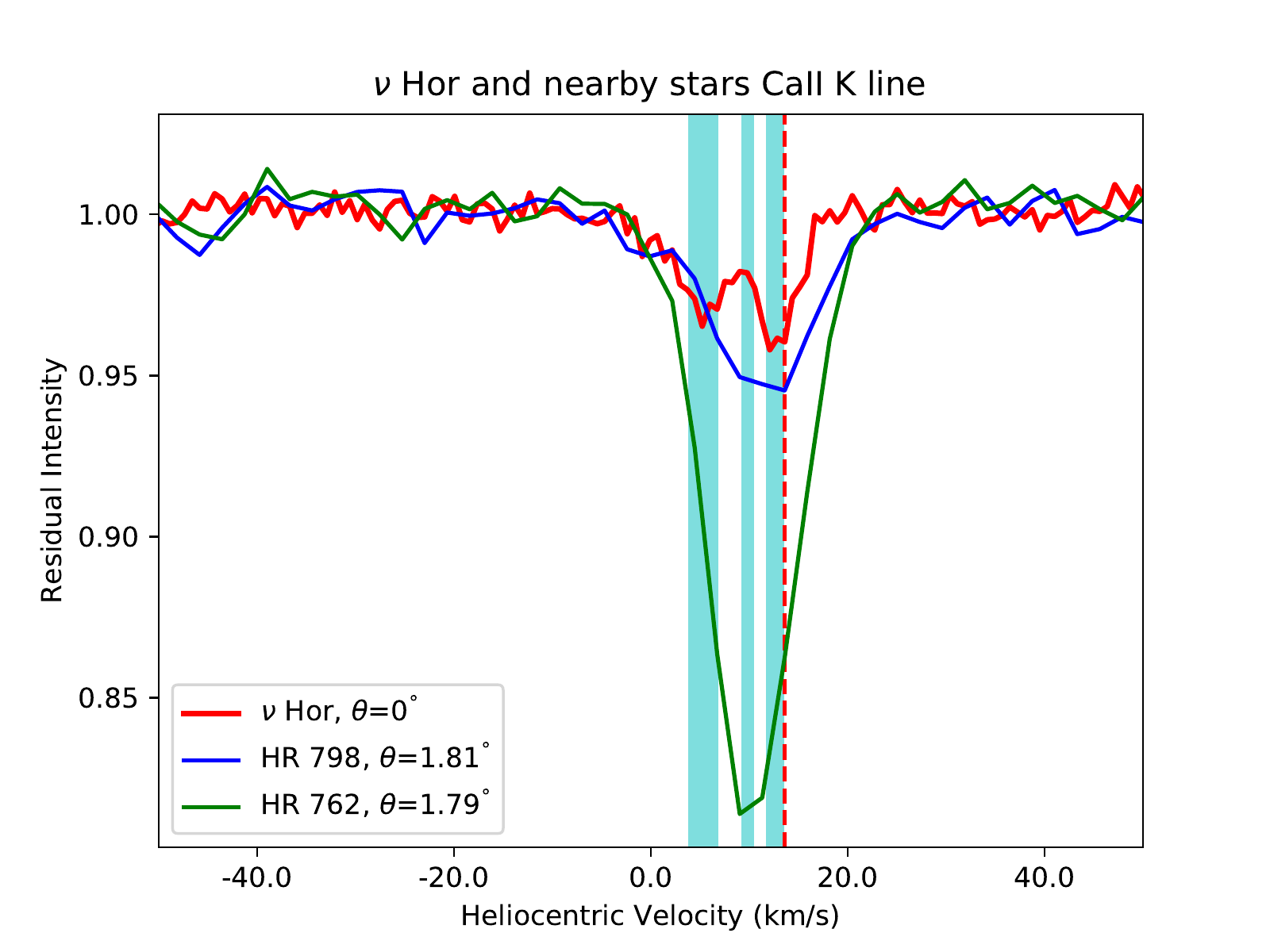}
\end{figure*}

\begin{figure*}  %-> funciona bien con pdf
\includegraphics[width=0.49\textwidth]{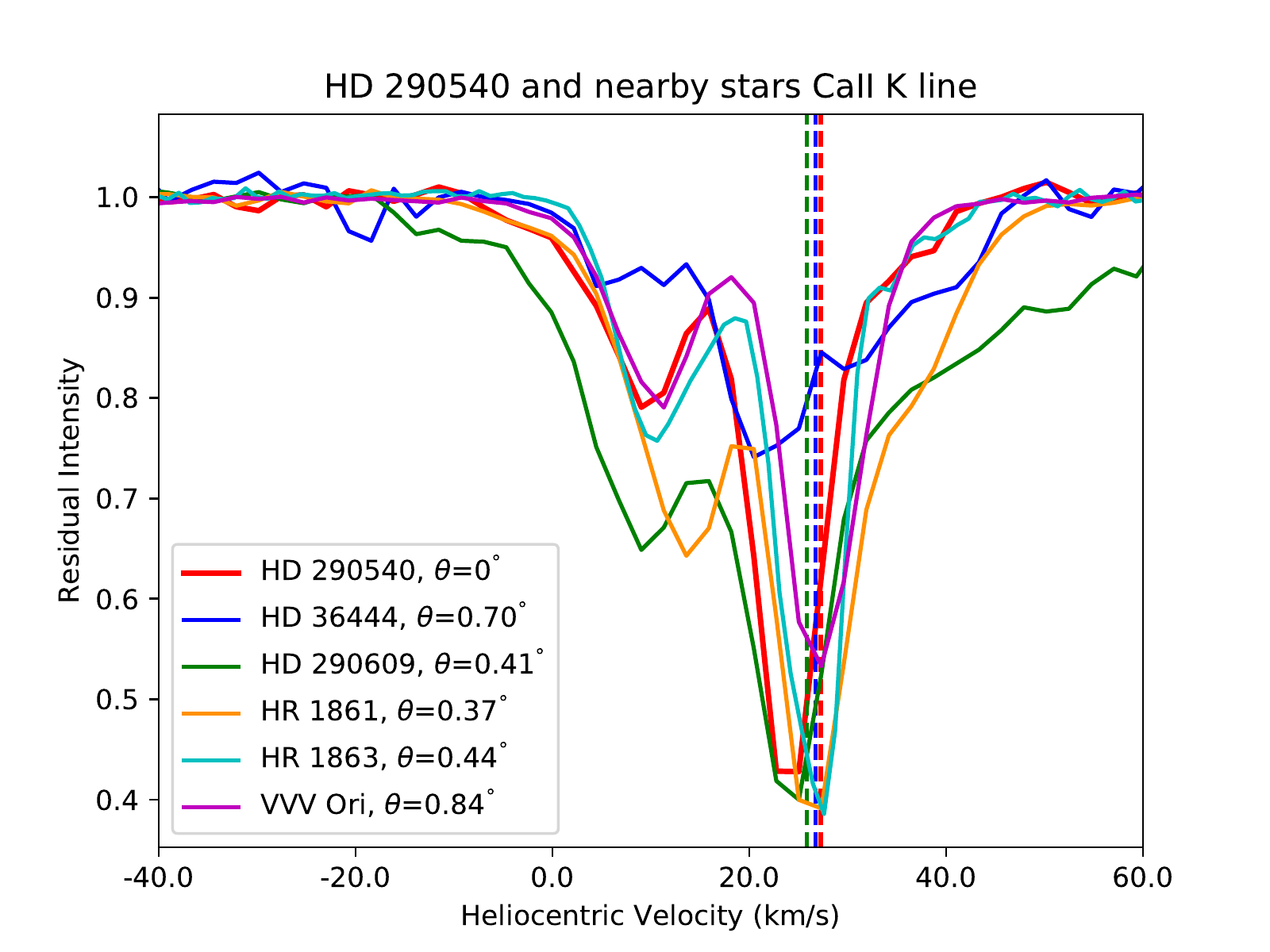}
\includegraphics[width=0.49\textwidth]{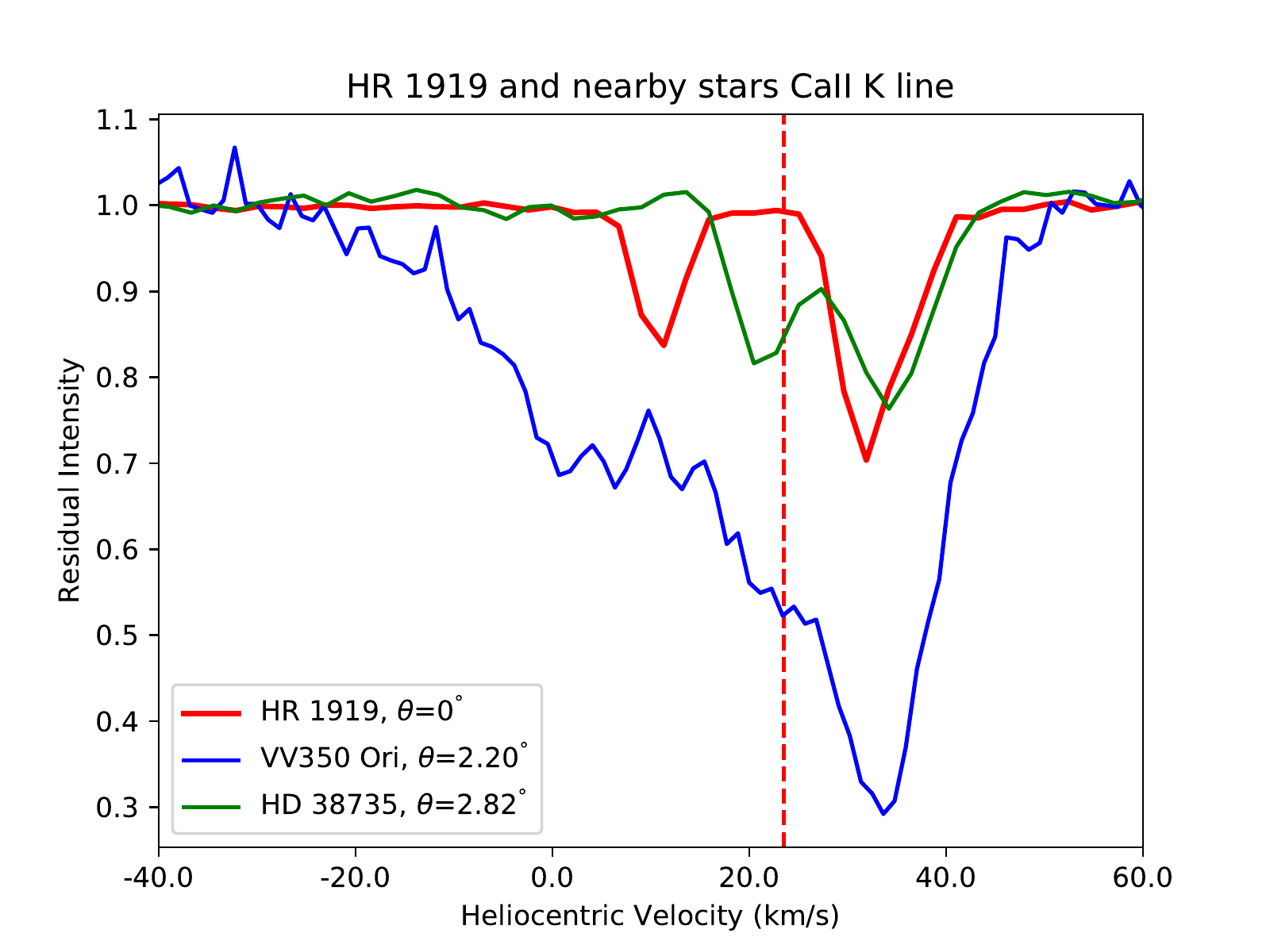}
\end{figure*}

\begin{figure*}  %-> funciona bien con pdf
\includegraphics[width=0.49\textwidth]{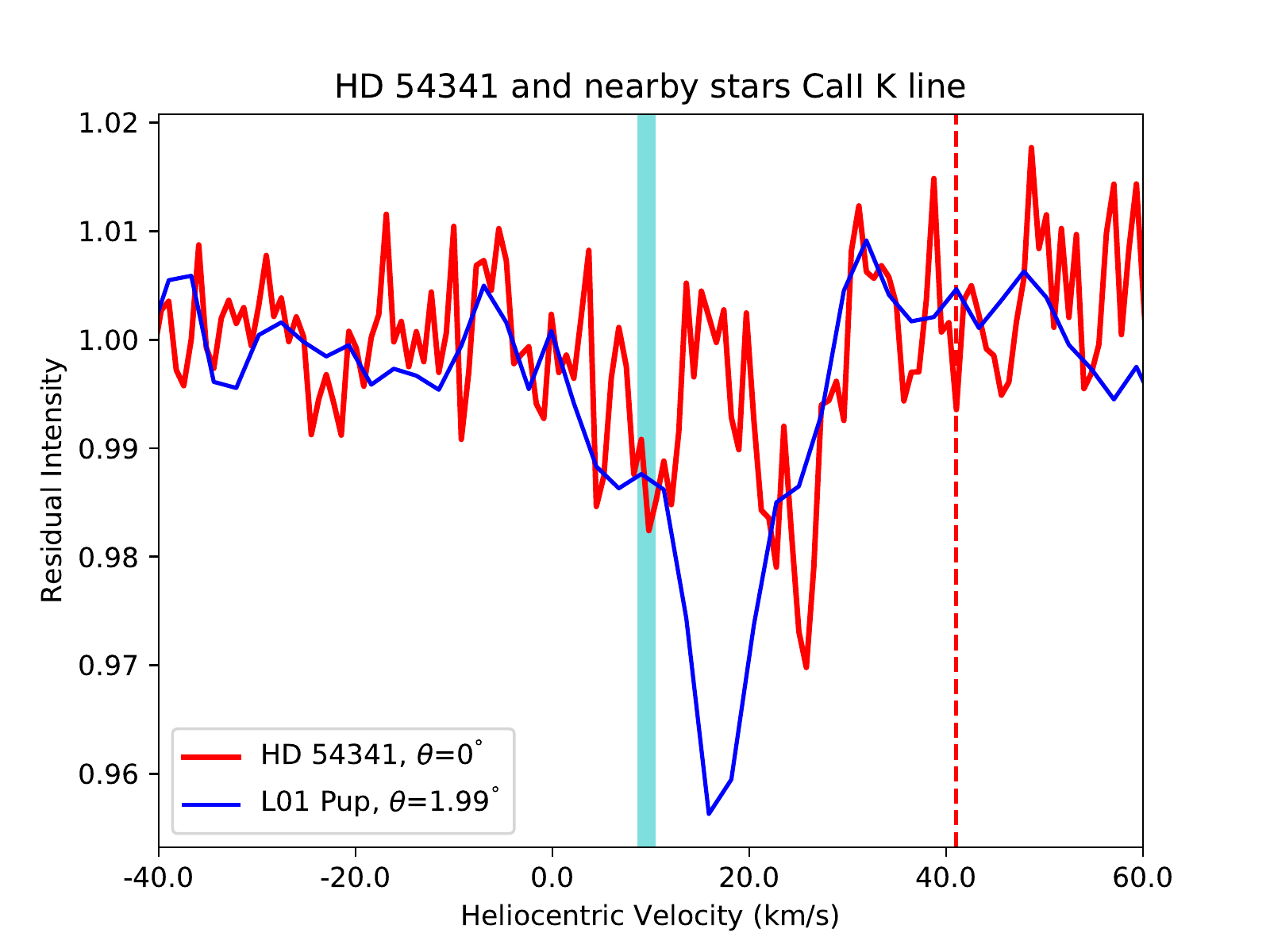}
\includegraphics[width=0.49\textwidth]{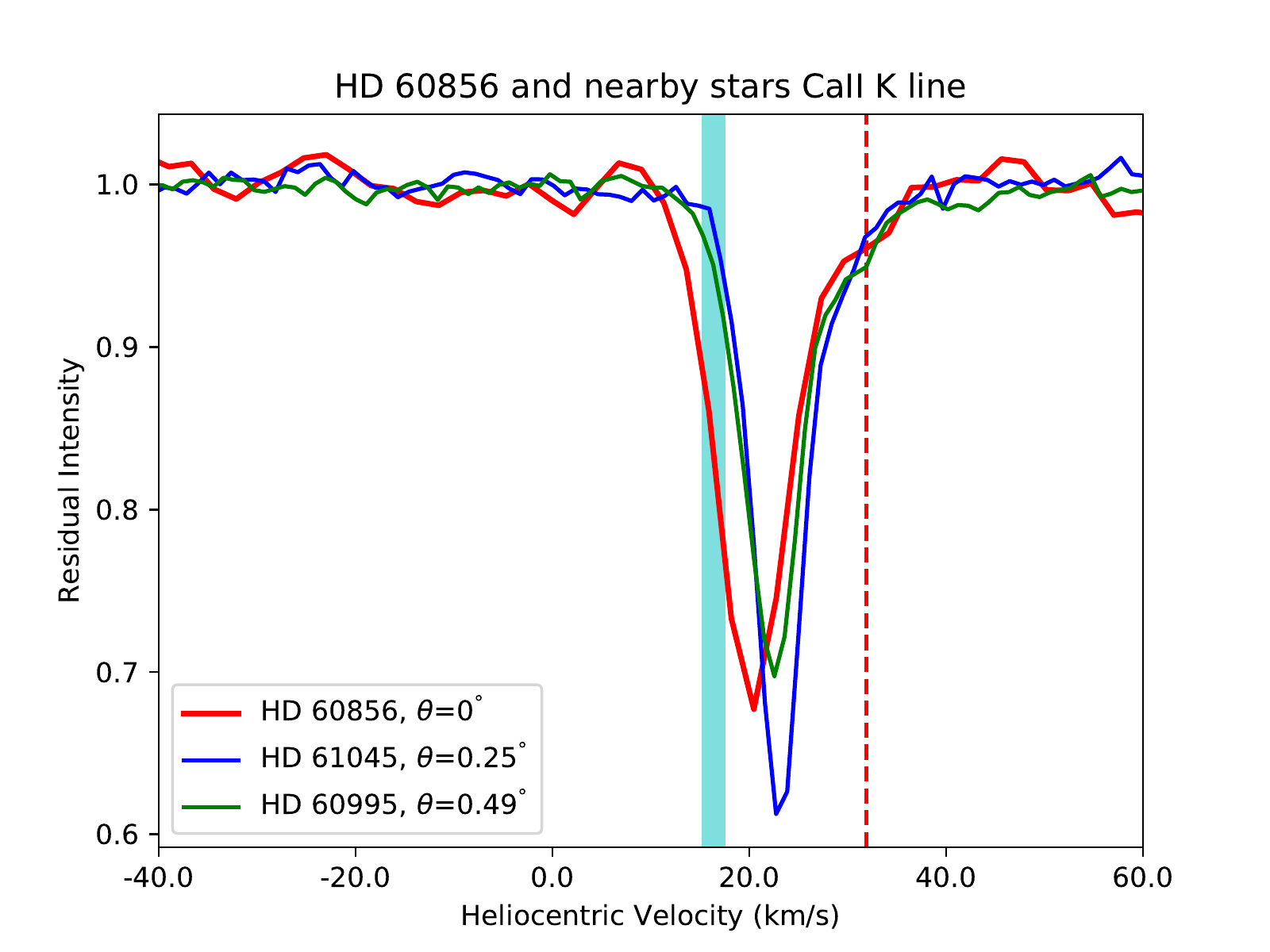}
\caption{Nearby stars around $\beta$03\,Tuc, $\nu$ Hor, HD\,290540 (along with HD\,36444 and HD\,290609), HR1919, HD\,54341 and HD\,60856. Dashed line marks the estimated radial velocity of the star and cyan lines mark the velocity of the traversing clouds in the line of sight with their respective errors as their line widths.}
\label{fig:near1}
\end{figure*}

\begin{figure*}  %-> funciona bien con pdf
\includegraphics[width=0.49\textwidth]{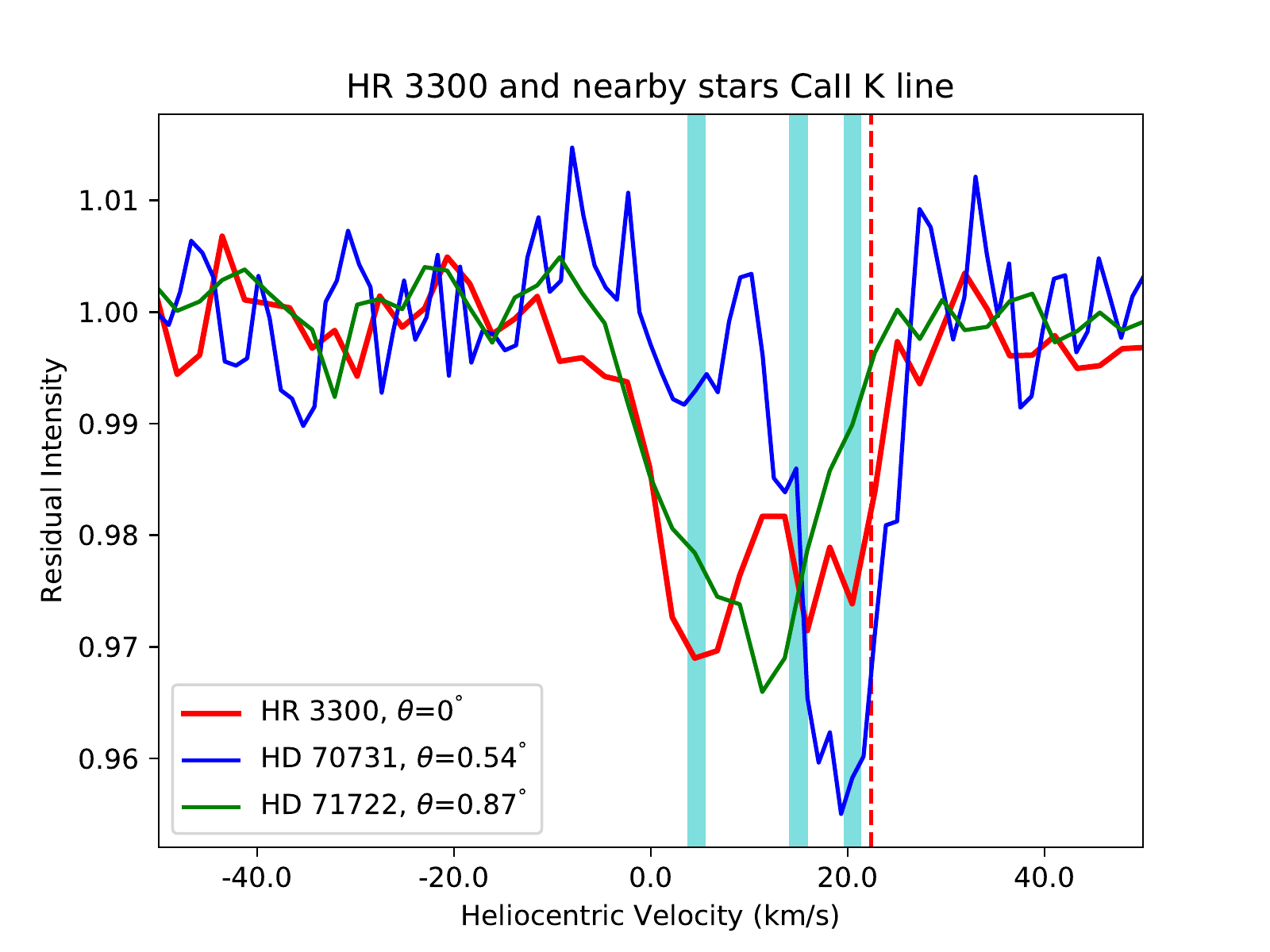}
\includegraphics[width=0.49\textwidth]{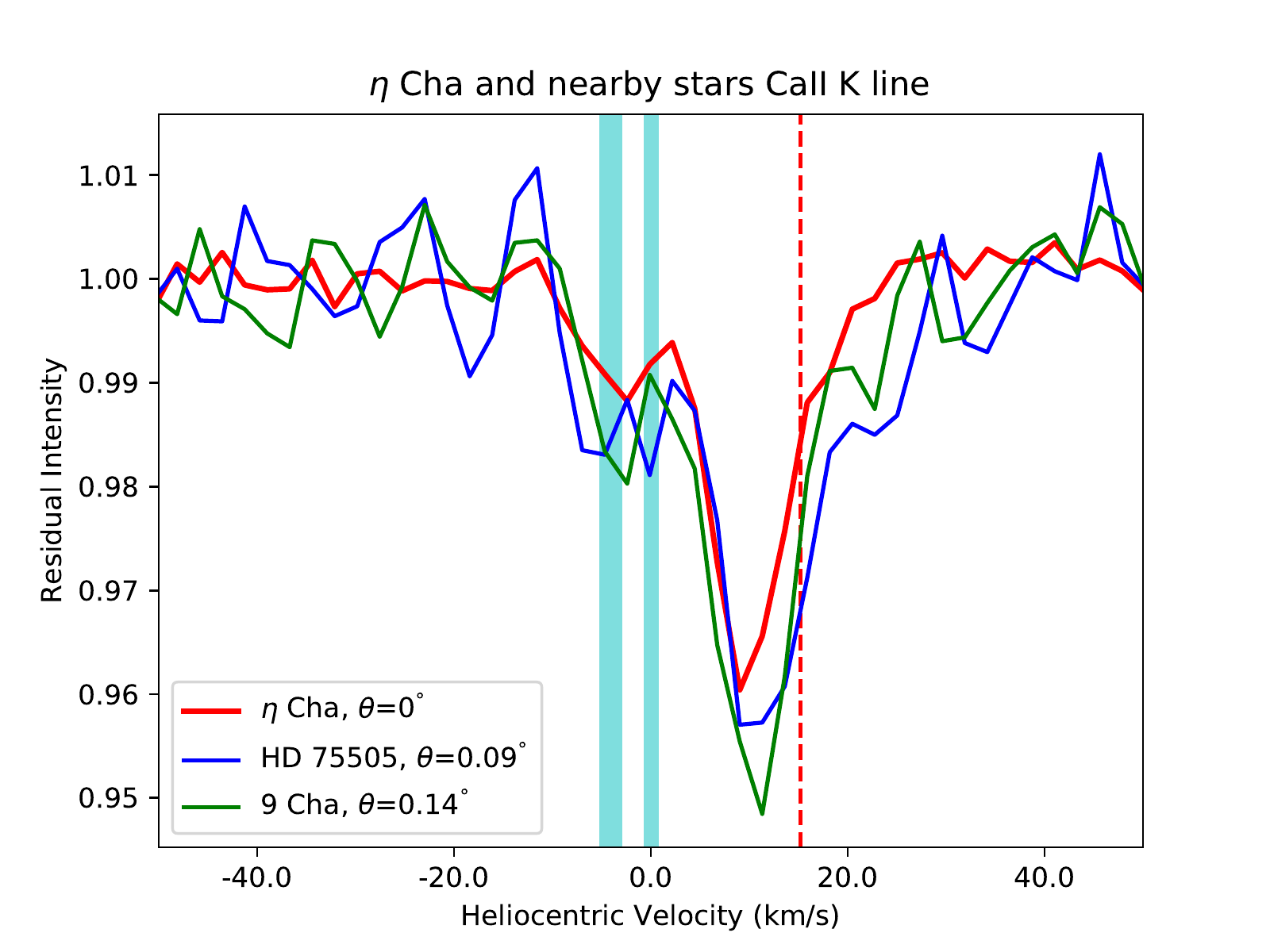}
\end{figure*}

\begin{figure*}  %-> funciona bien con pdf
\includegraphics[width=0.49\textwidth]{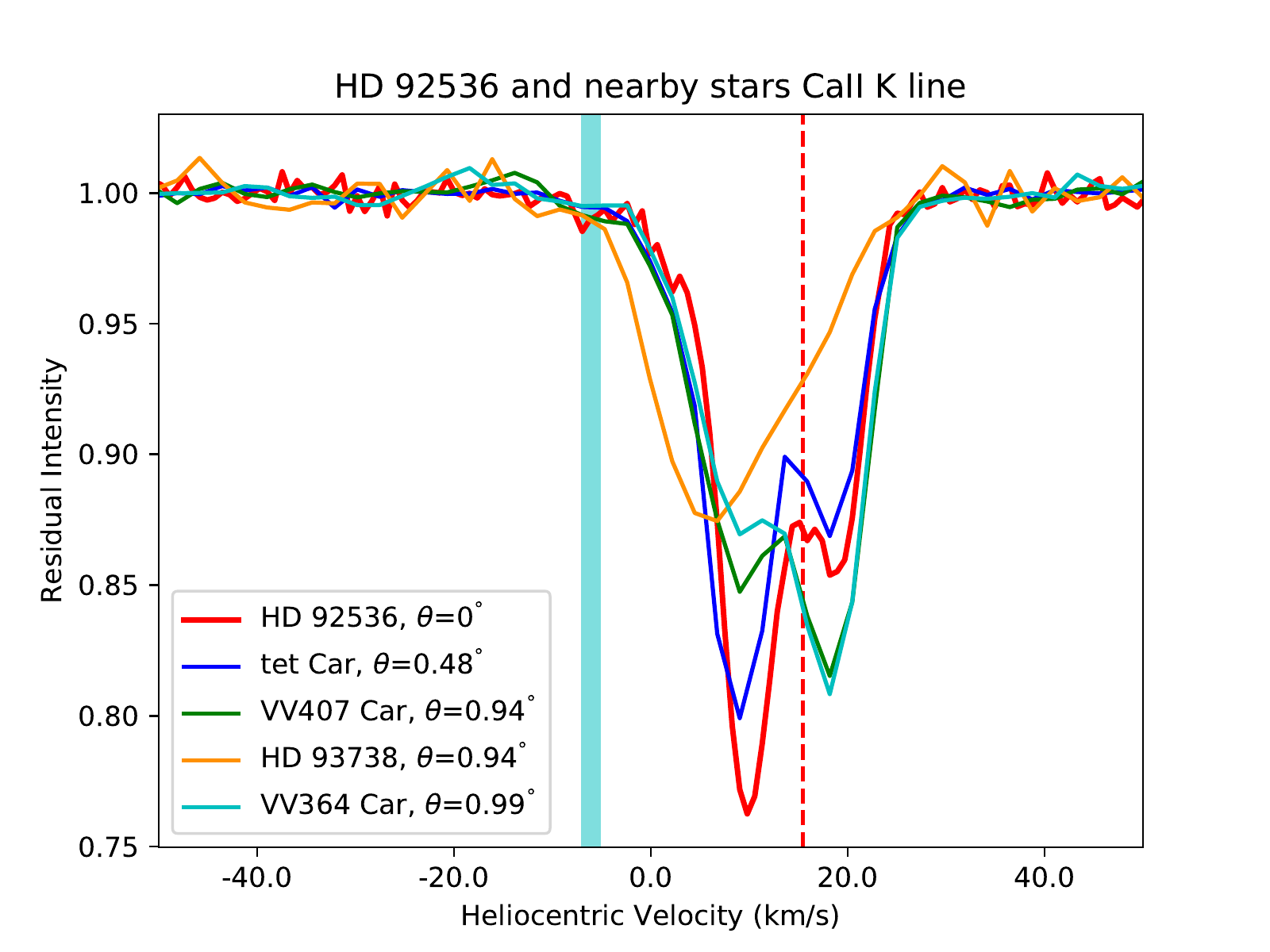}
\includegraphics[width=0.49\textwidth]{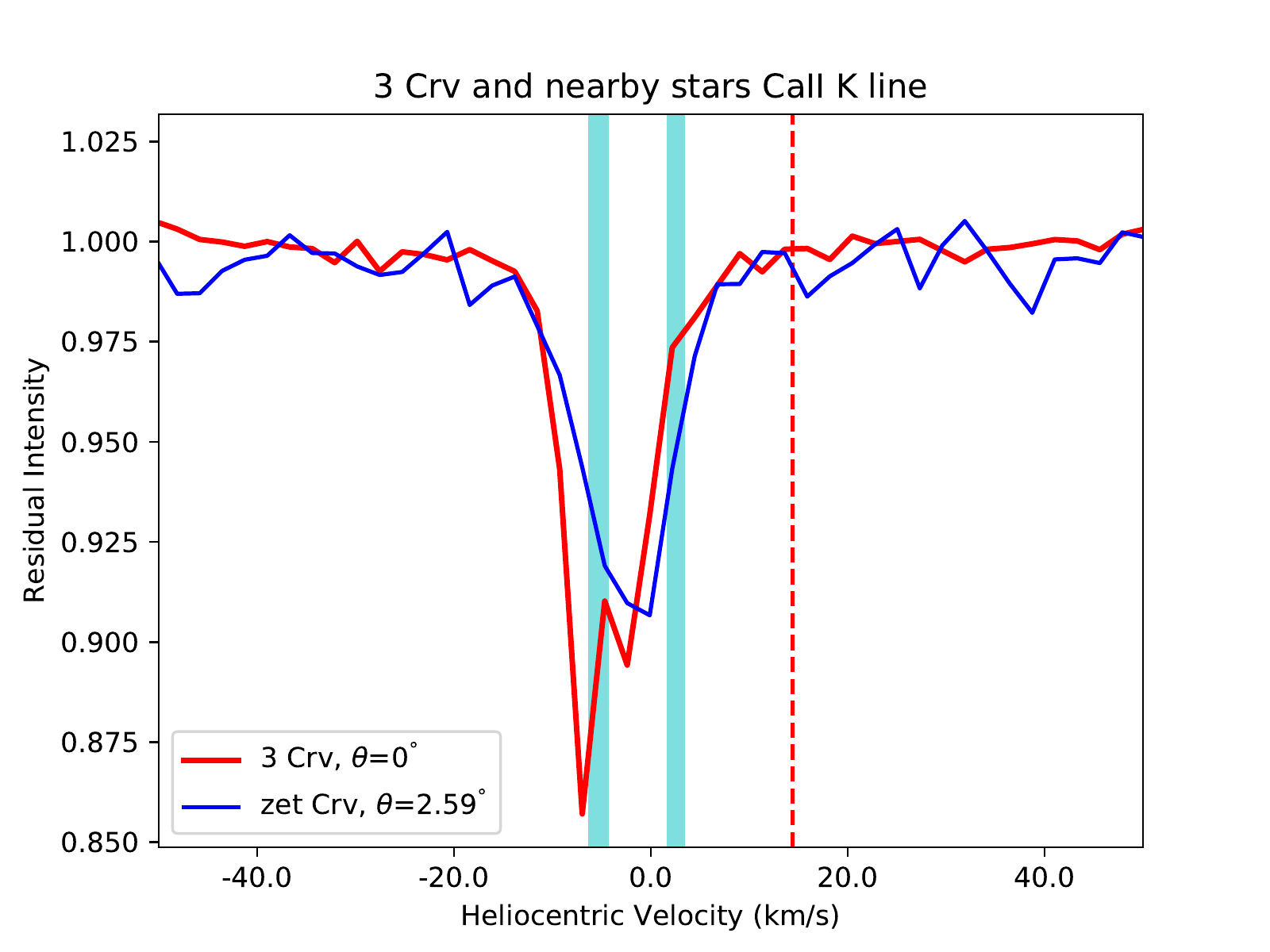}
\end{figure*}

\begin{figure*}  %-> funciona bien con pdf
\includegraphics[width=0.49\textwidth]{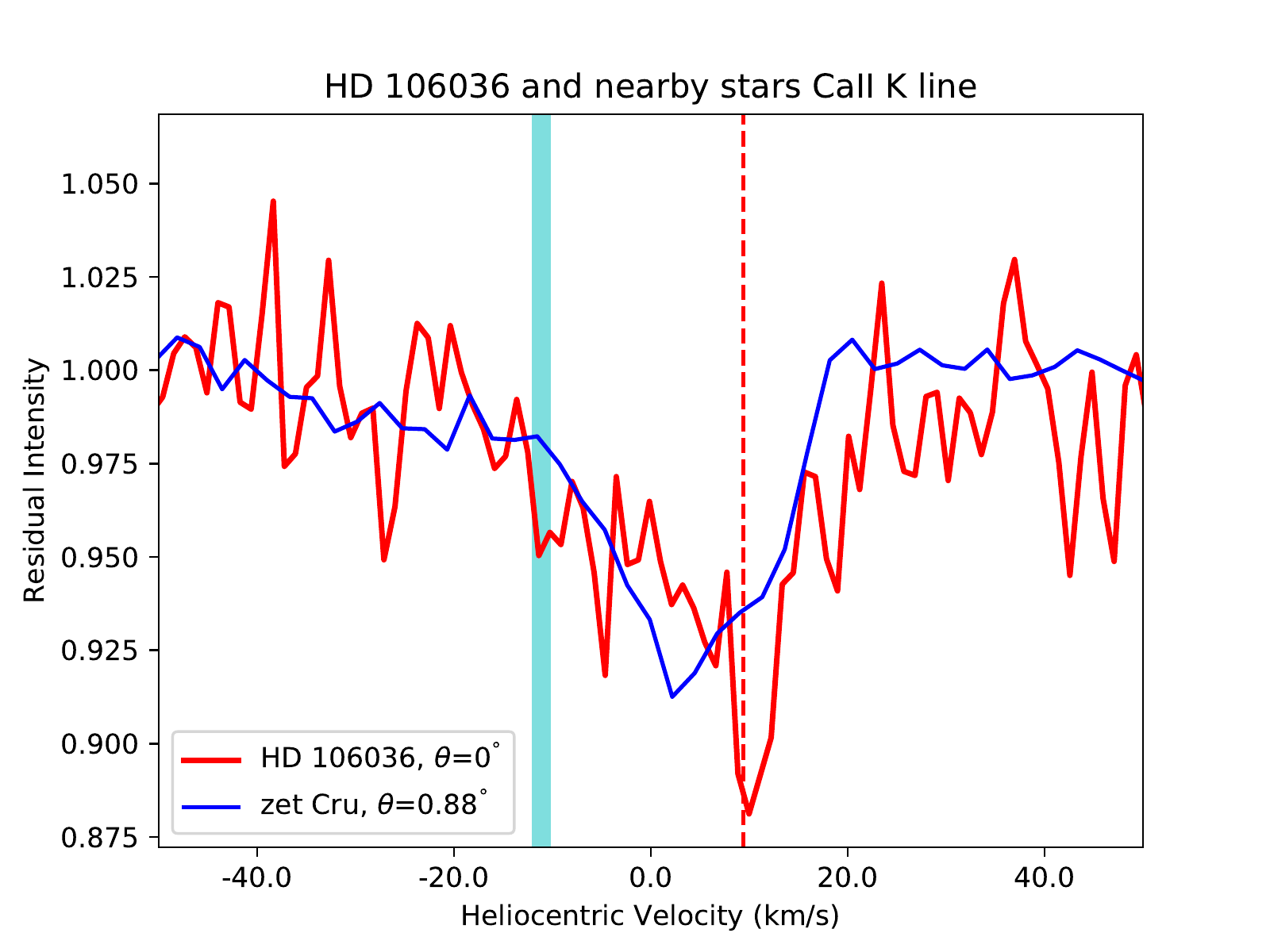}
\includegraphics[width=0.49\textwidth]{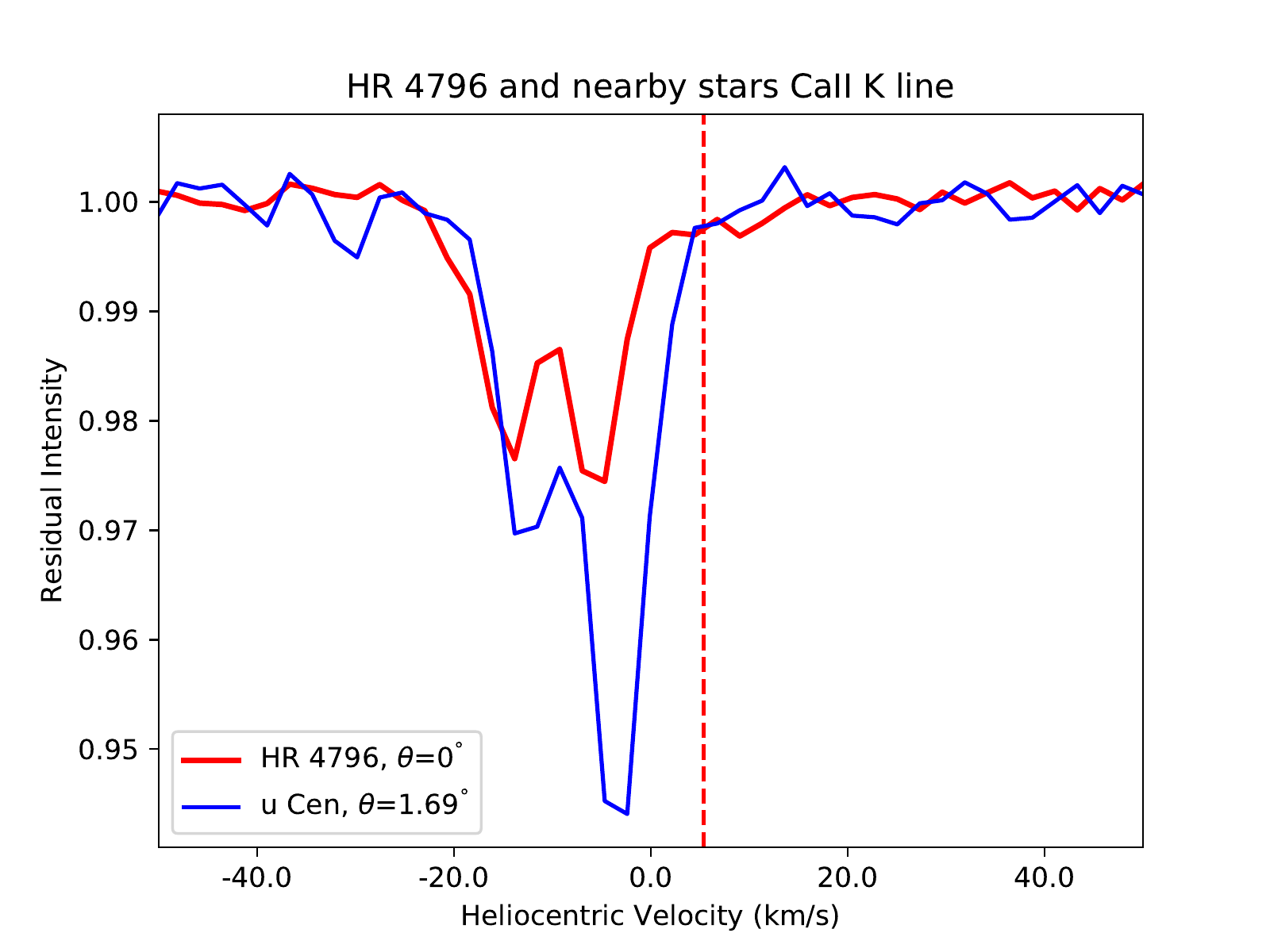}
\caption{Nearby stars around  HR 3300, $\eta$ Cha, HD\,92536, 3 Crv, HD\,106036 and HR\,4796. Dashed line marks the estimated radial velocity of the star and cyan lines mark the velocity of the traversing clouds in the line of sight with their respective errors as their line widths.}
\label{fig:near2}
\end{figure*}

\begin{figure*}  %-> funciona bien con pdf
\includegraphics[width=0.49\textwidth]{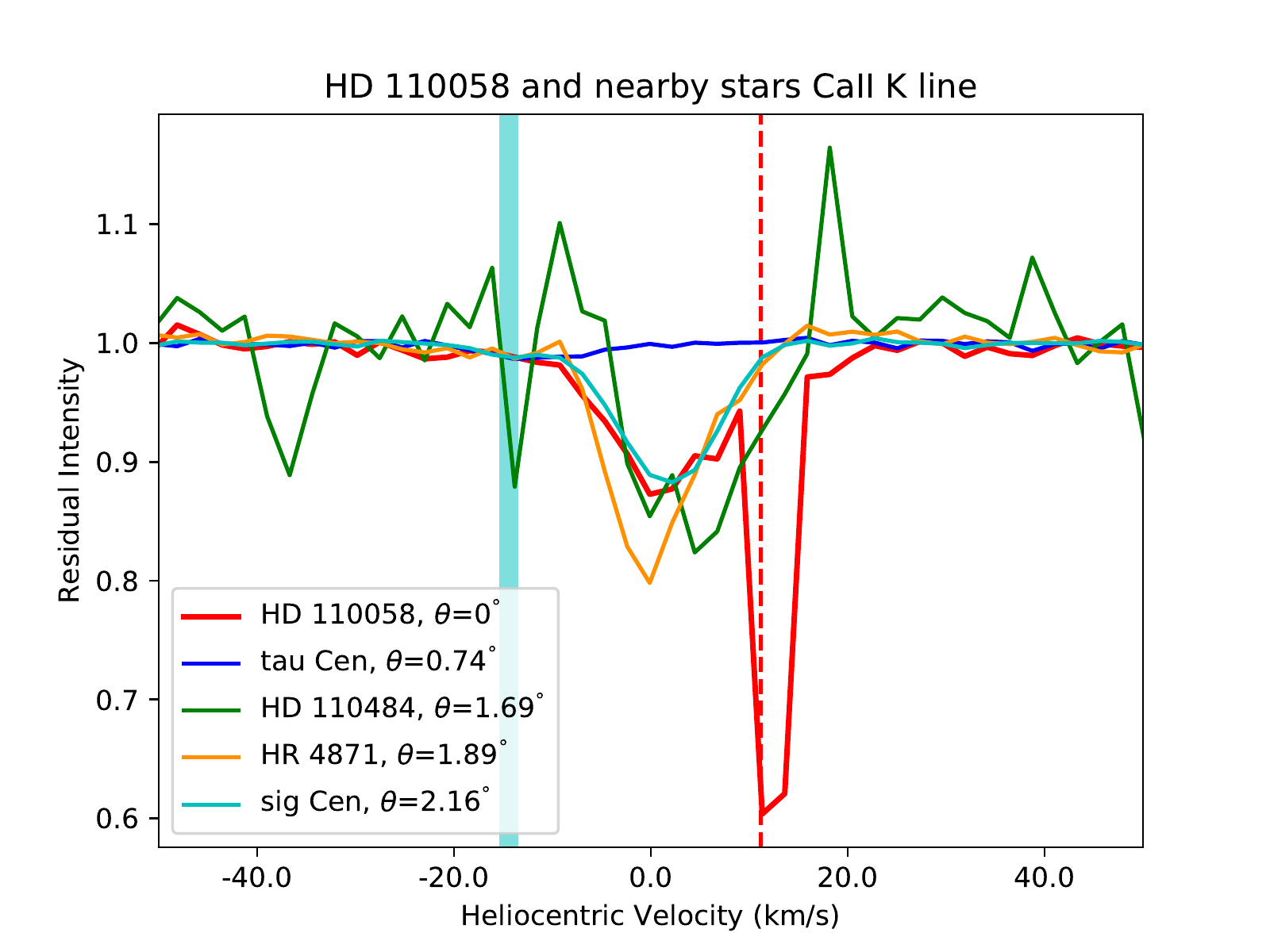}
\includegraphics[width=0.49\textwidth]{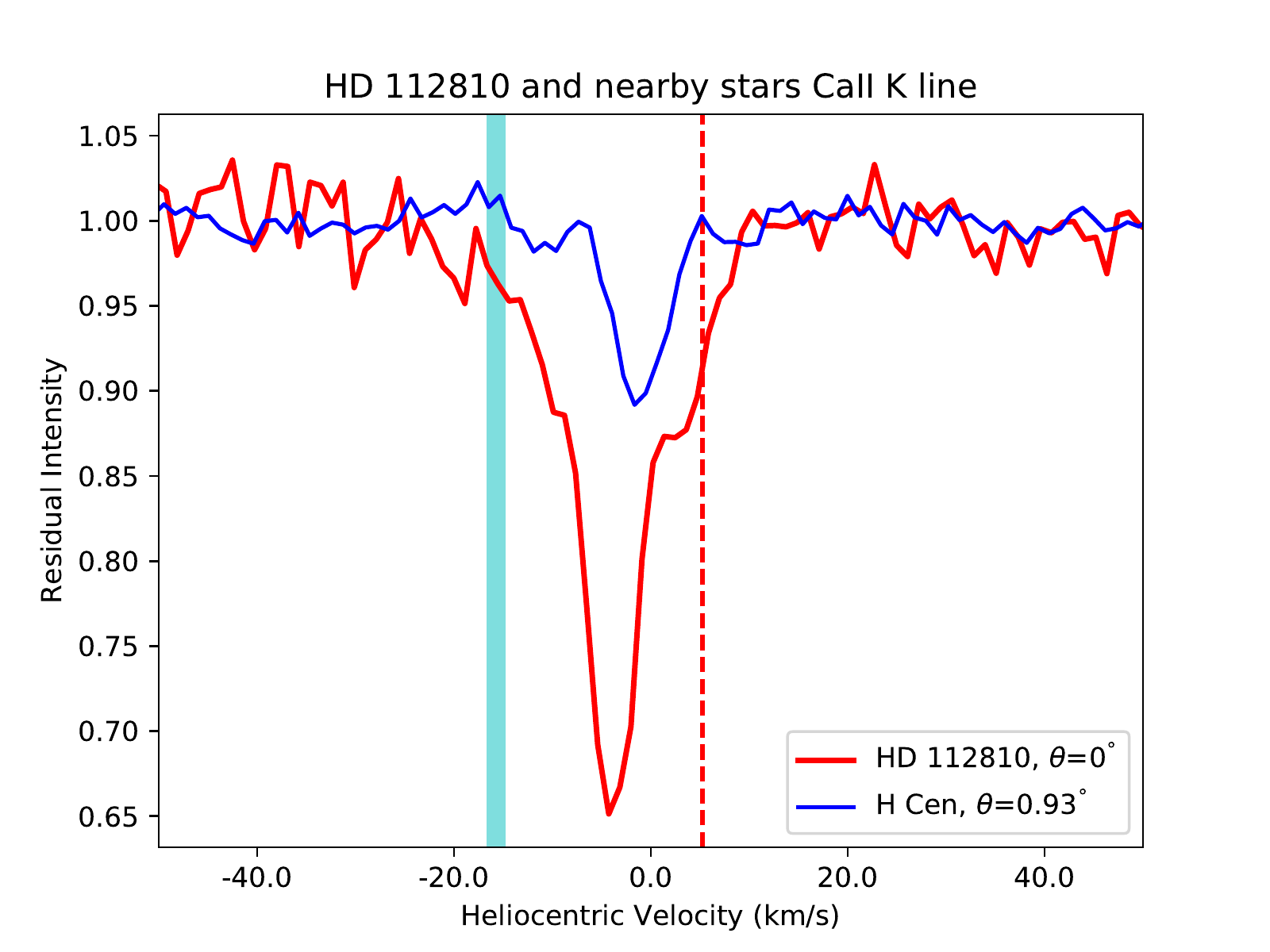}
\end{figure*}

\begin{figure*}  %-> funciona bien con pdf
\includegraphics[width=0.49\textwidth]{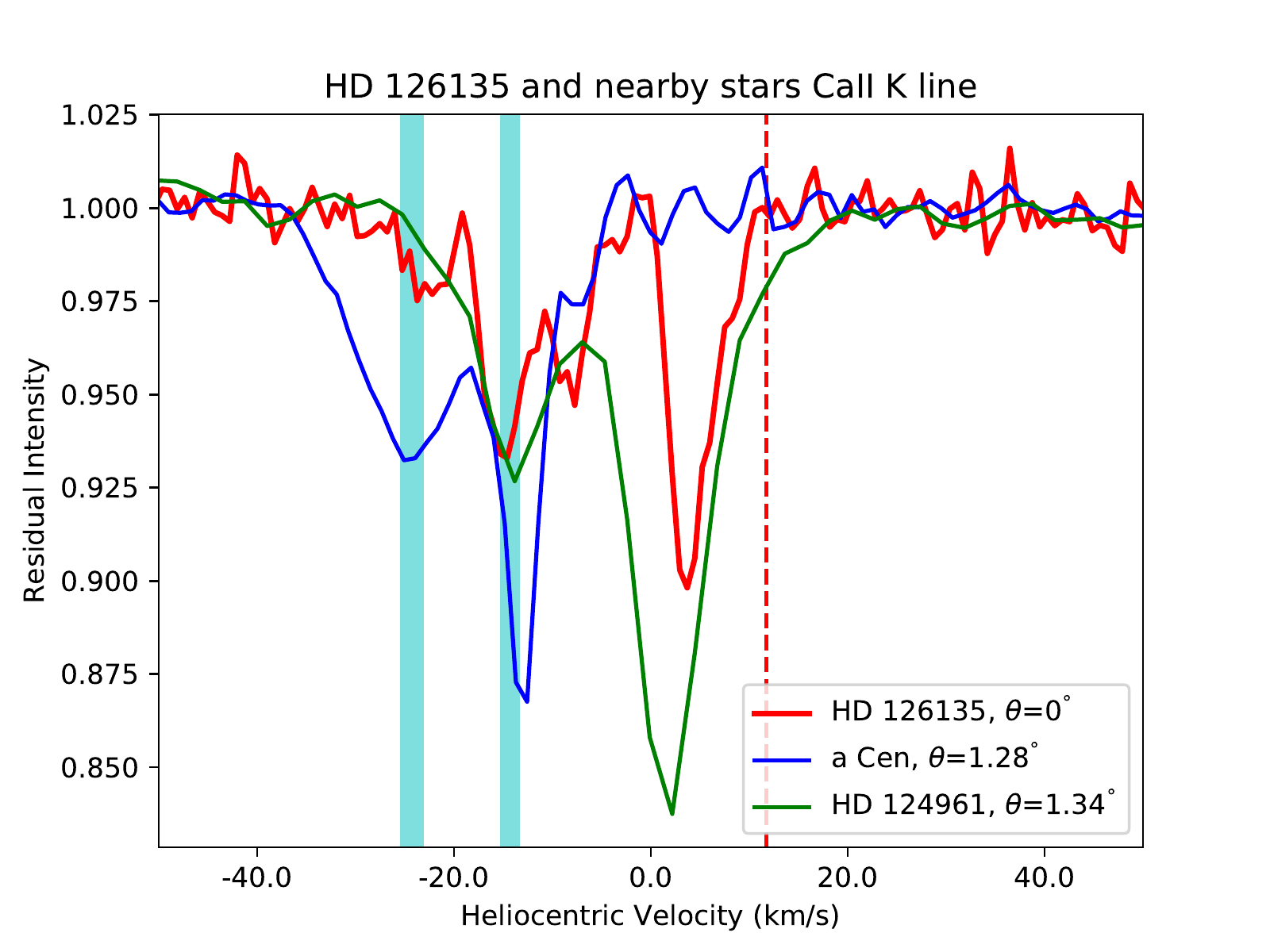}
\includegraphics[width=0.49\textwidth]{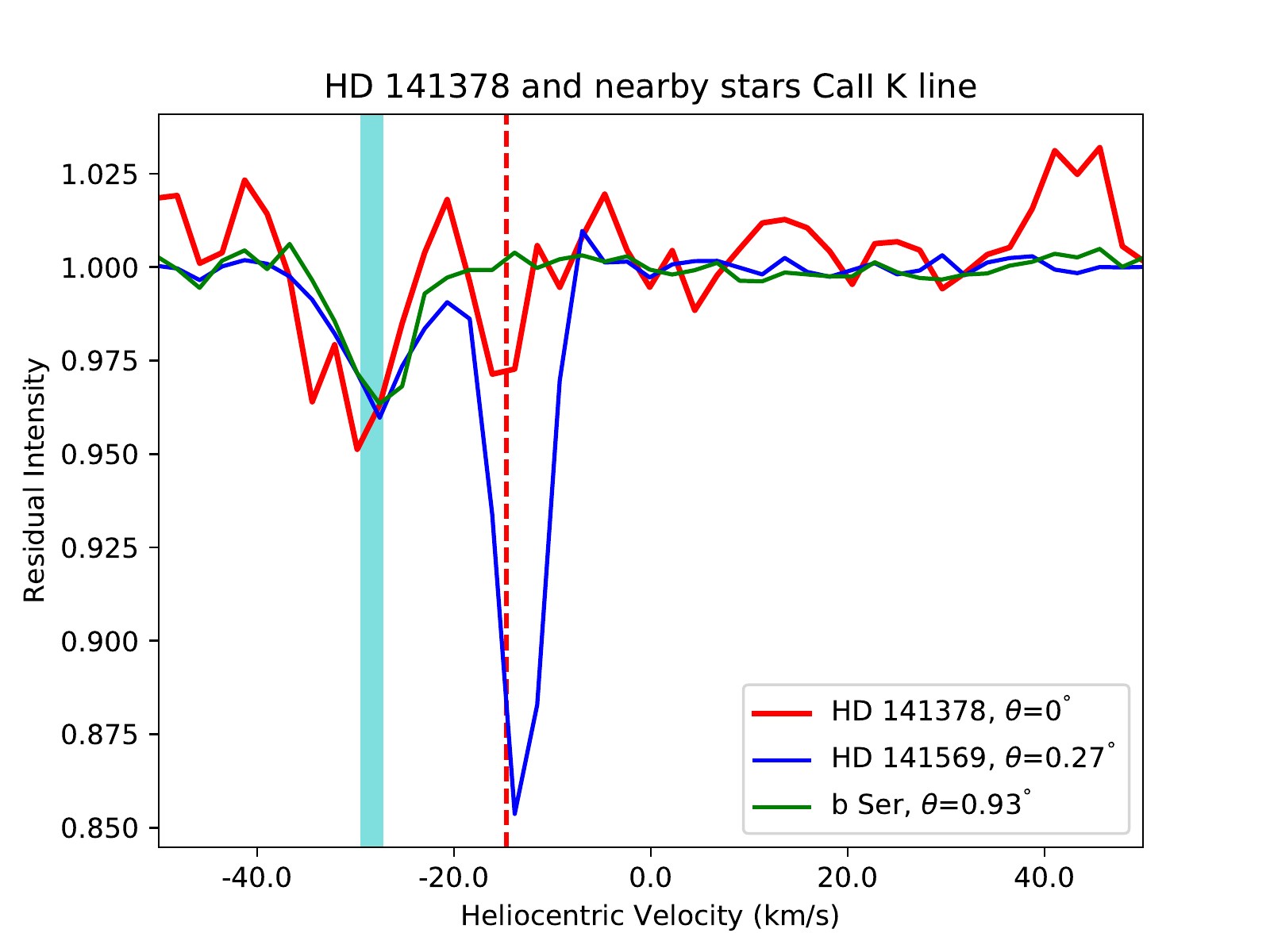}
\end{figure*}

\begin{figure*}  %-> funciona bien con pdf
\includegraphics[width=0.49\textwidth]{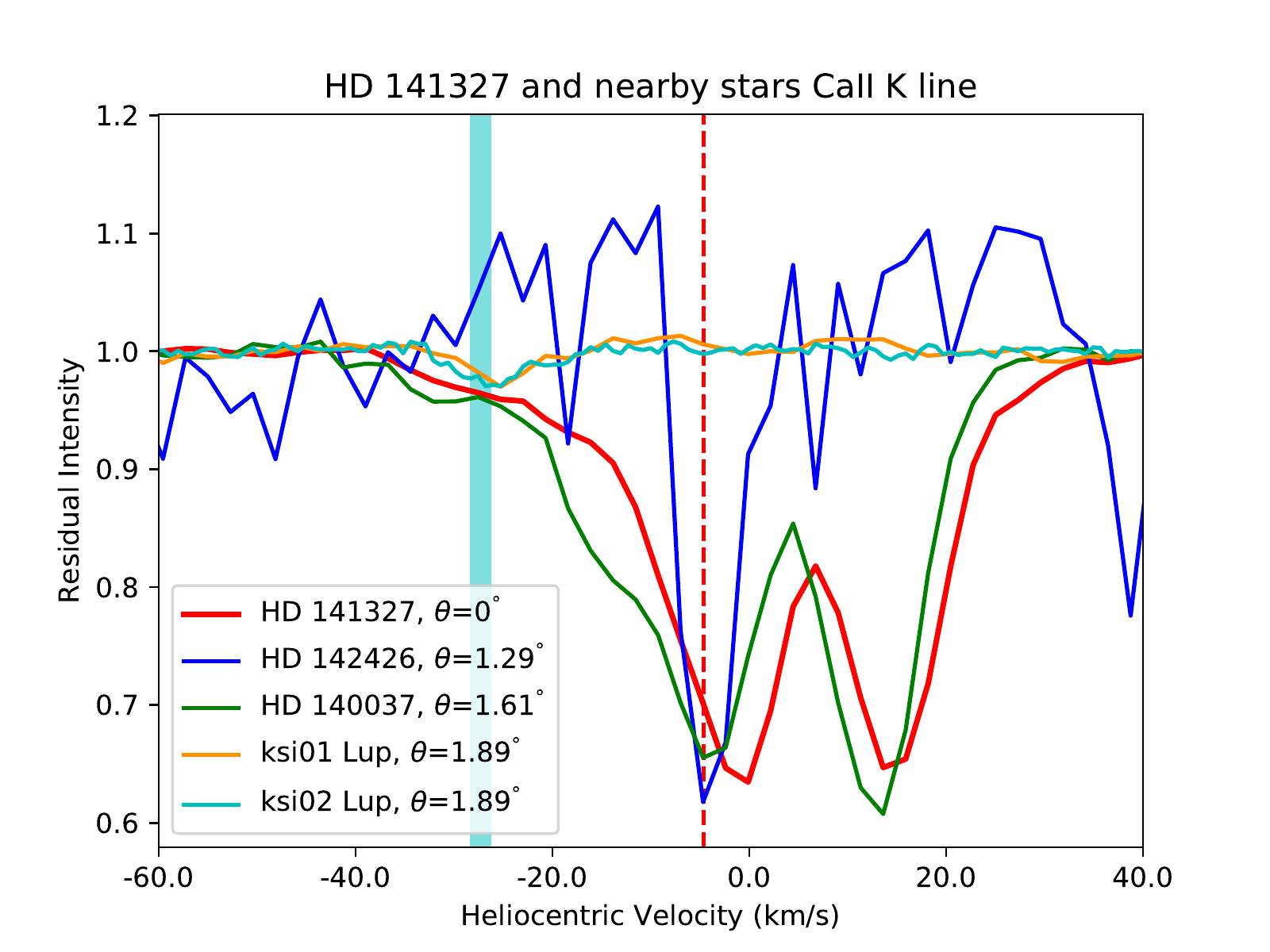}
\includegraphics[width=0.49\textwidth]{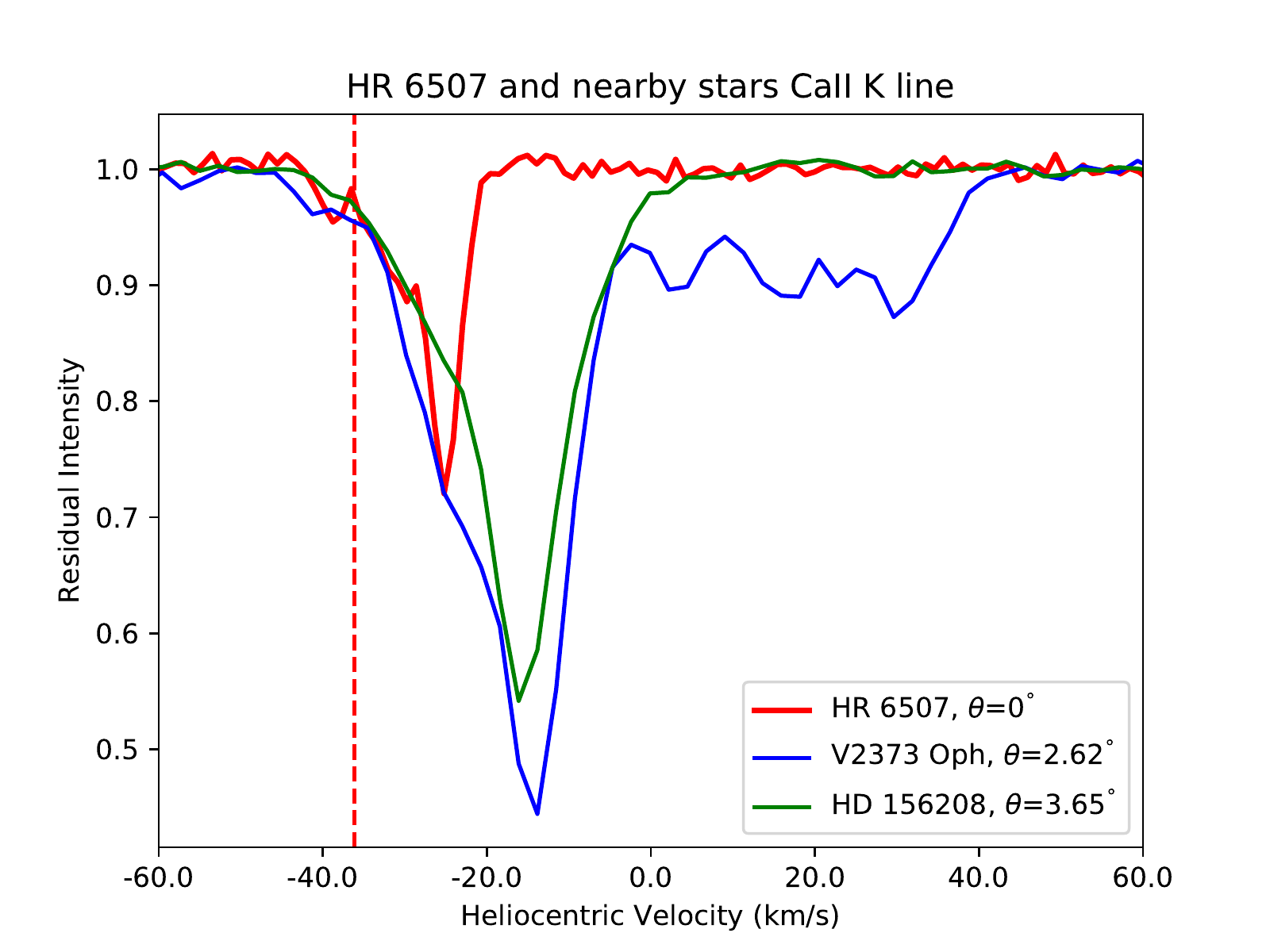}
\caption{Nearby stars around HD\,110058, HD\,112810, HD\,126135, HD\,141378, HD\,141327 and HR\,6507. Dashed line marks the estimated radial velocity of the star and cyan lines mark the velocity of the traversing clouds in the line of sight with their respective errors as their line widths. }
\label{fig:near3}
\end{figure*}

\begin{figure*}  %-> funciona bien con pdf
\includegraphics[width=0.49\textwidth]{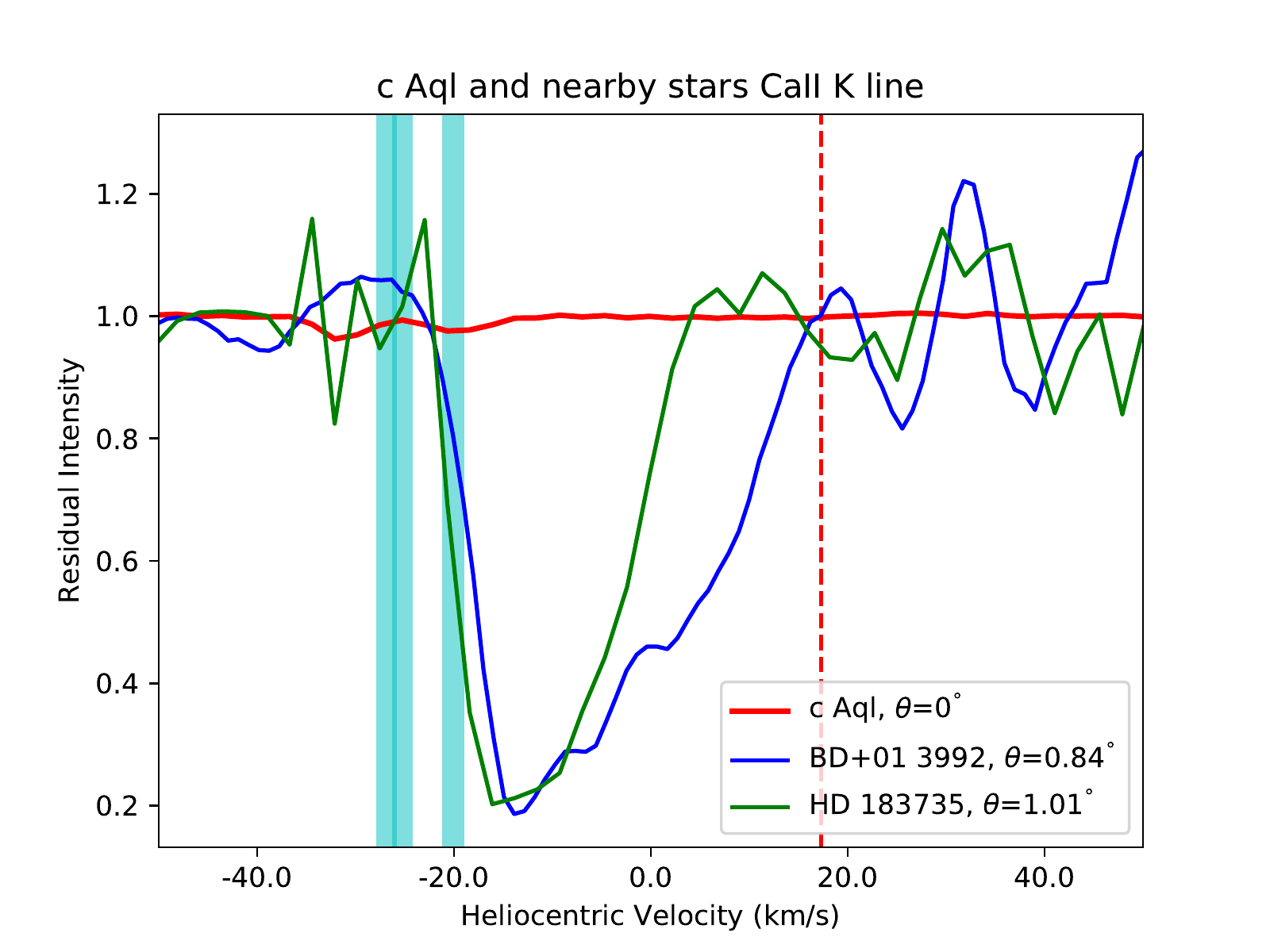}
\caption{Nearby stars around c Aql. Dashed line marks the estimated radial velocity of the star and cyan lines mark the velocity of the traversing clouds in the line of sight with their respective errors as their line widths.}
\label{fig:near4}
\end{figure*}

\onecolumn
 
\begin{center}
\begin{longtable}{lcccc}

	\caption{All nearby stars used in the analysis per each object of science, their number of spectra, instrument and ESO program ID. Note that HD\,290540, HD\,36444 and HD\,290609 are each other's nearby stars and in addition have three other common nearby stars.}
	\label{tab:nearby} \\
		\hline
		Object & Nearby star & Number of Spectra & Instrument & Program ID \\
		\hline
		\endfirsthead
		
		 	\multicolumn{5}{c}{{\bfseries \tablename\ \thetable{} -- continued from previous page}} \\
 		\hline 
		Object              & Nearby star & Number of Spectra & Instrument & Program ID \\
 	\hline
 	\endhead
 	
 	\hline \multicolumn{5}{r}{{Continued on next page}} \\ 
	\endfoot
	
	\hline 
	\endlastfoot
$\beta$03\,Tuc & $\beta$01\,Tuc & 6 & HARPS & 073.C-0733(E) \\ 
 &  & 2 & HARPS & 075.C-0689(A) \\ 
 &  & 2 & HARPS & 077.C-0295(A) \\ 
 &  & 2 & HARPS & 077.C-0295(B) \\ 
 &  & 3 & FEROS & 094.A-9012(A) \\ 
$\nu$\,Hor & HR\,798 & 1 & FEROS & 179.C-0197(B) \\ 
 & HR\,762 & 4 & FEROS & 088.C-0498(A) \\ 
HD\,290540, HD\,36444 and HD\,290609 & HR\,1861 & 1 & FEROS & 074.B-0455(A) \\ 
 & HR\,1863 & 12 & UVES & 266.D-5655(A) \\ 
 & VVV\,Ori & 98 & UVES & 194.C-0833(C) \\ 
 &  & 6 & FEROS & 096.A-9030(A) \\ 
 &  & 2 & FEROS & 096.A-9024(A) \\ 
HR\,1919 & VV350\,Ori & 2 & UVES & 082.C-0831(A) \\ 
 & HD\,38735 & 79 & FEROS & 084.C-1008(A) \\ 
 &  & 24 & FEROS & 084.A-9004(B) \\ 
 &  & 4 & FEROS & 091.D-0414(B) \\ 
HD\,54341 & L01\,Pup & 1 & FEROS & 088.A-9003(A) \\ 
HD\,60856 & HD\,61045 & 8 & UVES & 072.D-0410(A) \\ 
 & HD\,60995 & 6 & UVES & 098.C-0463(A) \\ 
HR\,3300 & HD\,70731 & 7 & UVES & 093.D-0852(A) \\ 
 & HD\,71722 & 5 & HARPS & 094.C-0946(A) \\ 
 &  & 12 & FEROS & 094.A-9012(A) \\ 
$\eta$\,Cha & HD\,75505 & 7 & FEROS & 084.A-9003(A) \\ 
 &  & 1 & FEROS & 086.A-9006(A) \\ 
 & 9\,Cha & 310 & FEROS & 078.D-0549(A) \\ 
 &  & 3 & FEROS & 084.A-9003(A) \\ 
HD\,92536 & tet\,Car & 1 & FEROS & 073.D-0291(A) \\ 
 &  & 1 & FEROS & 074.D-0300(A) \\ 
 &  & 15 & UVES & 076.C-0503(A) \\ 
 &  & 80 & UVES & 077.C-0547(A) \\ 
 &  & 1 & FEROS & 078.D-0080(A) \\ 
 &  & 125 & UVES & 194.C-0833(A) \\ 
 & VV407\,Car & 9 & FEROS & 086.D-0449(A) \\ 
 & HD\,93738 & 2 & FEROS & 096.A-9018(A) \\ 
 & VV364\,Car & 6 & FEROS & 086.D-0449(A) \\ 
3\,Crv & zet\,Crv & 1 & FEROS & 179.C-0197(D) \\ 
HD\,106036 & zet\,Cru & 1 & FEROS & 090.D-0358(A) \\ 
HR\,4796 & u\,Cen & 30 & FEROS & 60.A-9700(A) \\ 
 &  & 121 & HARPS & 60.A-9036(A) \\ 
 &  & 120 & HARPS & 60.A-9700(G) \\ 
HD\,110058 & tau\,Cen & 8 & HARPS & 076.C-0279(A) \\ 
 &  & 4 & HARPS & 076.C-0279(C) \\ 
 &  & 1 & FEROS & 078.D-0080(A) \\ 
 &  & 20 & UVES & 087.D-0010(A) \\ 
 & HD\,110484 & 3 & FEROS & 083.C-0139(A) \\ 
 & HR\,4871 & 1 & FEROS & 078.D-0080(A) \\ 
 &  & 1 & FEROS & 087.C-0227(C) \\ 
 &  & 13 & HARPS & 088.C-0353(A) \\ 
 &  & 13 & HARPS & 089.C-0006(A) \\ 
 & sig\,Cen & 1 & FEROS & 082.B-0484(A) \\ 
 &  & 40 & FEROS & 084.B-0029(A) \\ 
HD\,112810 & H\,Cen & 10 & UVES & 266.D-5655(A) \\ 
 &  & 2 & HARPS & 185.D-0056(A) \\ 
 &  & 1 & HARPS & 185.D-0056(C) \\ 
HD\,126135 & a\,Cen & 14 & UVES & 266.D-5655(A) \\ 
 &  & 64 & UVES & 073.D-0504(A) \\ 
 &  & 3 & HARPS & 075.C-0234(A) \\ 
 &  & 3 & HARPS & 079.C-0170(A) \\ 
 &  & 18 & UVES & 081.C-0475(A) \\ 
 &  & 4 & UVES & 097.D-0035(A) \\ 
 & HD\,124961 & 2 & FEROS & 072.D-0021(A) \\ 
 &  & 2 & FEROS & 073.D-0049(A) \\ 
 &  & 2 & FEROS & 082.D-0061(A) \\ 
HD\,141378 & HD\,141569 & 1 & UVES & 075.C-0637(A) \\ 
 &  & 109 & UVES & 079.C-0789(A) \\ 
 &  & 1 & FEROS & 083.A-9003(A) \\ 
 &  & 7 & FEROS & 085.A-9027(B) \\ 
 & b\,Ser & 10 & UVES & 076.B-0055(A) \\ 
 &  & 8 & HARPS & 077.C-0295(A) \\ 
 &  & 2 & HARPS & 077.C-0295(C) \\ 
 &  & 2 & FEROS & 083.A-9014(A) \\ 
 &  & 2 & FEROS & 083.A-9011(B) \\ 
 &  & 1 & FEROS & 083.A-9014(B) \\ 
 &  & 3 & FEROS & 084.A-9011(B) \\ 
 &  & 3 & FEROS & 085.A-9027(G) \\ 
HD\,141327 & HD\,142426 & 6 & FEROS & 089.D-0097(B) \\ 
 &  & 2 & FEROS & 090.D-0061(B) \\ 
 &  & 2 & FEROS & 091.D-0145(A) \\ 
 & HD\,140037 & 1 & FEROS & 179.C-0197(C) \\ 
 &  & 1 & FEROS & 091.C-0713(A) \\ 
 & ksi01\,Lup & 11 & FEROS & 075.D-0342(A) \\ 
 &  & 6 & HARPS & 075.C-0689(A) \\ 
 &  & 2 & HARPS & 075.C-0689(B) \\ 
 &  & 2 & HARPS & 077.C-0295(D) \\ 
 & ksi02\,Lup & 2 & HARPS & 075.C-0689(B) \\ 
 &  & 2 & HARPS & 077.C-0295(D) \\ 
 &  & 2 & HARPS & 077.C-0295(C) \\ 
 &  & 12 & HARPS & 184.C-0815(F) \\ 
HR\,6051 & HD\,146029 & 1 & FEROS & 179.C-0197(A) \\ 
 & HD\,144822 & 1 & FEROS & 077.C-0138(A) \\ 
HR\,6507 & V2373\,Oph & 9 & FEROS & 091.D-0122(A) \\ 
 & HD\,156208 & 8 & FEROS & 081.D-2002(A) \\ 
c\,Aql & BD+01\,3992 & 10 & UVES & 293.D-5036(A) \\ 
 & HD\,183735 & 1 & FEROS & 083.D-0034(A) \\ 
		\hline
\end{longtable}
\end{center} 

%\twocolumn

%%%%%%%%%%%%%%%%%%%%%%%%%%%%%%%%%%%%%%%%%%%%%%%%%%

% Don't change these lines
\bsp	% typesetting comment
\label{lastpage}
\end{document}